\documentclass{JFM-FLM_Au}
\usepackage{graphicx}
\usepackage{amsmath}
\usepackage[version=4]{mhchem}
\usepackage{siunitx}
\usepackage{longtable,tabularx}
\usepackage{ulem}

\renewcommand\eqref[1]{Equation~\ref{#1}}

\newcommand{\pts}{\left(}
\newcommand{\ptd}{\right)}

\newcommand{\funz}[1]{\!\pts #1 \ptd}
\newcommand{\pt}[1]{\pts #1 \ptd}

\usepackage{graphicx} 
\usepackage{multirow} 
\newcommand*{\fref}[1]{{figure~\ref{#1}}} 
\newcommand{\red}[1]{{\color{black}{#1}}}
\usepackage{enumitem}

\lefttitle{F. Scarano, A. Paduano and F. Avallone}
\righttitle{Journal of Fluid Mechanics}

\title{Noise dissipation mechanisms of an acoustic liner under grazing flow}

\author{Francesco Scarano\aff{1}, Angelo Paduano\aff{1} \and Francesco Avallone\aff{1}}

\affiliation{\aff{1}Department of Mechanical and Aerospace Engineering (DIMEAS), Politecnico di Torino, Corso Duca degli Abruzzi, 24, 10129 Turin, Italy}

\corresau{Francesco Scarano, \email{francesco.scarano@polito.it}}

\begin{document}
\maketitle

\begin{abstract}
High-fidelity lattice–Boltzmann very-large-eddy simulations are performed to describe the noise dissipation mechanisms in \red{a single cavity} acoustic liner subjected to grazing turbulent flow at a \red{centreline} Mach number of 0.3 and plane acoustic waves. The study examines the effects of sound pressure level (ranging from 130 to 160 dB) and \red{source} frequency, as well as of the direction of acoustic-wave propagation relative to the grazing flow. The acoustic energy dissipation mechanisms are the viscous losses \red{within the shear layer forming} along the internal walls of the orifice and the vortex-shedding. The latter is quantified through Howe’s energy corollary.
In the absence of grazing flow, acoustic energy is dissipated almost equally during both inflow and outflow phases, with vortex shedding dominating at high SPL and viscous losses at low SPL. The introduction of a grazing flow alters the flow topology; in particular, the shear layer past the orifice generates a quasi-steady vortex that confines the acoustic-induced flow to the downstream half of the orifice. This topological change alters the two noise dissipation mechanisms: viscous losses increase at low SPL because the grazing flow pushes the fluid toward the downstream lip of the orifice; vortex shedding becomes phase dependent, dissipating acoustic energy during the inflow phase and generating acoustic energy during the outflow phase. This explains why the net acoustic dissipation decreases in the presence of grazing flow, highlighting the crucial role of near-wall flow topology on liner performances.
\end{abstract}

\begin{keywords}

\end{keywords}


\section{Introduction}
\noindent
Acoustic liners are extensively used in aircraft engines as a passive noise control devices to reduce noise emissions \citep{Motsinger1991, Winkler2021}; they are usually installed in the intake of engines and in the core jet section.  
The noise source in aircraft engines consists of two main components: a tonal component at the blade-passing frequency (BPF), and a broadband component generated by turbulence impingement, which arises from the close proximity of the fan to the stator stage \citep{Mallat1989ARepresentation, Hughes2011TheIntegration, Casalino2018TurbofanMethod}.
Recent development of ultra high bypass ratio engines, characterized by a larger fan diameter compared to traditional high bypass ratio engines, has significantly increased the contribution of fan noise to the overall engine noise.

The simplest acoustic liner is the single-degree-of-freedom (SDOF) one, which comprises a cavity backing and a perforated face-sheet \citep{Motsinger19914Treatment}. In the absence of grazing turbulent flow and high-intensity acoustic waves, a conventional SDOF liner behaves like a Helmholtz resonator. The presence of acoustic waves at a frequency close to that of resonance excites an acoustic-induced flow within the orifice that leads to acoustic dissipation \citep{tam_microfluid_2000}.
The resonant frequency of the liner is typically tuned to coincide with the fan's BPF or its harmonics, making SDOF liners particularly suitable for fan noise attenuation. The response of acoustic liners is typically characterized through the impedance, which depends on geometric parameters, sound pressure level (SPL), and flow conditions \citep{Bonomo2023AProfiles}.

The physics of acoustic liners and their dissipation mechanism is well-established when they are exposed solely to acoustic waves \citep{Melling1973THELEVELS, tam_microfluid_2000}, however a gap persists in the understanding and modelling noise dissipation when the liners operate under real working conditions, i.e. in the presence of both acoustic wave and grazing turbulent flow at high Mach numbers and SPL. 
In the absence of grazing flow, impedance can be directly linked with the absorption coefficient and thus energy dissipation, however this relation does not hold when the liner works in the presence of grazing flow and at high SPL \citep{Tam2010}.

In the absence of grazing flow, energy dissipation in a SDOF liner occurs through viscous losses and via vortex shedding mechanisms. For moderate SPLs, typically below $140~\mathrm{dB}$, viscous dissipation dominates; it occurs along the internal surfaces of the orifices where laminar boundary layers develop \citep{tam_microfluid_2000}. 
As the SPL increases and the operating regime transitions into a fully non-linear one, the acoustic-induced flow topology is characterized by the emergence of turbulent jets and the shedding of vortices at the orifice entrances \citep{Zhang_Bodony_2012, tam_microfluid_2000}. Here, acoustic energy is transformed into turbulent kinetic energy associated with the rotational motion of vortices, which is ultimately dissipated as heat through viscous processes \citep{tam_microfluid_2000}. According to \citet{tam_microfluid_2000}, vortex shedding is amplified near the liner’s resonant frequency but remains largely unaffected by the angle of incidence of the acoustic waves. Recent experimental investigations without grazing flow, such as that by \citet{tang_piv_2024}, have revealed the presence of multi-scale vortex structures driven by an acoustic excitation at high SPL.

A detailed understanding of the acoustic response of liners subjected to high-speed turbulent grazing flow and acoustic waves remains incomplete. Under these conditions, liners operate always in the non-linear regime despite the SPL at which they are exposed, and the coupling between flow and acoustics becomes increasingly complex. In particular, the spatial and temporal evolution of the velocity field inside the orifices deviates substantially from the no-flow case \citep{Zhang2016}. 
The interaction between the acoustic-induced flow field and the grazing flow was visualized for the first time by  \citet{KennethBaumeister1975NASAOrifice}. They identified the presence of a vortex at the upstream side of the orifice neck, leading to a reduction in the effective inflow area. 
Subsequent computational studies have explored the flow physics inside the orifice in greater detail. Initial investigations focused on simplified configurations without grazing flow \citep{tam_microfluid_2000}, and later extended to include grazing flow conditions. 

\citet{Shahzad2023DirectLiners} performed direct numerical simulations of a turbulent grazing flow over a lined wall in the absence of acoustic forcing, revealing substantial modifications of the near-wall flow topology compared to a smooth surface. This was linked with an increase in drag.
Fully three-dimensional numerical simulations examining the interaction between acoustic waves and grazing flow over an acoustic liner were performed by \citet{Zhang2016}. The simulations elucidated the intricate coupling between the flow field and the liner’s acoustic response, and how this coupling depends on the state of the incoming boundary layer, laminar or turbulent. Their results showed that the influence of boundary-layer parameters is more pronounced at low SPL.
In addition, they showed the ejection of vorticity both outside the liner and into the liner cavity through the orifice, an effect that becomes increasingly pronounced at high SPL, where the influence of the grazing flow weakens.
Using micro–particle image velocimetry, \citet{Leon2019} investigated how a turbulent grazing flow interacts with a conventional acoustic liner under acoustic excitation. At low SPL, the grazing flow remained essentially unaffected, whereas at higher SPL the acoustic forcing produced mean-flow distortions above the orifice and synthetic-jet-like motions penetrating deeply into the grazing flow.
The local state of the turbulent grazing flow, in turn, strongly influences the spatial decay of the SPL along a lined duct.
Recent studies by \citet{paduano2025impact} confirmed that the SPL decays less in the presence of flow over an acoustic liner, suggesting that the underlying noise dissipation mechanisms are locally modified by the \red{near-wall flow features}.

Despite extensive research efforts, a comprehensive description of the flow dynamics within the liner's orifices and a reliable quantification of acoustic dissipation mechanisms under grazing flow are still lacking. This challenge arises mainly from the difficulty of resolving fine-scale flow features, such as the vorticity field and wall shear stresses on the perforated plate. High-fidelity numerical simulations, complemented by controlled experiments, therefore offer a powerful approach to investigate these phenomena and deepen our understanding of the noise dissipation mechanisms.

In this study, we aim to quantify the contributions of both vortex shedding and viscous dissipation to the overall acoustic energy dissipation. 
We compare configurations with and without grazing flow and explore the effect of varying SPL, source frequency and acoustic propagation direction, covering both the linear and non-linear regimes. The main research questions guiding this work are:
\begin{itemize}
    \item How does the presence of grazing flow alter the dissipation of acoustic energy in an acoustic liner?
    \item How does the change in flow topology inside the orifice influence the dissipation mechanism?
    \item How do the relative contributions of viscous effects at the orifice walls and vortex shedding vary in the presence of grazing flow, and what is the impact of changing the SPL?
\end{itemize}

The dissipation by viscous effects at the orifice walls is evaluated following the approach proposed by \citet{tam_microfluid_2000} while the dissipation by vortex shedding is evaluated applying Howe's energy corollary \citep{howe_dissipation_1980}.
The Howe's energy corollary has been previously applied to highlight the mechanisms of generation and absorption of sound in air-jet instruments relying on both numerical \citep{tabata_three-dimensional_2021} and experimental \citep{yoshikawa_experimental_2012} data.

The analysis are applied to high-fidelity data obtained with lattice-Boltzmann very-large-eddy simulations (LB/VLES) of a fully resolved liner configuration, consisting of a single cavity and a perforated plate with multiple orifices. This configuration reduces computational cost while focusing on the dissipation mechanisms of interest. 
By analyzing a single orifice within one cavity, we can isolate the direct effect of the grazing flow itself from the additional complexities introduced by the development of the flow over a full liner \citep{paduano2025impact, Shahzad_2025AIAA}.
The liner geometry replicates that used in previous experimental campaigns conducted in the Grazing Flow Impedance Tube (GFIT) facility at NASA Langley Research Center \citep{Jones2010}, as well as in the numerical studies by \citet{Zhang2016}.

The paper is organised as follows. Section~2 describes the data reduction methodology, the technique to evaluate the acoustic-induced velocity, and the methods used to quantify dissipation due to vortex shedding and viscous effects. Section~3 outlines the numerical setup, solver details, geometry, and the simulation test matrix.  Section~4 focuses on the flow field inside the orifices, examining both the acoustic-induced velocity distribution inside the orifice and the shear layer development at the orifice mouth. Section~5 presents the \red{parametric} analysis of the dissipation mechanisms \red{when varying the SPL value}, \red{while Section 6 investigates the effect of changing source frequency, propagation direction and the interference between two consecutive orifices}. Section~7 concludes the paper.

\section{Data reduction methodology}
\label{sec:methodology}

\red{

\subsection{Nomenclature for velocity components}
\noindent

Considering the time-resolved nature of the analysis, the instantaneous
velocity field is denoted by $\boldsymbol{u}'$ and is decomposed as
\begin{equation}
\boldsymbol{u}' = \boldsymbol{U} + \boldsymbol{u},
\end{equation}
where $\boldsymbol{U}$ is the mean velocity and $\boldsymbol{u}$ represents
the fluctuating component. The instantaneous velocity components
$(u',v',w')$ in the three spatial directions $(x,y,z)$.
Accordingly, $U, V, W$ are the time-averaged velocity components, and $u, v, w$ the fluctuating
components following the Reynolds decomposition.

In the presence of acoustic forcing, the fluctuating velocity $\boldsymbol{u}$
includes two contributions: the turbulent (aerodynamic) fluctuations and the
acoustic-induced velocity $\boldsymbol{u}_{\mathrm{ac}}$, which arises from the periodic forcing \citep{Scarano_decompositions}.

\subsection{Acoustic dissipation by viscous effects}
\noindent

The rate of mechanical energy dissipation due to viscosity is evaluated from
the instantaneous velocity field following the approach of
\citet{tam_microfluid_2000}. In a three-dimensional framework, the local
volumetric viscous dissipation density is defined as
\begin{equation}
\Phi(\boldsymbol{x},t)
=
\sigma_{ij}(\boldsymbol{x},t)
\frac{\partial u'_i}{\partial x_j},
\label{eq:viscous_density}
\end{equation}
where the viscous stress tensor is
\begin{equation}
\sigma_{ij}(\boldsymbol{x},t)
=
\mu
\left(
\frac{\partial u'_i}{\partial x_j}
+
\frac{\partial u'_j}{\partial x_i}
\right).
\label{eq:stress_tensor}
\end{equation}
where $\mu$ is the dynamic viscosity, and $i,j=1,2,3$ are the indices of the three spatial directions and $u'_i$ is the  instantaneous velocity component and $\boldsymbol{x}= (x,y,z)$ denotes the spatial position vector.
The quantity $\Phi(\boldsymbol{x},t)$ has units of power density
($\mathrm{W/m^3}$) and represents the irreversible conversion of mechanical
energy into internal energy due to viscous effects.

The imposed acoustic forcing induces an oscillatory motion within the orifice, which in turn generates time-dependent velocity gradients within the near-wall region of the orifice; these oscillations are responsible for the additional viscous dissipation associated with the acoustic field.
To isolate the acoustically induced viscous losses, and exclude the viscous dissipation associated solely to the aerodynamic development of the turbulent grazing flow, the integration volume is restricted to a near-wall region adjacent to the orifice side walls, extending up to approximately 30 wall units (corresponding to $\Delta x/d < 0.09$). 
This choice follows the formulation adopted for resonant liners, where the portion of acoustic energy dissipated into heat by viscous effects is primarily attributed to near-wall processes
within the orifice neck \citep{tam_microfluid_2000}. The adopted control
volume therefore captures both the near-wall gradients provided by the
wall model and the viscous stresses within the boundary layer inside the orifice,
including turbulent contributions injected from the grazing flow.

The three-dimensional integration domain is shown in Appendix \ref{appD}.
The viscous dissipation rate in a control volume $V$ enclosing the
orifice and the region in proximity of the orifice internal walls is therefore
\begin{equation}
D(t)
=
\iiint_V
\Phi(\boldsymbol{x},t)
\,\mathrm{d}V.
\label{eq:viscous_volume}
\end{equation}

Dissipation occurring in the bulk flow through the orifice, associated with the subsequent
decay of vortical structures through the turbulent cascade, is not
directly included in the viscous term. Its energetic contribution is
instead represented through the vortex-shedding analysis based on Howe’s
corollary, which quantifies the transfer of acoustic energy into
vortical motion.

The phase-dependent viscous dissipation rate is obtained by phase-averaging
$D(t)$ over one acoustic period and is denoted as $D(\phi)$. The net viscous
energy dissipated per unit time over one oscillation period $T$ is
\begin{equation}
E_{\mathrm{viscous}}
=
\frac{1}{T}
\int_0^T
D(\phi)\,\mathrm{d}\phi.
\label{eq:viscous_energy}
\end{equation}

All viscous dissipation related quantities are normalized in wall units using the
friction velocity $u_\tau=\sqrt{\tau_w/\rho}$, where $\tau_w$ is the wall shear stress and $\rho$ is the fluid density, and the viscous scale
$\delta_\nu = \nu/u_\tau$, where \( \nu \) the kinematic viscosity. 
These quantities are computed for the smooth baseline case without acoustic.
The local volumetric dissipation density is
normalized as
\begin{equation}
\Phi^{+}(\phi)
=
\frac{\Phi}{\rho u_\tau^4/\nu},
\end{equation}
while the volume integrated dissipation rate is normalized as
\begin{equation}
D^{+}(\phi)
=
\frac{D}{\rho \nu^2 u_\tau}.
\end{equation}

\subsection{Acoustic energy conversion into vortex shedding}
\noindent

The conversion of acoustic energy into vortical motion is quantified using
Howe's energy corollary
\citep{howe_contributions_1975, howe_dissipation_1980, howe_absorption_1984}.
The local volumetric power density transferred from the acoustic field to
the vortical field is given by
\begin{equation}
\Pi_g(\boldsymbol{x},t)
=
\rho
\left(
\boldsymbol{\omega}
\times
\boldsymbol{u}'
\right)
\cdot
\boldsymbol{u}_{\mathrm{ac}},
\label{eq:howe_density}
\end{equation}
where $\boldsymbol{\omega}$ is the vorticity,
and $\boldsymbol{u}_{\mathrm{ac}}$ is the acoustic-induced velocity fluctuation field and $\boldsymbol{u'}$ is the instantaneous velocity vector.

The total acoustic energy conversion rate within the control volume $V$
is defined as
\begin{equation}
\Pi(t)
=
\iiint_V
\Pi_g(\boldsymbol{x},t)
\,\mathrm{d}V.
\label{eq:howe_volume}
\end{equation}
The control volume is chosen such to include the shear layer forming at
the orifice mouth, the orifice and cavity regions, and the portion of the
grazing boundary layer interacting directly with the orifice. This ensures
that the acoustic–vortical interaction associated with the dominant
structures is fully captured.

The phase-dependent quantity \red{$\Pi^+(\phi)$} is obtained by
phase-averaging over one acoustic cycle. The net energy transferred per
unit time over one period $T$ is
\begin{equation}
E_{\mathrm{shedding}}
=
\frac{1}{T}
\int_0^T
\Pi(\phi)\,\mathrm{d}\phi.
\label{eq:howe_energy}
\end{equation}

The local power density converted into vortical field is normalized in wall units as
\begin{equation}
\Pi_g^{+}
=
\frac{\Pi_g}{\rho u_\tau^4/\nu},
\end{equation}
while the integrated conversion rate is normalized as
\begin{equation}
\Pi^{+}
=
\frac{\Pi}{\rho \nu^2 u_\tau}.
\end{equation}

A positive value of \red{$\Pi^+_g$} or $\Pi^+$ indicates absorption of acoustic energy
by the vortical field, whereas a negative value indicates acoustic energy
generation. 

}
\red{
\subsection{Acoustic-induced velocity estimation}

In order to evaluate Howe’s energy corollary, it is necessary to isolate the
velocity component induced by the imposed acoustic forcing from the
fluctuations generated by the turbulent grazing flow. The acoustic-induced velocity field is extracted using Spectral proper
orthogonal decomposition (SPOD) \citep{schmidt_guide_2020}, following the
methodology detailed in \citet{Scarano_decompositions}.

SPOD is applied using as input the fluctuating component of the velocity field. SPOD
provides a set of frequency-dependent orthogonal modes ranked by their
spectral energy content. The acoustic-induced velocity is reconstructed by
selecting the leading SPOD mode associated with the known forcing frequency
and performing a narrow band-pass reconstruction centred at that frequency.
This procedure isolates the velocity component that is coherent and
phase-locked with the imposed acoustic excitation.

As demonstrated in \citet{Scarano_decompositions}, for datasets characterised
by a single dominant acoustic forcing and in the absence of additional tonal
sources (e.g.\ self-sustained tones), SPOD yields results equivalent to those
obtained with the recently developed canonical correlation decomposition (CCD) by \cite{Lyu2024},
which identifies the acoustic velocity as the component that best
correlates with the forcing signal. Under these conditions, both approaches
provide a consistent estimate of the acoustically driven velocity field.

In many liner studies
(e.g.\ \cite{Zhang2012, Leon2019}), the acoustic velocity is obtained by
phase-locking the data using a triple-decomposition approach. However, in
the presence of grazing turbulent flow and high SPLs, the
SPOD-based approach has been shown to provide a more robust separation
between coherent acoustic motion and broadband turbulent fluctuations
\citep{Scarano_decompositions}, without assuming linearity or irrotationality.

It is important to clarify that the term ``acoustic-induced velocity'' refers
here to the velocity component coherent with the forcing frequency, rather
than to a strictly irrotational field.
A Helmholtz decomposition or a reconstruction based on the acoustic
Euler equations \citep{schoder_postprocessing_2020,unnikrishnan_pressure_2020} would require full three-dimensional information on the
velocity divergence and appropriate boundary conditions. Such requirements
are not satisfied for the two-dimensional extracted planes considered in the
parametric analysis, and the underlying assumptions are not strictly valid in
strongly nonlinear flow–acoustic interaction regimes. The SPOD-based
approach therefore provides a physically consistent and practically robust
method to extract the velocity component relevant to the acoustic
energy-transfer analysis carried out in this study.

}

\section{Numerical setup and database description}
\label{Sec:setup}
\subsection{Numerical solver}
\noindent

\red{The simulations are performed using a lattice Boltzmann method
(LBM) solver combined with a very large eddy Simulation (VLES) approach. This methodology is selected as a compromise between linearized
frequency-domain methods and fully resolved simulations.
Classical liner simulations, based on linearized governing equations, are very efficient for predicting the frequency-dependent liner response in terms of impedance under small-amplitude perturbations. However, by construction they neglect SPL-induced nonlinearities and therefore are not appropriate when the liner is exposed to high SPL. In addition, they commonly assume a prescribed base flow (often a steady sheared profile \citep{Myers1980, GABARD201630}) that is only weakly influenced by the liner surface and does not fully represent the mutual coupling between the grazing boundary layer and the liner treatment \citep{Tam2010}. 

At the opposite end, highly resolved approaches
such as direct numerical simulation (DNS) resolve the full nonlinear
flow–acoustic interaction, but their computational cost is prohibitive for
the Reynolds numbers and parametric space of practical interest.

From a practical standpoint, the LBM framework offers low numerical
dissipation for acoustic wave propagation, efficient parallelization, and automatic Cartesian meshing. These
features, combined with the VLES turbulence treatment, make it possible to
perform parametric studies over a wide range of SPLs and
frequencies at a tractable computational cost, while retaining the essential
physics of the flow–acoustic interaction.
}

The commercial software 3DS Simulia PowerFLOW 6-2019R4 is used to compute the flow \red{and acoustic} field. 
PowerFLOW simulation technology is based on a LB method with collision relaxation time and distribution function dynamically calibrated to the time scales of slow turbulent structures modelled through a turbulence transport model.
The solver has already been validated for canonical honeycomb liner configurations, both in the absence of flow \citep{Mann2013,Hazir2017} and  with grazing flow \citep{Manjunath2018,Avallone2019}. The same solver has also been used to simulate the effect of a SDOF liner installed on the nacelle of the NASA Source Diagnostic Test engine configuration \citep{Casalino2017}, and the predicted sound attenuation has recently been confirmed by another simulation carried out by using a different high-fidelity flow solver \citep{Shur2021}.


The LB scheme is based on an expansion of the distribution function $f\funz{{\boldsymbol{x}},{\bf \xi},t}$, say the probability density of finding particles at location ${\boldsymbol{x}}$, advected at velocity ${\bf \xi}$ at time $t$, solution of the Boltzmann equation, in a series of Hermite polynomials \citep{Shan2006a}. These constitute an orthogonal basis, which is particularly suited to describe a flow in the kinetic space. Indeed, the first four coefficients, from $0^{\text{th}}$ to $3^{\text{rd}}$ order, of the expansion of the Maxwellian distribution function $f^{(0)}$ at equilibrium are algebraically related to the moments of macroscopic flow, say mass, momentum, energy/momentum fluxes, and heat fluxes. An interesting property of a Hermite expansion is that the series can be truncated at a given order without altering the low-order coefficients; therefore an expansion of $f$ truncated at the order $N\!>\!3$ provides a unique representation of the macroscopic hydrodynamic status of a fluid.

A key component of PowerFLOW LB scheme is the usage of a regularized collision operator $\Omega_i$ in the non-dimensional lattice Boltzmann equation 
\begin{equation}
    f_i\funz{{\boldsymbol{x}}+{\bf \xi}_i,t+1}\!=\!f_i\funz{{\boldsymbol{x}},t}\!+\!\Omega_i,
\end{equation}
projected along the discrete particle velocity ${\bf \xi}_i$. Following \citet{Zhang2006}, the LB equation can be equivalently written as
\begin{equation}
 f_i\funz{{\boldsymbol{x}}+{\bf \xi}_i,t+1}\!=\!f^{(0)}_i\funz{{\boldsymbol{x}},t}\!+\!f^{(1)}_i\funz{{\boldsymbol{x}},t}\!+\!\Omega_i,   
\end{equation}
where $f^{(1)}_i$ is the perturbation. In conditions that are not very far from equilibrium, the collision operator is linearly related to the perturbation through coefficients that are negatively/inversely proportional to relaxation time $\tau_{ij}$ of the collision process along the discrete velocity direction $i$ due to chaotic motion along the direction $j$, say 
\begin{equation}
 f_i\funz{{\boldsymbol{x}}+{\bf \xi}_i,t+1}\!=\!f^{(0)}_i\funz{{\boldsymbol{x}},t}\!+\!\sum_j\pt{\delta_{ij}-1/\tau_{ij}}f^{(1)}_i\funz{{\boldsymbol{x}},t}.   
\end{equation}
If the perturbation is expanded in Hermite series, this starts by the second order term and can be truncated at the third order term to recover the macroscopic fluid status. The resulting regularized collision operator will therefore include only terms proportional to the second order Hermite polynomials, accounting for energy and momentum fluxes, and third order terms polynomials, accounting for heat fluxes. Finally, following \citet{Chen2014}, applying Galilean invariance to the collision operator results in a two-term regularized form, in which the two terms account for energy/momentum fluxes and heat fluxes, respectively, with corresponding relaxation times related to the macroscopic fluid viscosity and thermal conductivity.

Another important component of the present flow simulation methodology is related to turbulence modeling, which is key to tackle high Reynolds number flows. The way turbulence is accounted for in PowerFLOW is by modifying the relaxation time in the collision operator by considering the time scales related to the turbulent motion and to the strain rate and rotation of the resolved flow field. Moreover, the amount of turbulent kinetic energy is used to define the equilibrium state of the gas. \red{As discussed by \citet{Chen2004}, the expansion of kinetic theory from particles to
eddies leads to the fundamental observation that the Reynolds stresses, which are a
consequence of chaotic turbulent motion, have a nonlinear structure and are better
suited to represent turbulence in states far from equilibrium, such as in the
presence of distortion, shear, separation, curvature, or rotation. At first order,
the kinetic expansion recovers the classical eddy-viscosity form
\begin{equation}
\sigma_{ij}^{(1)}
=
2\nu_t S_{ij},
\label{eq:first_order_reynolds_stress}
\end{equation}
where
\begin{equation}
S_{ij}
=
\frac{1}{2}
\left(
\frac{\partial u'_i}{\partial x_j}
+
\frac{\partial u'_j}{\partial x_i}
\right)
\label{eq:strain_rate_tensor}
\end{equation}
is the resolved strain-rate tensor, and the turbulent viscosity can be expressed as
\begin{equation}
\nu_t
=
\frac{2}{3}k\tau_t.
\label{eq:turbulent_viscosity_relaxation_time}
\end{equation}
If the turbulent relaxation time is estimated from a two-equation model, such as the
RNG \(k\)--\(\epsilon\) model, one may write
\begin{equation}
\tau_t
=
\frac{3}{2}C_\mu\frac{k}{\epsilon},
\qquad
\nu_t
=
C_\mu\frac{k^2}{\epsilon}.
\label{eq:rng_turbulent_time_scale}
\end{equation}

However, although the relaxation time is computed using the two-equation
\(k\)--\(\epsilon\) re-normalization group, RNG, model in PowerFLOW
\citep{Yakhot1992, Teixeira1998}, this model is not employed to compute an
equivalent eddy viscosity in the same sense as in traditional Reynolds-Averaged
Navier--Stokes, RANS, formulations. In standard RANS models, \(\nu_t\) is used to
explicitly close the Reynolds stresses in the averaged momentum equations. In the
LB/VLES formulation, instead, the turbulence model is used to dynamically recalibrate
the Boltzmann relaxation process to the characteristic time scales of the unresolved
turbulent motion.

The key advantage of this kinetic interpretation is that the Reynolds stresses are
not restricted to a purely linear eddy-viscosity form. As shown by \citet{Chen2004},
a second-order expansion of the kinetic model gives
\begin{equation}
\sigma_{ij}
\approx
-\frac{2}{3}k\delta_{ij}
+ 2\nu_t S_{ij}
- 2\nu_t
\frac{D}{Dt}
\left(
\tau_t S_{ij}
\right)
- 6\frac{\nu_t^2}{k}
\left(
S_{ik}S_{kj}
-
\frac{1}{3}\delta_{ij}S_{kl}S_{kl}
\right)
+ 3\frac{\nu_t^2}{k}
\left(
S_{ik}\Omega_{kj}
+
S_{jk}\Omega_{ki}
\right),
\label{eq:second_order_reynolds_stress}
\end{equation}
where
\begin{equation}
\Omega_{ij}
=
\frac{1}{2}
\left(
\frac{\partial u'_i}{\partial x_j}
-
\frac{\partial u'_j}{\partial x_i}
\right)
\label{eq:rotation_rate_tensor}
\end{equation}
is the resolved rotation-rate tensor, and
\begin{equation}
\frac{D}{Dt}
=
\frac{\partial}{\partial t}
+
\mathbf{u'}\cdot\nabla
\label{eq:material_derivative}
\end{equation}
is the material derivative following the resolved flow. 
\eqref{eq:second_order_reynolds_stress} shows that the turbulent stress contains not
only the classical linear contribution \(2\nu_t S_{ij}\), but also memory,
strain--strain, and strain--rotation effects. The memory term accounts for the finite
time required by the turbulent eddies to relax toward equilibrium, while the nonlinear
terms describe the response of the unresolved turbulent motion to strong deformation
and rotation of the resolved flow.

Hence, no Reynolds stresses are explicitly added to the governing equations. Rather,
their effect emerges implicitly through the non-equilibrium momentum exchange
produced by the modified collision operator.}
\red{In the present study, the LBM equations are discretized on a Cartesian grid whose volume elements are voxels. The adopted numerical scheme is the D3Q19 lattice, where “D3” denotes three spatial dimensions and “Q19” the number of discrete velocity directions \citep{Qian1992LatticeEquation}. Accordingly, the total number of degrees of freedom (DOFs) is directly related to the number of voxels in the computational domain (and to the 19 distribution functions stored per voxel in the D3Q19 formulation).}

\red{

\subsection{Wall model}
The simulations employ a pressure-gradient-extended turbulent wall model (PGE-WM),
following the formulation of \citet{Teixeira1998}. This approach modifies the classical
logarithmic law-of-the-wall to account for the influence of streamwise pressure gradients
on the near-wall velocity distribution.

In the generalized formulation, the inner-scaled velocity is expressed as
\begin{equation}
    U^+ = \frac{1}{k} \ln\!\left(\frac{y^+}{A}\right) + B,
    \label{eq:lawofthewall}
\end{equation}
where $k$ and $B$ are the von Kármán constant and additive constant, respectively,
and $y^+=u_\tau y/\nu$ is the viscous wall-normal coordinate and $u_ \tau$ is the friction velocity.
The factor $A$ introduces a correction to the standard logarithmic profile in order to
capture the modification of the near-wall structure induced by pressure gradients.

Physically, the presence of a streamwise pressure gradient alters the development of the
boundary layer by modifying the local velocity gradient and effectively stretching or
compressing the inner-scaled profile. This effect is incorporated through the scaling
parameter $A$, defined as
\begin{align}
    A &= 1 + \frac{\beta\,\left|\frac{\mathrm{d}p}{\mathrm{d}s}\right|}{\tau_w},
    \qquad \text{if } \boldsymbol{\hat{u}'}\cdot\frac{\mathrm{d}p}{\mathrm{d}s} > 0, \\
    A &= 1, \qquad \text{otherwise},
\end{align}
where $\tau_w$ denotes the wall shear stress, $\mathrm{d}p/\mathrm{d}s$ is the streamwise
pressure gradient, $\boldsymbol{\hat{u}'}$ is the unit vector of the local velocity, and $\beta$ is a
length scale representative of the unresolved near-wall region. The condition
$\boldsymbol{\hat{u}'}\cdot\frac{\mathrm{d}p}{\mathrm{d}s} > 0$ identifies an adverse pressure gradient, for
which the pressure increases in the direction of the local near-wall flow. The use
of the unit vector ensures that this condition depends only on the direction of the
flow, and not on its magnitude. The magnitude of the correction is instead governed
by the non-dimensional ratio $\beta |\frac{\mathrm{d}p}{\mathrm{d}s}|/\tau_w$, which compares the
pressure-gradient contribution across the unresolved near-wall region with the
local wall shear stress.

This formulation enables the wall model to dynamically adjust to adverse or favourable
pressure-gradient conditions while preserving the structure of the logarithmic law in
the absence of pressure gradients.
Confidence in the numerical framework is provided by the work by \cite{paduano2025impact}, where
the same LB–VLES approach is applied to a more complex configuration involving a full
acoustic liner. That study demonstrates that this numerical framework accurately reproduces experimental
measurements of SPL decay and acoustic impedance \citep{Bonomo2023AProfiles}, indicating that the
overall noise dissipation properties are well captured.

}

\subsection{Computational domain \red{and liner geometry}}
\begin{figure}
    \centering
    \includegraphics[width=\textwidth]{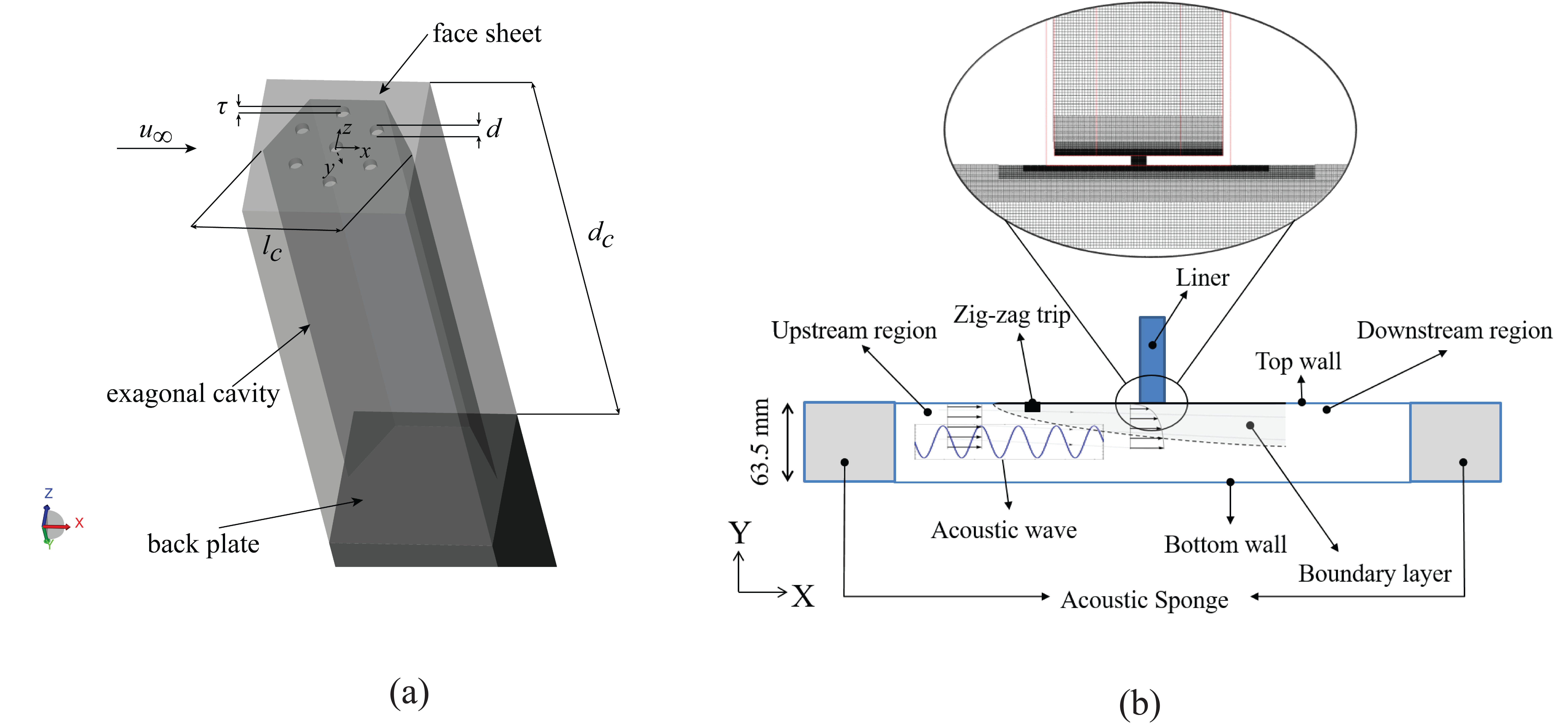}
    \caption{(a) Schematic of the cavity with representation of the coordinate reference system. The $\textit{y}$ axis is oriented towards the inside of the cavity. (b) Schematic of the computational setup with the grid in a plane crossing the central orifice.}
    \label{fig:Geometry_setup}
\end{figure}

\noindent
The acoustic liner geometry is similar as the one experimentally investigated by \citet{Jones2004} in the GFIT facility at NASA Langley Research Center at several free-stream Mach numbers and computationally investigated by \citet{Zhang2016} at free-stream Mach number equal to $0.5$. The liner is shown in Figure \ref{fig:Geometry_setup}. A rigid face sheet of thickness $\tau\!=\!0.64$~mm is perforated with cylindrical holes of diameter $d\!=\!0.99$~mm, which corresponds to a length-to-diameter ratio of $0.65$. A single honeycomb cavity with $7$ orifices is studied, thus resulting in a porosity of $\sigma\!=\!6.4\%$ very close to the reference study ($\sigma\!=\!8.7\%$). While in the experimental study the orifices are randomly located such that they can overlap neighbouring cavities, in this study, six orifices are placed at the center of each of the six equilateral triangles that form the hexagon and one at its centre. The cell depth is $d_c\!=\!38.10$~mm and the distance between the two opposite corners of the cell is $l_c\!=\!11.7$~mm. A rigid back plate closes the cell from below and the cell walls are rigid. 

The liner is placed along the top wall of a duct that has a rectangular cross section with height, $h_c$, equal to $63.5$~mm, as in the GFIT facility, while the width is restricted to $12$~mm to reduce the computational cost.

The acoustic liner is grazed by a turbulent flow with free stream Mach number equal to $M\!=\!0.3$. 
\red{The friction velocity upstream of the liner is equal to \SI{3.4}{m/s}. This value will be used in the rest of the paper to normalize the quantities in wall units.}
Transition to turbulence is forced using a zig-zag strip at $1750$~mm upstream of the liner. The zig-zag strip is $1$~mm thick, it has length and wavelength equal to $10$~mm and angle of $60^\circ$. The location of the zig-zag trip has been selected such to replicate the time-average turbulent boundary layer profile measured in the GFIT \citep{Jones2010} \red{(see \fref{fig:validation_Mprofile_deltastar} in the Appendix \ref{appA})}.
The velocity profile upstream of the liner has a displacement thickness $\delta^*$ equal to $2.3 \times 10^{-2}$~m and momentum thickness \red{$\theta$} equal to $1.96 \times 10^{-3}$~m. The flow has a Reynolds number based on the momentum thickness, \red{$Re_{\theta}$, equal to $14000$.} 
\red{Previous experimental and numerical studies have shown that a key parameter
controlling the interaction between grazing flow and liner acoustic response is the displacement
thickness $\delta^*$ of the impinging boundary layer (e.g.\ \cite{Quintino2025}). The value
of $\delta^*$ directly affects the shear-layer topology at the orifice mouth, the effective
blockage of the orifices, and consequently the acoustic impedance and dissipation mechanisms
of the liner. For this reason, the present study focuses on a configuration for which
$\delta^*$ is matched to the GFIT measurements. This configuration has already been investigated
experimentally and numerically \cite{Zhang2012, Zhang2016, Jones2010}, providing a well-established benchmark
for comparison.}

The streamwise coordinate, \( x \), is aligned with the grazing flow direction (from left to right in the figures), the spanwise coordinate is \( z \), and the wall-normal coordinate is \( y \). The origin of \( y \) is located on the outer surface of the perforated plate exposed to the flow, with positive \( y \) directed into the cavity.

Periodic boundary conditions are applied on the side walls, no-slip boundary condition on the top wall and slip boundary condition on the bottom wall, \red{to reduce the computational costs while replicating the impinging flow conditions
of the GFIT at NASA Langley}. 
At the inlet, free stream velocity corresponding to the free-stream Mach number is assigned while pressure boundary condition is set at the outlet. Additional acoustic sponge regions, where viscosity is increased, are placed ad the inlet and outlet of the computational domain to dampen the reflection of acoustic waves.


\red{

\subsection{Simulation strategy}
The simulation approach is based on two steps: first, the spatially developing boundary layer is obtained. Then, after the transient, when the aerodynamic field is converged, an instantaneous flow field was saved and modified by overlaying a plane acoustic wave with specified frequency and amplitude depending on the test case. This modified flow field served as the initial condition for the acoustic simulations.
In the second step, the computation is still performed in the time domain using the same compressible LB solver employed in the first step. The scheme is advanced explicitly in lattice units (LU), with $\Delta t_{LU}=1$ and $\Delta x_{LU}=1$ (one streaming step per time step). For the present lattice-to-physical mapping, this corresponds to a physical time step of $\Delta t = 5.714832 \times 10^{-7}\,\mathrm{s}$.

This two-step approach explicitly accounts for the interaction between acoustic waves and unsteady turbulence \citep{paduano2025impact, Avallone2021Acoustic-inducedLayer}, unlike approaches where the acoustic perturbation is linearised around the mean flow solution \citep{Tam2010}. Moreover, this methodology captures the inherently nonlinear response of the liner when exposed to a grazing acoustic wave at high SPL, since the interaction between the unsteady turbulence and the imposed acoustic field is directly resolved.
The main drawback of this approach is the requirement for the computational domain to be sufficiently long to accommodate at least ten acoustic wavelengths of the lowest frequency of interest. However, this method significantly reduces computational costs when evaluating multiple configurations, making it an efficient choice for parametric studies.

}

\red{
\subsection{Mesh resolution}

The LB/VLES equations are solved on a Cartesian mesh which is automatically generated. 
A variable-resolution (VR) scheme was employed to discretize the flow domain. The
grid resolution varies by a factor of two between adjacent VR regions. A total of four VR
regions are adopted, with progressively coarser resolution away from the liner. By construction, in a time step, particles are advected exactly from one point to the other points of the lattice stencils. Therefore, the local time step varies by a factor $2$ in adjacent VRs. Bounce-back boundary condition for no-slip walls and the specular reflection for frictionless walls are ensured thanks to a generalized volumetric formulation for the intersection of arbitrary-oriented surface elements and the volume elements \citep{Chen1998a}. 

 The mesh is structured to be finest near the wall and gradually
coarser moving toward the centre of the domain outside the turbulent boundary layer.
The finest grid resolution was applied over the entire face sheet, inside the orifices, and
within the backing cavity.
Inside the orifice, the minimum grid spacing 
expressed in viscous units, this yields
\[
\Delta x^{+}_{\min} = \Delta y^{+}_{\min} = \Delta z^{+}_{\min} \approx 5.5,
\]
corresponding to an effective resolution of $40$~voxels/mm ($\approx 40$~voxels/$d$), which is identical in all three spatial directions. This value is similar to the value suggested by \citet{Manjunath2018} in the absence of acoustic waves $(\approx 42$~voxels/$d)$. 
Outside the boundary layer, in the outer flow region, the maximum grid spacing corresponds
to
\[
\Delta x^{+}_{\max} = \Delta y^{+}_{\max} = \Delta z^{+}_{\max} \approx 42.
\]
A schematic of the computational setup is reported in Figure \ref{fig:Geometry_setup}, where an example of the computational grid close to cavity is shown.
The mesh convergence study is reported in the Appendix \ref{appA}.

}

\subsection{Test cases}
The test matrix is summarized in Table \ref{tab:Table_test_matrix}.  
Five sets of simulations are considered.  

\begin{table}
    \centering
    
\begin{tabular}{l l c c c c } 

     &\bf{Case} & \bf{SPL [dB] } & \bf{$f$ [Hz]}   &  \bf{ $M$} & \bf{ac. prop. dir. }\\[3pt]
   
      1&     Validation against GFIT   & -   & - & 0.3 & $-$\\
    2& Validation against GFIT    & 130   & 1800, 2200, 3000 & 0.0 & $x^+$\\
     3& Effect of SPL  & 130,140,150,160   & 2200 & 0; 0.3& $x^+$ \\
       4& Effect of source frequency &150   & \red{1200,1600}, 1800, 2200, 3000& 0; 0.3& $x^+$ \\
        5& Effect of flow direction &150   & 2200& 0.3& $x^+; x^-$ \\
6&\red{Interference between subsequent orifices} &150   & 2200& 0.3& $x^+$ \\
7& \red{3D analysis} &150   & 2200& 0;0.3& $x^+$ \\
\end{tabular}  
    \caption{List of the simulations \red{and analysis} carried out}
    \label{tab:Table_test_matrix}
\end{table}

The first test case considers the simulation with grazing flow only and is used to validate the incoming flow conditions against the results reported by \cite{Jones2004}. 
The second test case corresponds to the condition without grazing flow, in which the SPL, expressed in decibels using a reference pressure of $20 \times 10^{-6}$~Pa, is fixed at 130 dB while the source frequency is varied. This configuration serves as a validation for the acoustic response and allows direct comparison of the impedance results with those obtained in the GFIT facility. The grid convergence assessment, along with the validation of the incoming flow conditions and the impedance in the absence of grazing flow, are presented in Appendix~\ref{appA}.

The third test case constitutes the core of the present analysis. The frequency of the grazing acoustic wave is fixed at $2200$~Hz, which is higher than the resonant frequency in the absence of flow but close to the resonant frequency in the presence of grazing flow. 
\red{In the absence of grazing flow, the resonant frequency can be written following \cite{PantonResonantresonators1975} as:
\begin{equation}
f_{0} = \frac{a_0}{2 \pi} \sqrt{\frac{S}{V_c (\tau + \tau^*) + P}},
\label{eq:resonance_frequency}
\end{equation}
where \( a_0 \) is the speed of sound, \( S \) denotes the orifice area, and \( V_c = A d_c \) is the cavity volume, and $A$ is the cavity area. The terms \( P = \frac{1}{3} d_c^2 A \) and \( \tau^* \approx 0.8\sqrt{S/\pi} \) represent end corrections. These terms account for the additional oscillating fluid mass associated not only with the fluid inside the orifice neck, but also with a small portion of fluid within the cavity and in the external region near the orifice.
The formula gives a value which is approximately equal to 1600 Hz in the absence of grazing flow while the value rises in the presence of grazing flow as found numerically and experimentally by \cite{paduano2025impact, Quintino2025, Bonomo2023AProfiles,Spillere2020, Yu2008ValidationData}.
This shift is attributed to a modification of the effective fluid volume oscillating within the orifice, which in turn alters the end corrections.}

Both with and without grazing turbulent flow at $M\!=\!0.3$, the SPL is varied systematically. This enables the investigation of dissipation mechanisms across linear and nonlinear regimes, and how these are modified by the presence of grazing flow.  

The fourth test case explores the effect of changing the source frequency at a fixed SPL of $150$~dB, for both configurations with and without grazing flow. This dataset provides insight into the frequency dependence of the dissipation mechanisms.  

The fifth test case addresses the role of acoustic propagation direction relative to the grazing flow. In the $x^+$ configuration, the acoustic waves propagate in the same direction as the grazing flow, whereas in the $x^-$ configuration they propagate in the opposite direction. The latter scenario is particularly relevant, as it closely resembles the operating condition of acoustic liners installed in engine nacelles, where the grazing flow enters from the upstream side while the dominant acoustic waves are generated downstream by the fan and travel in the upstream direction.

\red{
All these analysis are conducted by extracting data on a two-dimensional streamwise plane encompassing a region that includes the central orifice of the cavity, together with a portion of the cavity and the flow above and below it. 
The domain considered extends over approximately $3 \times 4.5$ orifice diameters.
The limited streamwise extent of the computational domain, together with the periodic spanwise boundary condition, ensures that three-dimensional effects are negligible, thus making the 2D field analysis representative of the orifice dissipation mechanisms.

To investigate potential orifice–orifice interaction effects, the analysis
is extended for one representative configuration (SPL = 150~dB in the
presence of grazing flow) by extracting an additional streamwise-aligned
plane containing two consecutive orifices. This analysis,
reported in Section~\ref{sec:2orifices}, examines how the modification of
the grazing boundary layer induced by the upstream orifice alters the shear
layer over the downstream orifice, and consequently affects the
acoustic-induced velocity and local dissipation mechanisms.

Finally, a full three-dimensional analysis is performed for one
representative configuration (150 dB, both in the presence and absence of
grazing flow) to validate the dissipation methodologies. The 3D results,
presented in Section~\ref{sec:3D_analysis}, confirm the consistency of the two-dimensional formulation. The comparison demonstrates that both the
physical interpretation and the trends identified in the two-dimensional
analysis are consistent with the fully three-dimensional evaluation.
While the 3D analysis provides methodological validation, the 2D
formulation enables a systematic parametric investigation over a wide
range of acoustic amplitudes and frequencies at a reasonable computational
cost.}

\section{Flow field within the orifice}
\label{sec:flowfield}

\subsection{Acoustic-induced velocity profiles within the orifice}
\noindent

To characterise the interaction between the acoustic plane waves and the grazing flow, we focus on the \red{wall-normal} acoustic-induced velocity fluctuation \red{in wall units,  $v_{\text{ac}}^{+} = v_{\text{ac}}/u_\tau$}, which is directly linked to the conversion and dissipation of acoustic energy \citep{tam_microfluid_2000} \red{and the periodic exchange of mass through the orifice of the liner \citep{Zhang2016a}.} 
The phase of the acoustic oscillation is indicated by \( \phi \). The inflow phase (\( \phi = \pi/2 \)) corresponds to the instant of maximum mass flux entering the cavity, while the outflow phase (\( \phi = 3\pi/2 \)) denotes the instant of maximum mass flux exiting the cavity.

\begin{figure}[htpb]
    \centering
    \includegraphics[width=1\textwidth]{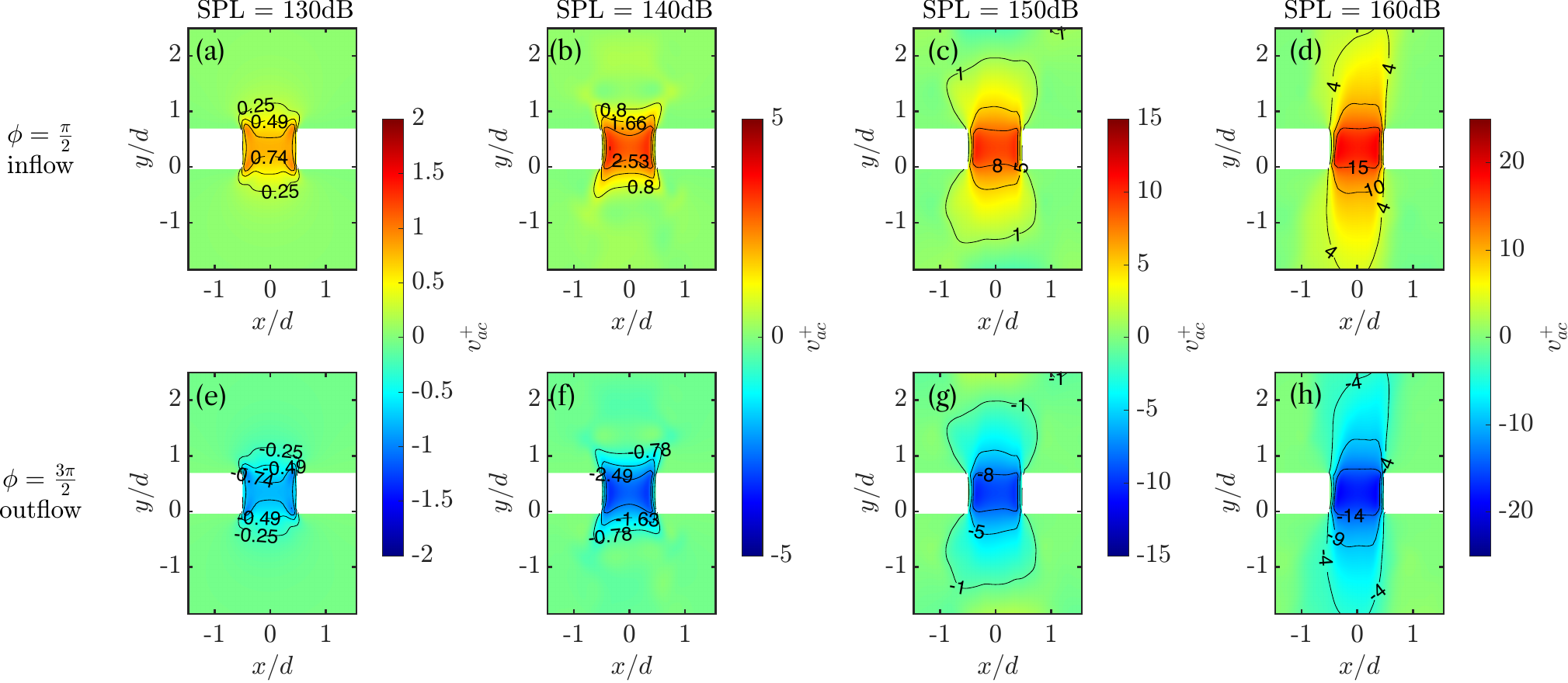}
    \caption{Contour of the wall-normal acoustic-induced velocity \red{in wall units, $v^+_{\text{ac}}$}, at the inflow ($\phi=\pi/2$) and outflow ($\phi=3\pi/2$) phases, effect of the SPL (left to right column), no flow condition. }
    \label{fig:contour_vac_noflow}
\end{figure}

\begin{figure}[htpb]
    \centering
    \includegraphics[width=1\textwidth]{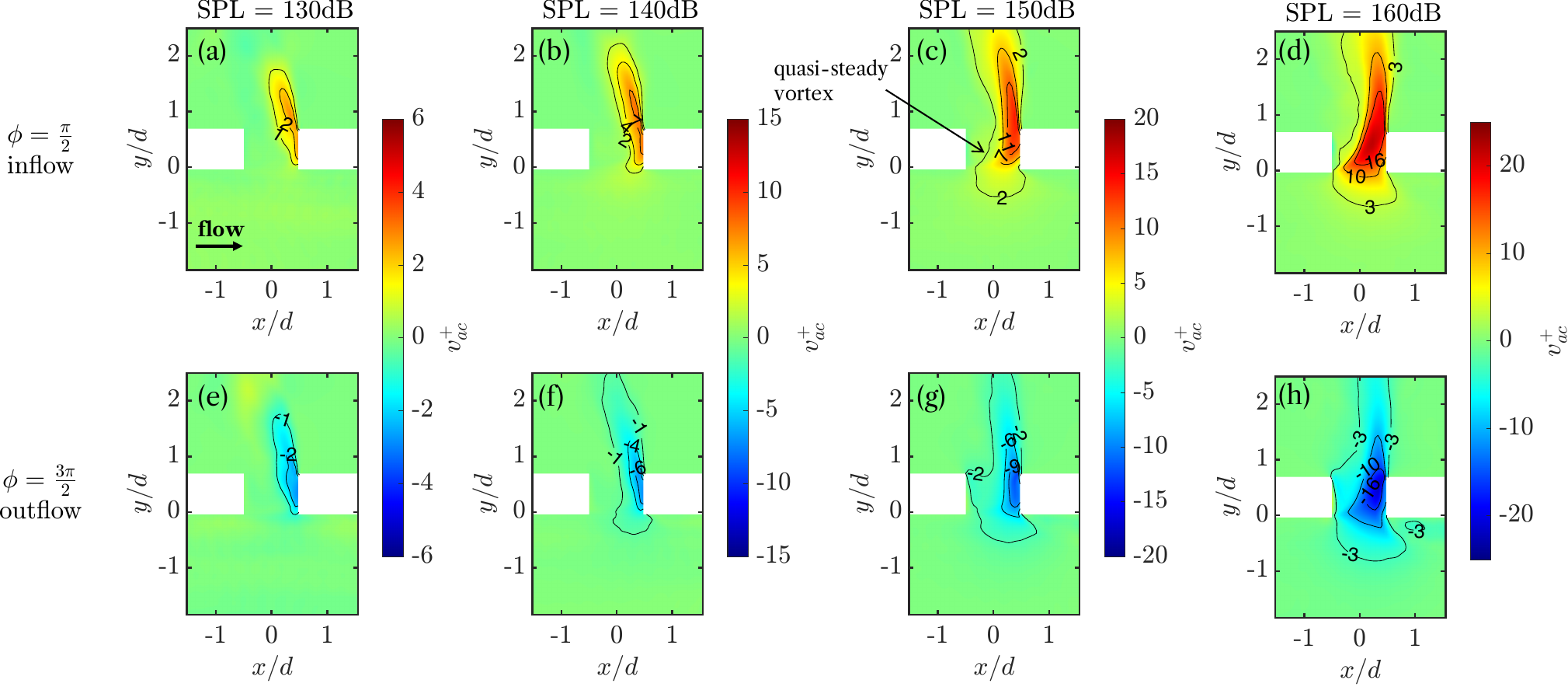}
    \caption{Contour of the wall-normal acoustic-induced velocity \red{in wall units, $v^+_{\text{ac}}$}, at the inflow ($\phi=\pi/2$) and outflow ($\phi=3\pi/2$) phases, effect of the SPL (left to right column),  $M=0.3$.}
    \label{fig:contour_vac_M03}
\end{figure}

Figure \ref{fig:contour_vac_noflow} and \fref{fig:contour_vac_M03} show the contours of the acoustic-induced velocity during the inflow and outflow phases, for the cases without and with grazing flow, respectively. The effect of the SPL is reported in each figure. 

For the no-flow case, the acoustic-induced velocity shows high amplitude on the entire orifice extent, and the inflow and outflow phases are almost perfectly symmetric. 
As the SPL increases, the distribution of the acoustic-induced velocity becomes broader, extending deeper into the cavity ($y/d>1$) during the inflow phase and farther into the flow region ($y/d<-0.6$) during the outflow phase. For instance, at SPL $=130\,\mathrm{dB}$ (\fref{fig:contour_vac_noflow} (a-e)), the velocity field exhibits high velocity amplitude only in the vicinity of the orifice, while it attenuates rapidly within approximately one diameter in the wall-normal direction; at higher SPL, the region of influence extends farther from the wall. A boundary layer region at the orifice side walls is also visible which is consistent to what reported by \citet{Zhang2012}. 
\red{The magnitude of $v^+_{ac}$ increases with the SPL. Its maximum value over the orifice cross section is a fraction of the friction velocity at 130 dB and increases up to 15 times larger than $u_{\tau}$ at 160 dB.}

When grazing flow is introduced, the flow topology of the acoustic-induced velocity changes markedly, in agreement with \citet{Zhang2016} and \citet{paduano2025impact}. The acoustic-induced velocity becomes concentrated in the downstream half of the orifice, displaying a jetting-like pattern during the inflow and outflow phases, while the upstream half is occupied by a quasi-steady vortex resulting from the vena contracta effect. The formation of this quasi-steady vortex reduces the effective porosity of the orifice, leading to increased flow blockage and a corresponding shift of the liner’s resonant frequency toward higher values. This aspect will be further discussed in the section dedicated to the frequency-dependent analysis of the dissipation mechanisms.
In addition, the increased blockage induced by the grazing flow, and the consequent reduction in effective porosity, may contribute to the rise in acoustic resistance, as discussed in Appendix~\ref{appA}, where the impedance calculations are presented.

As the SPL increases (\fref{fig:contour_vac_M03} (c-g and d-h)), the extent of this quasi-steady vortex diminishes, and peaks of acoustic-induced velocity are localized farther into the cavity. However, unlike the no-flow case, the inflow and outflow phases are largely asymmetric due to the shear imposed by the grazing flow, which acts as a barrier especially in the outflow phase. At high SPL, the amplitude of the pressure fluctuations generates a vertical velocity \red{almost twenty times larger than the friction velocity and} comparable with the grazing flow free-stream velocity. This velocity affects the near-wall region more strongly in the outflow phase (\fref{fig:contour_vac_M03} (h)).

\begin{figure}[htpb]
    \centering
    \includegraphics[width=1\textwidth]{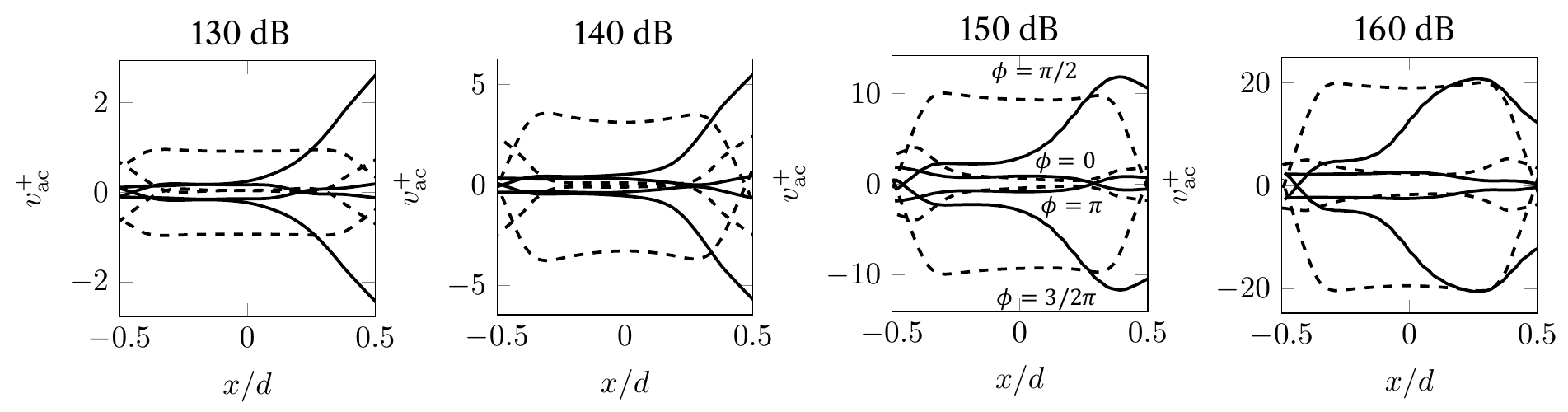}
    \caption{Spatial distribution along the diameter of the non-dimensional acoustic-induced vertical velocity \red{$v^+_{ac}$} as a function of the SPL at half face-sheet thickness. Various phases are reported, dashed line is the no flow case and solid line is $M=0.3$ case.}
    \label{fig:Acoustic_induced_v_component}
\end{figure}

The $v^+_{ac}$ taken at mid-height of the orifice are shown in \fref{fig:Acoustic_induced_v_component} for the grazing flow (continuous lines) and the no-flow (dashed lines) cases at $y/\tau=0.5$ at four phases. 
In the presence of flow, the velocity does not vanish at the walls because of the finite extent of the exported computational domain in PowerFLOW and of the wall model. Nevertheless, the near-wall velocity gradients, computed through the wall model applied at the orifice walls, are directly provided by the solver and are used for the evaluation of viscous dissipation.

In the absence of flow, the velocity profiles are nearly symmetric across the orifice, with small streamwise asymmetries arising from the grazing nature of the acoustic wave.
Remarkably, in the presence of grazing flow, at low SPL (\fref{fig:Acoustic_induced_v_component} (a,b)) the peak of the acoustic-induced velocity is about twice that of the no-flow case. This suggests that the grazing flow energy adds to the acoustic energy, pushing flow into the downstream half of the orifice at velocities higher than the ones obtained in the absence of grazing flow. This means that the flow contribution dominates at low SPL; this scenario is consistent with the observed broadband amplification of SPL by the turbulent grazing flow found by \citet{RONCEN2025119058} and a transfer of energy from the broad frequencies associated with the turbulence field to the tonal frequency of the acoustic field. When SPL exceeds $140\,\mathrm{dB}$, the \red{maximum value of $v^+_\text{ac}$ in the presence of grazing flow approaches the maximum of the case without grazing flow}, indicating that the acoustic forcing overcomes the effect of the grazing flow.

With grazing flow, the maximum and minimum of the \red{$v^+_{ac}$} profiles are consistently located in the downstream half of the orifice, around $x/d \approx 0.8$, due to the quasi-steady vortex which acts as a barrier to the acoustic waves. This spatial asymmetry could be responsible for the increased acoustic resistance by effectively reducing the orifice cross-section. As the SPL increases, the location of the maximum \red{$v_{ac}^+$} shifts toward the orifice centre and the distribution along the orifice diameter becomes more symmetric, indicating an increased acoustic-induced mass flow. 

Overall, the effect of the grazing flow is to concentrate the acoustic-induced velocity in the downstream corner, and to increase the peak value relative to the no-flow case at low SPL. The effective porosity, linked to the area available for the acoustic-induced flow, increases with SPL as the acoustic forcing becomes dominant, reducing the blockage effect of the quasi-steady vortex.

\subsection{Shear layer at the orifice mouth}
\noindent

\begin{figure}[htpb]
    \centering
    \includegraphics[width=1\textwidth]{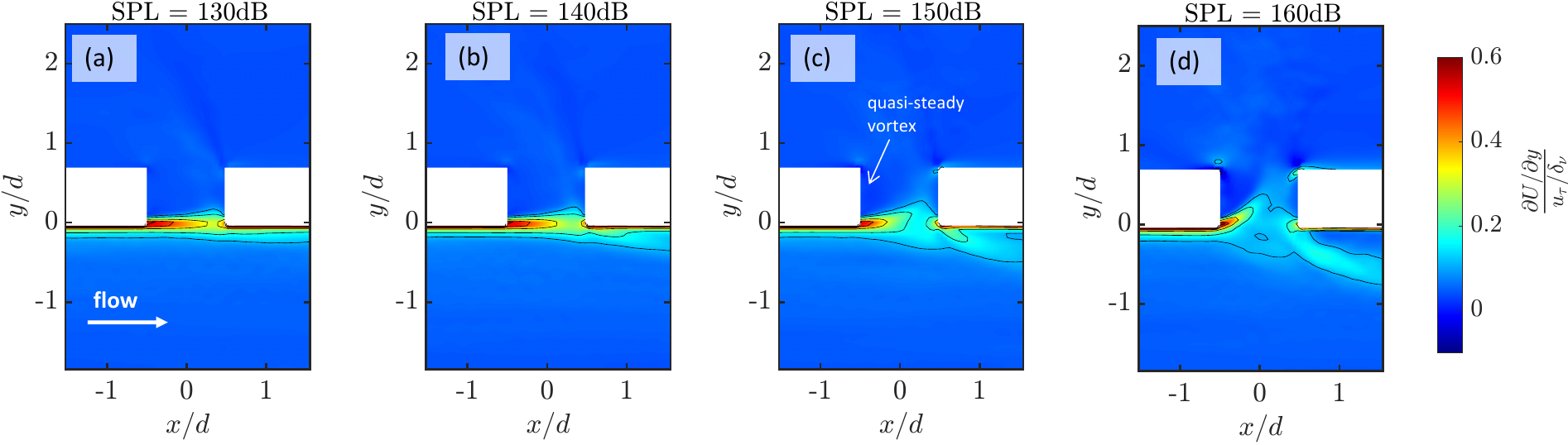}
    \caption{Contour of the \red{normalized shear} forming at the mouth of the orifice in the presence of grazing flow at $M=0.3$, effect of the SPL.}
    \label{fig:Utau}
\end{figure}

The development of the shear layer over the orifice of the acoustic liner, induced by the near-wall mean velocity gradient imposed by the grazing flow, is examined through \red{the contour of the normalized shear
\begin{equation}
\frac{\partial U/\partial y}{u_{\tau}/\delta_\nu},
\end{equation}
shown in \fref{fig:Utau}, where $\delta_v $ is the viscous scale.} 
The grazing flow acts as a barrier that limits the penetration of the acoustic-induced flow into the orifice, while variations in the SPL modify the shear layer above it.  Only the grazing flow configuration is shown, since the \red{near-wall mean streamwise velocity gradient} is meaningful exclusively in the presence of a mean flow over the orifice.

At low SPL (\fref{fig:Utau} (a-b)), the shear layer blocks the acoustic-induced velocity from entering the orifice over the majority of the diameter extension. The \red{strength of the shear} remains high across most of the orifice diameter, except for a small downstream region where it slightly decreases. This confirms that the grazing flow constrains the acoustic-driven inflow, localizing it only near the downstream corner. The majority of the orifice is occupied by a quasi-steady vortex \citep{Baumeister1975}; at SPL \(=130\,\mathrm{dB}\), this vortex extends over nearly the entire orifice diameter (\fref{fig:Utau} (a)). This flow pattern is consistent with the contour of the acoustic-induced velocity reported in \fref{fig:contour_vac_M03} (a).

As the SPL increases, the size of the quasi-steady vortex progressively diminishes. At \(140\,\mathrm{dB}\) (\fref{fig:Utau} (b)), approximately half of the cavity depth shows a reduction in \red{the normalized shear}, suggesting that the acoustic excitation partially overcomes the shear-layer blockage. At \(150\,\mathrm{dB}\) (\fref{fig:Utau} (c)), the strong shear region is further reduced: a significant portion of the orifice cross-section exhibits lower \red{magnitude of the velocity gradient}. 
The larger extent and magnitude of the acoustic-induced velocity during the inflow and outflow (see \fref{fig:contour_vac_M03} (c)) leads to a disruption of the shear layer. This in turn affects the near-wall flow topology on the grazing flow side. The contours, in fact, reveals wake-like features downstream of the orifice, indicating that the combination of the orifice geometry and the high acoustic-induced flow modifies the near-wall flow development downstream of the orifice.

At 160 dB the disruptive effect of the acoustic-induced flow on the shear layer over the orifice is further enhanced (\fref{fig:Utau} (d)).
The \red{region over the orifice having high velocity gradient magnitude} is reduced and the shear bends into the orifice. The size of the quasi-steady vortex is consequently reduced. 
This results in a stronger and more extended acoustic-induced flow in and out of the cavity compared to lower SPLs. Moreover, the wake-like behaviour downstream of the orifice becomes stronger, suggesting the onset of local flow separation downstream of the orifice.

Similar observations \red{on the effect of SPL on the shear layer were reported by \citet{Leon2019Near-wallFlow}}, who documented vortex shedding and the ejection of vortices from the cavity into the grazing flow. Their study, \red{conducted outside the orifice, close to the wall, using micro-PIV measurements,} proposed that when the ratio between the peak acoustic-induced velocity and the \red{friction velocity} exceeds approximately two, the effect of the acoustic forcing becomes sufficiently strong to generate vortices ejected in the grazing flow; otherwise, the influence on the grazing flow remains limited. 

\red{
In \fref{fig:contour_vac_M03}, iso-levels of $v_{ac}^+$ are reported. At 130 and 140 dB, values exceeding the threshold of two
identified by \citet{Leon2019Near-wallFlow} are confined within the
orifice.
No significant region with $v_{\text{ac}}^{+} \geq 2$ extends into the
boundary layer.
As the SPL increases beyond 140 dB, the region where
$v_{\text{ac}}^{+} \geq 2$ expands beyond the orifice mouth and penetrates
into the grazing boundary layer. This region becomes substantially larger
at 160 dB.
Therefore, for SPL values above 140 dB, the ratio exceeds the threshold
suggested by \citet{Leon2019Near-wallFlow} over a spatially extended region,
indicating that the acoustic forcing is sufficiently strong to influence
the surrounding grazing flow, in addition to modifying the flow inside the
orifice.
}

The quantitative impact of the observed changes in flow topology and the development of the shear layer under grazing flow on the noise absorption mechanism is assessed in the following section through a detailed analysis of acoustic dissipation.

\red{
\section{Parametric analysis of acoustic dissipation for a single orifice}

In this section, we present the parametric study of the acoustic dissipation
mechanisms based on the two-dimensional (per unit span) formulation,
evaluated on a streamwise slice containing the orifice. This approach
enables a systematic investigation over the range of SPLs,
frequencies, and flow conditions considered.

A full three-dimensional evaluation is performed for one representative
configuration (150~dB, with and without grazing flow) and is reported in
Section~\ref{sec:3D_analysis}. The comparison confirms that the trends and
physical interpretation derived from the two-dimensional analysis are
consistent with the fully three-dimensional results.

}

\subsection{Acoustic dissipation by vortex shedding}
\noindent

\begin{figure}[htpb]
    \centering
    \includegraphics[width=1\textwidth]{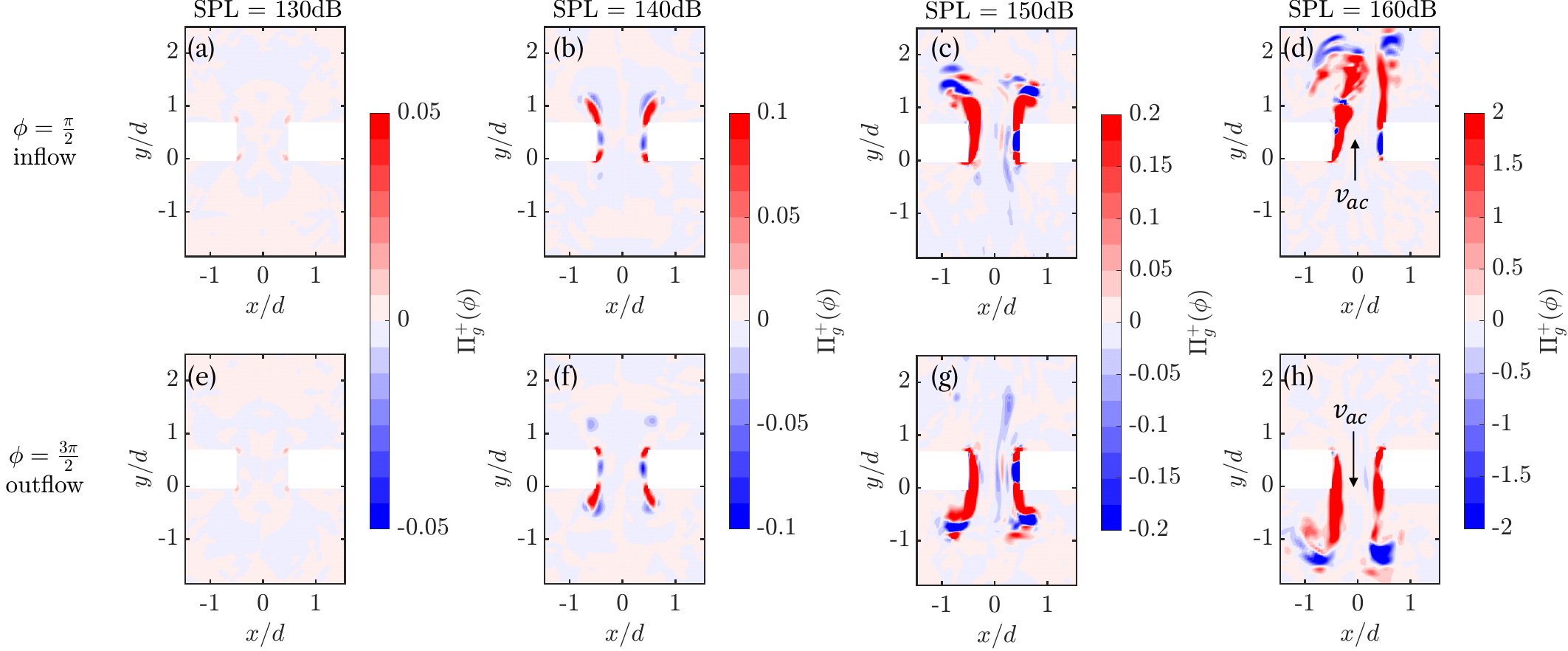}
    \caption{Contour of the power density \red{$\Pi^+_g$ in viscous units} transferred from the acoustic field to the vortical field during the inflow and outflow phases when varying the SPL. No-flow, forcing frequency equal to \SI{2200}{Hz}.}
    \label{fig:Pi_M0}
\end{figure}

\begin{figure}[htpb]
    \centering
    \includegraphics[width=1\textwidth]{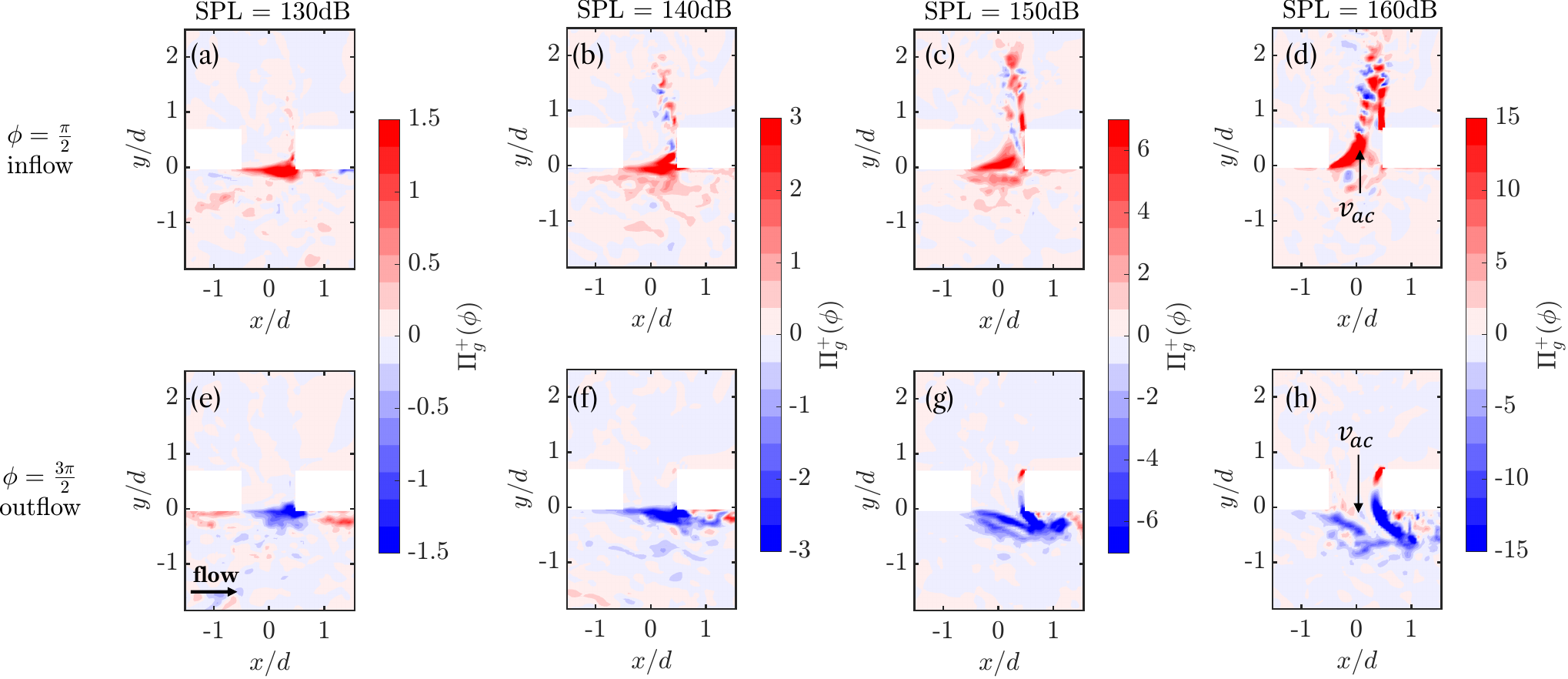}
    \caption{Contour of the power density \red{$\Pi^+_g$ in viscous units} transferred from the acoustic field to the vortical field during the inflow and outflow phases when varying the SPL. $M=0.3$, forcing frequency equal to \SI{2200}{Hz}.}
    \label{fig:Pi_M03}
\end{figure}

The contours of the power density transferred from the acoustic field to the vortical field \red{normalized in viscous units}, \red{$\Pi^+_g(\phi)$}, at different SPL levels are shown in \fref{fig:Pi_M0} and \fref{fig:Pi_M03} for the no-flow condition and in the presence of grazing flow at $M=0.3$, respectively. In these plots, regions contributing to acoustic dissipation by vortex shedding appear in red, while regions associated with acoustic generation are shown in blue. Two characteristic phases are shown: the inflow phase $(\phi=\pi/2)$ and the outflow phase $(\phi=3\pi/2)$.

In the absence of the grazing flow (\fref{fig:Pi_M0}), at the lower SPL levels, 130 and 140\,dB, the power density transferred from the acoustic field to the vortical field is negligible.
This observation is consistent with the observations reported by \citet{tam_microfluid_2000} who showed that for SPL below 150 dB the contribution of vortex shedding to the overall dissipation is minimal. As the SPL increases to 150\,dB, positive regions of \red{$\Pi^+_g$} associated with vortical structures appear at the orifice mouth during both the inflow and outflow phases. 
At 160\,dB, these patterns becomes clearly visible: vortices form at the orifice mouth and roll up either inside the cavity (inflow) or into outside it (outflow). These regions, identifiable both in the inflow and in the outflow, contribute to the rate of acoustic energy dissipation by vortex shedding.

When the grazing flow is introduced, the dissipation mechanism changes significantly. Even at low SPL, the magnitude of \red{$\Pi^+_g(\phi)$} increases compared to the no-flow case, suggesting that turbulence structures and shear-layer vortices enhance the transfer of acoustic energy into vorticity. The contours of \red{$\Pi^+_g$} follow the topology dictated by the friction and the acoustic-induced velocity.
The regions with non-negligible values of \red{$\Pi^+_g$} are localized in fact in the downstream half of the orifice due to the presence of the quasi-steady vortex in the upstream half.
The magnitude of \red{$\Pi^+_g$} and the portion of the domain, which contributes to the conversion of acoustic energy, increase with SPL. 
During the inflow phase, the positive contribution of \red{$\Pi^+_g$} is linked with a vortex generated at the upstream lip of the orifice. This vortex rolls up and penetrates farther into the cavity as the SPL increases.
In contrast, during the outflow, \red{$\Pi^+_g$} assumes negative values. 
The orifice behaves locally as a noise source: the outward acoustic jet perturbs the grazing flow, leading to the formation of vortical structures that are convected downstream. At 160 dB the negative contribution extends farther in the grazing flow following a wake-like pattern downstream of the orifice.

\begin{figure}[htpb]
    \centering
    \includegraphics[width=1\textwidth]{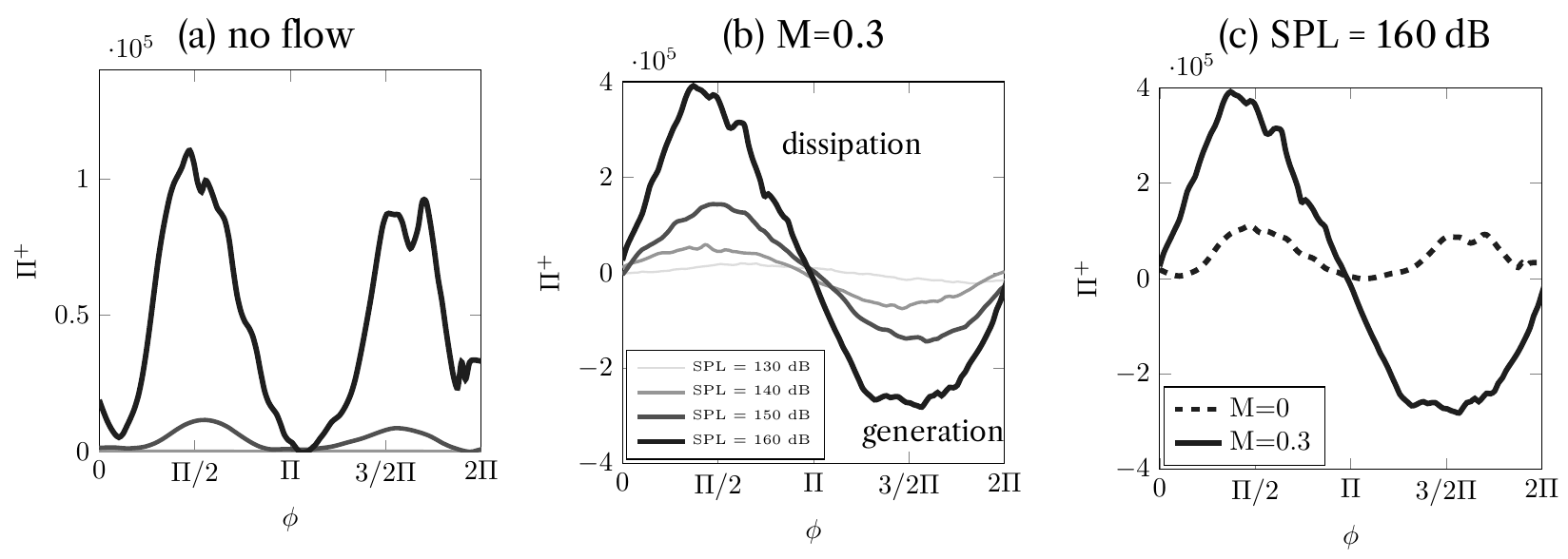}
    \caption{Phase averaged acoustic energy dissipation rate by vortex shedding \red{in viscous units} as a function of the phase, $\Pi^+(\phi)$, (a) for the no-flow case and (b) for the $M=0.3$ case when varying the SPL (darker and thicker lines indicate higher SPL); (c) comparison of no-flow and $M=0.3$ case at 160 dB.}
    \label{fig:Pi_phi}
\end{figure}

The evolution of the rate of acoustic dissipation \red{in viscous units} integrated over the domain as a function of the phase, \red{$\Pi^+(\phi)$}, is presented in \fref{fig:Pi_phi}. 
In the absence of grazing flow (\fref{fig:Pi_phi} (a)), \red{$\Pi^+(\phi)$} exhibits two distinct positive peaks in the inflow and outflow phases. 
This indicates that dissipation by vortex shedding occurs in both phases. These peaks become more pronounced when increasing the SPL, whereas at 130 and 140\,dB the dissipation rate is negligible, as expected \citep{tam_microfluid_2000}.

The evolution of \red{$\Pi^+(\phi)$} in the presence of grazing flow is shown in \fref{fig:Pi_phi}(b). Similarly to the no-flow case, the amplitude of the dissipation rate increases with SPL. However, at all SPL levels, the system exhibits a positive dissipation rate during the inflow phase and a negative one during the outflow phase. As a result, the net acoustic dissipation over a full cycle is lower than in the no-flow condition, despite the higher instantaneous values of \red{$\Pi^+(\phi)$} observed during the inflow phase with grazing flow. This is clearly illustrated in \fref{fig:Pi_phi}(c) for SPL $=160\,\mathrm{dB}$, where the peak dissipation with grazing flow exceeds that of the no-flow case, yet the overall effect over a complete acoustic cycle is reduced due to the generation of acoustic energy during the outflow phase. This aspect will be further quantified in Section \ref{sec:budget}.

These findings suggest that, in the presence of grazing flow, the outflow phase contributes to the generation of acoustic energy because of the interaction of the outflow of acoustic-induced velocity with the turbulent grazing flow. The system thus exhibits a behaviour analogous to that of a jet in cross-flow, a configuration known to enhance noise radiation \citep{Camelier1976JetCrossflow, Stimpert1973CrossflowNoise}. Consequently, the overall amount of acoustic energy dissipated by the liner through vortex shedding decreases in the presence of grazing flow compared with the no-flow condition.

\subsection{Acoustic dissipation by viscous effects at the mouth of the orifice}

\begin{figure}[htpb]
    \centering
    \includegraphics[width=1\textwidth]{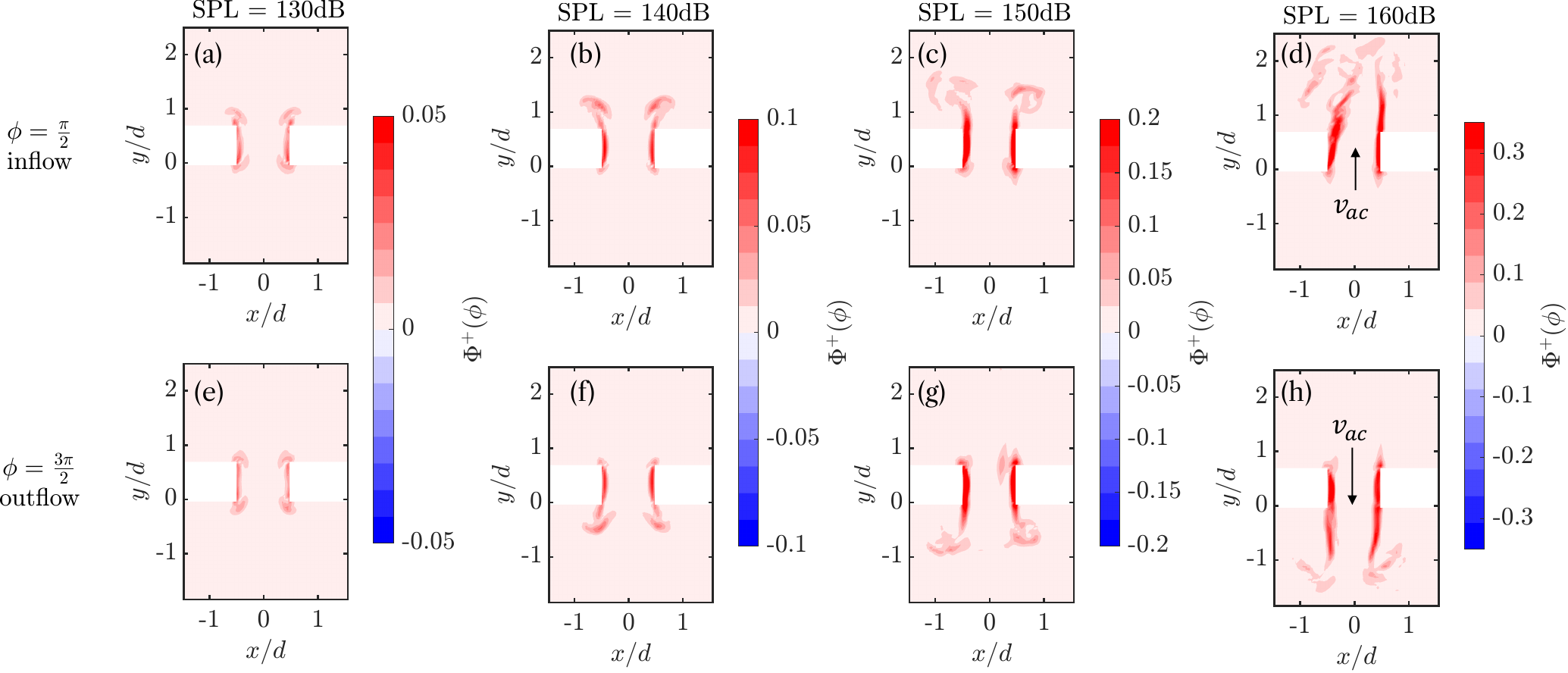}
    \caption{Contour of \red{the viscous dissipation density in viscous units, $\Phi^+$,} in the inflow and outflow phases when varying the SPL. $M=0$ case, forcing frequency equal to \SI{2200}{Hz}.}
    \label{fig:viscdiss_M0}
\end{figure}

Contours of \red{viscous dissipation density in wall units $\Phi^+(\phi)$}, are shown in \fref{fig:viscdiss_M0} for the no-flow condition. The contour scale changes for each subplot with SPL due to the increase in acoustic-induced velocity inside the orifice, which results in larger velocity gradients.
In the no-flow case, we observe that both the contour levels and the thickness of the boundary layer developing in the internal walls of the orifice increase with SPL. This trend reflects the rise in the amplitude of the acoustic-induced velocity. 

\begin{figure}[htpb]
    \centering
    \includegraphics[width=.7\textwidth]{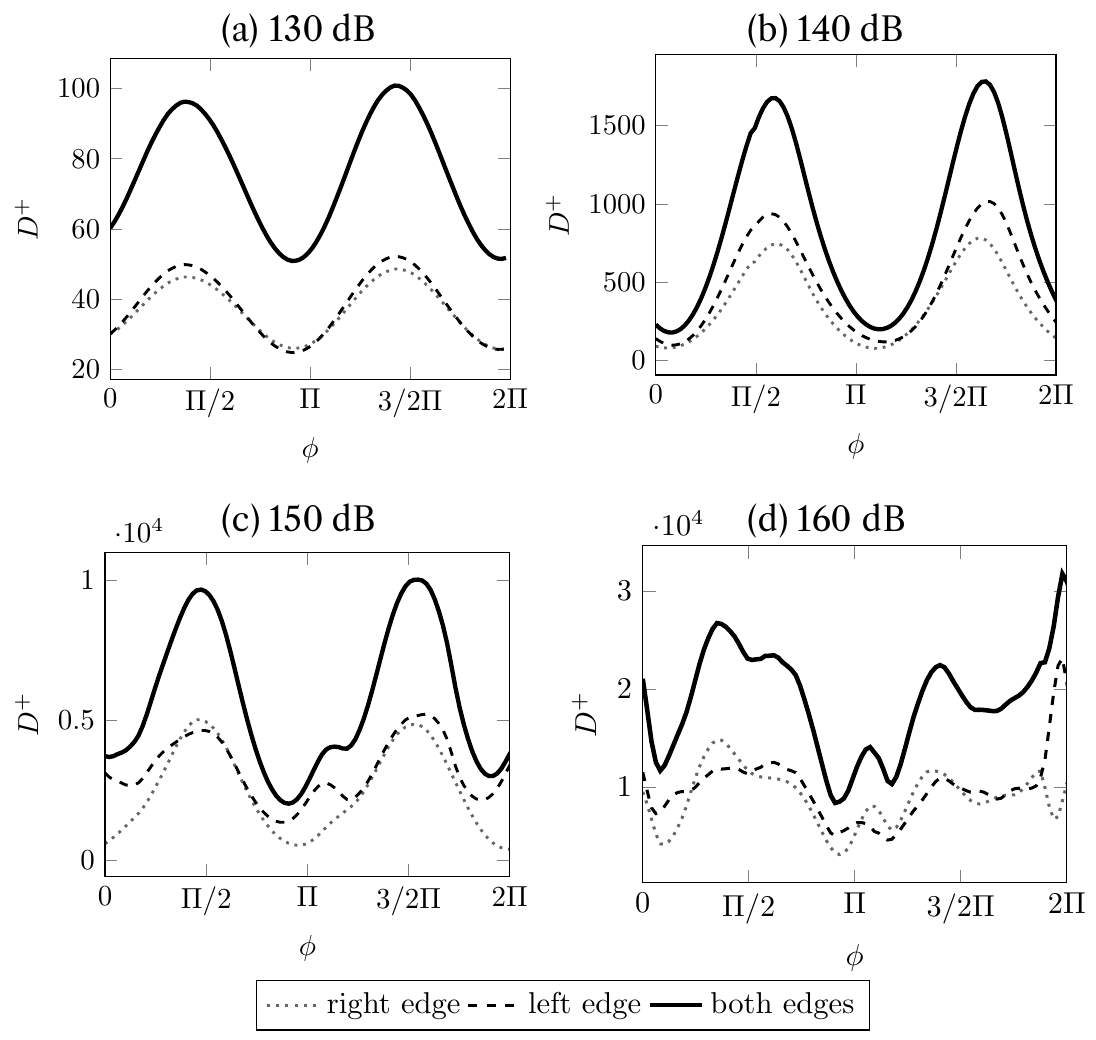}
    \caption{Phase averaged viscous dissipation rate \red{in viscous units, $D^+$} over one cycle at different SPL, left and right orifice edge contribution reported separately, no flow case, forcing frequency equal to 2200 Hz.}
    \label{fig:visc_leftrightM0}
\end{figure}

At all SPL levels, the viscous dissipation regions are nearly symmetric across the orifice, with slight asymmetries becoming more pronounced as SPL increases, attributed to the grazing nature of the incident acoustic wave. Figure~\ref{fig:visc_leftrightM0} shows the dissipation rate \red{in viscous units $D^+(\phi)$} as a function of phase, separated into contributions from the left and right sides of the orifice. At 130\,dB (\fref{fig:visc_leftrightM0}(a)), the two contributions are nearly identical, while at 140\,dB (\fref{fig:visc_leftrightM0}(b)), moderate differences emerge, likely due to a thicker boundary layer forming on the upstream side.

Regarding the phase dependence, in the absence of grazing flow, the inflow and outflow dissipation profiles remain largely symmetric. For SPL up to 150\,dB (\fref{fig:visc_leftrightM0}(a--c)), the dissipation exhibits two dominant lobes, indicating that viscous dissipation mainly occurs during the inflow and outflow phases. At 160\,dB, additional lobes appear, suggesting that non-linear effects and higher harmonic components of the acoustic-induced velocity contribute to the dissipation.

\begin{figure}
    \centering
    \includegraphics[width=1\textwidth]{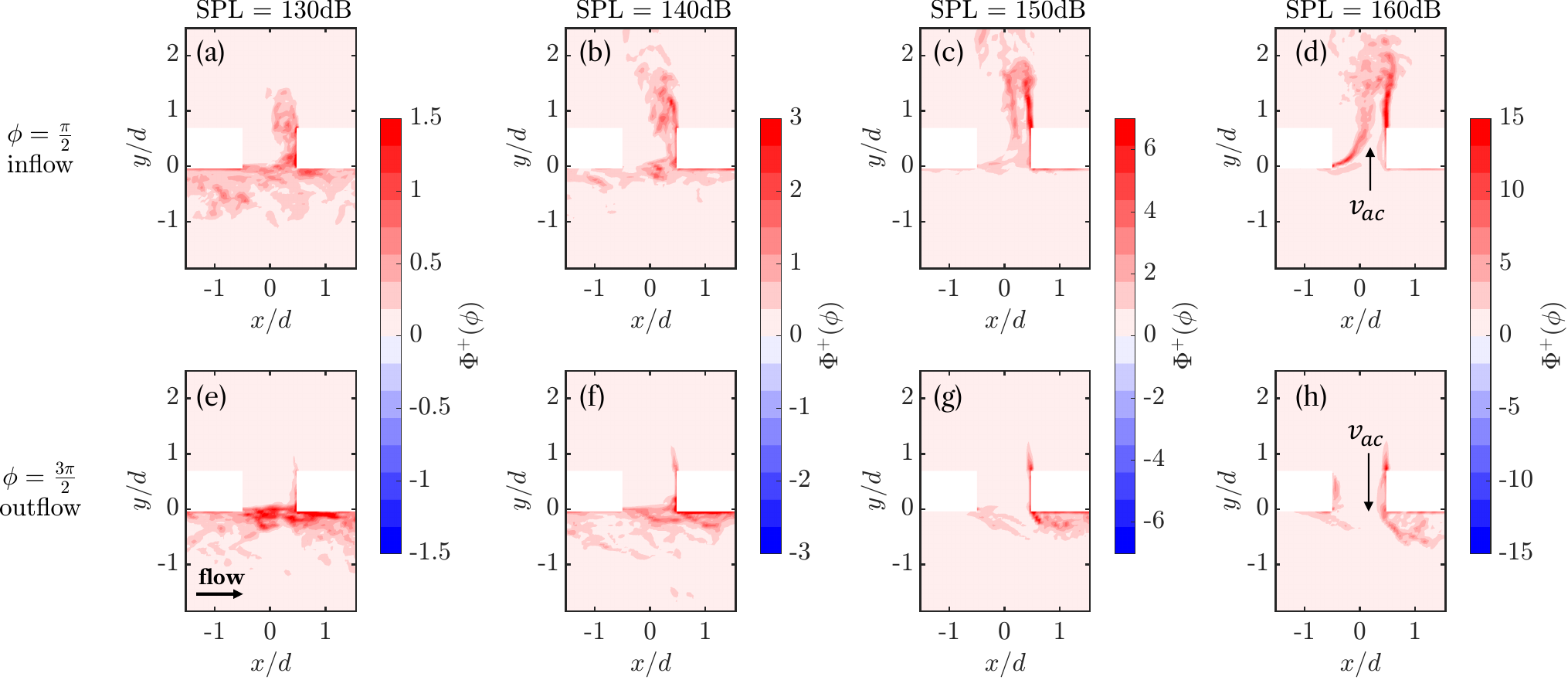}
    \caption{Contour of \red{the viscous dissipation density in viscous units $\Phi^+$} in inflow and outflow phases when varying the SPL. $M=0.3$ case, forcing frequency equal to \SI{2200}{Hz}.}
    \label{fig:viscdiss_M03}
\end{figure}

When the grazing flow is introduced (\fref{fig:viscdiss_M03}), several significant changes arise. First, at low SPL, the contour levels are higher than in the no-flow case, showing an increase in viscous dissipation driven by the flow itself. This increase stems from the grazing flow pushing the acoustic-induced flow preferentially towards the downstream (right) side of the orifice, as previously observed in the acoustic-induced velocity fields (\fref{fig:Acoustic_induced_v_component}).

A pronounced geometric asymmetry emerges: the dissipation is larger at the downstream side of the orifice, while the upstream side contributes negligibly. This is due to the quasi-steady vortex occupying the upstream half of the orifice, which blocks the flow through the orifice during the inflow and outflow phases. At higher SPL, a free-shear layer appears at the center of the orifice during the inflow phase (\fref{fig:viscdiss_M03}(c--d)). However, this contribution is not considered as wall viscous dissipation, as it corresponds to energy conversion into vorticity already accounted for in the vortex shedding dissipation.

\begin{figure}
    \centering
    \includegraphics[width=.7\textwidth]{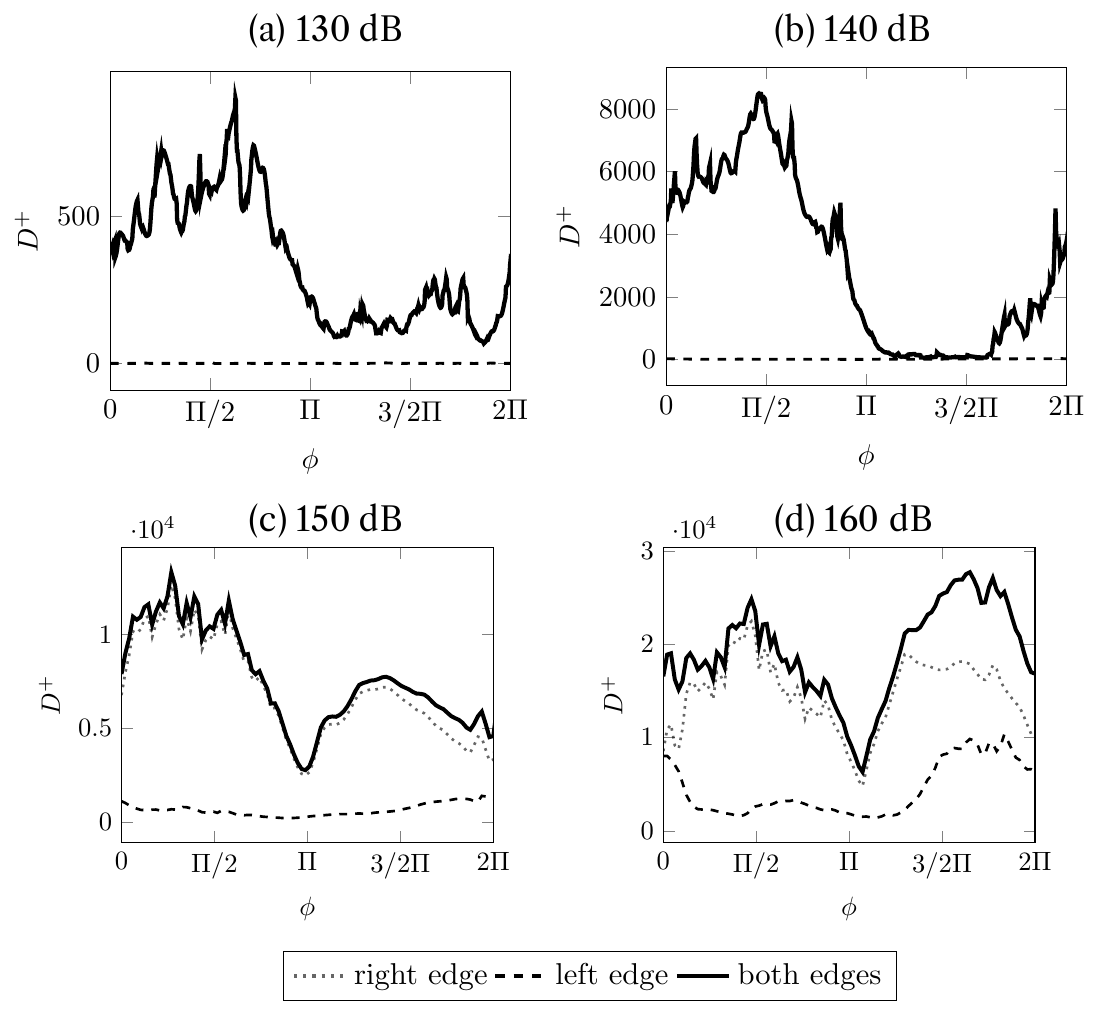}
    \caption{Phase averaged viscous dissipation rate \red{in viscous units, $D^+$} over one cycle at different SPL, left and right orifice edge contribution reported separately, M=0.3 condition, forcing frequency equal to 2200 Hz.}
    \label{fig:visc_leftrightM03}
\end{figure}

Figure~\ref{fig:visc_leftrightM03} quantifies separately the upstream and downstream wall contributions. The left side contribution remains negligible across SPL levels, except at 160\,dB, where it becomes significant during the outflow phase (\fref{fig:visc_leftrightM03}(d)). At this high SPL, the strong acoustic-induced outflow perturbs the quasi-steady vortex, allowing part of the flow to reach the upstream side and contributes to viscous dissipation.

\begin{figure}
    \centering
    \includegraphics[width=.7\textwidth]{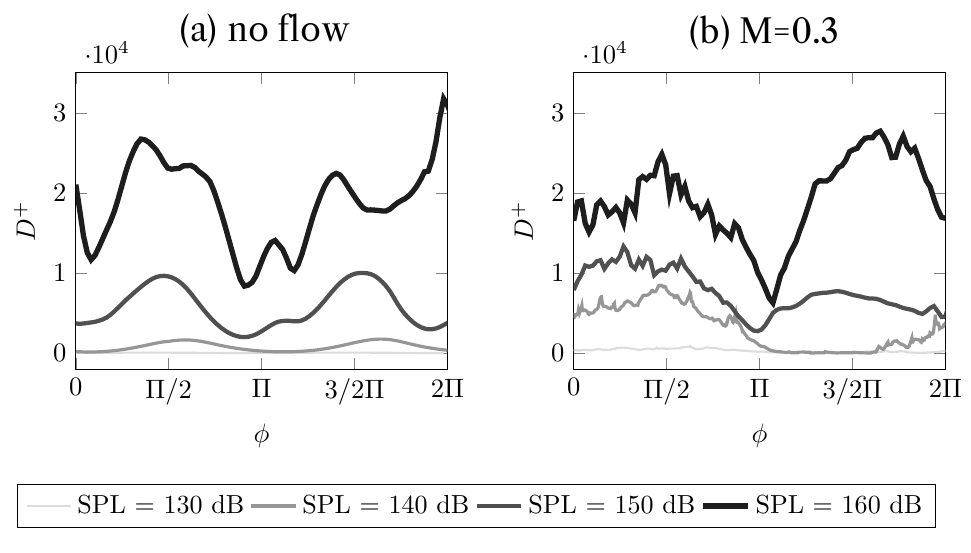}
    \caption{Phase averaged viscous dissipation rate \red{in viscous units, $D^+$} over one cycle at different SPL, contribution of both edges; (a) no flow condition, (b) M=0.3 condition, forcing frequency equal to 2200 Hz.}
    \label{fig:visc_SPL}
\end{figure}

Macroscopic differences in the viscous dissipation are evident in the phase evolution of the cumulative dissipation rate (left and right sides combined), summarised in \fref{fig:visc_SPL} for the cases with and without grazing flow. At low SPL, the dissipation rate with grazing flow is an order of magnitude higher than in the no-flow case (as can be seen by looking at the scale in \fref{fig:visc_leftrightM0}(a--b) with \fref{fig:visc_leftrightM03}(a--b)). The outflow phase contribution remains negligible at low SPL (\fref{fig:visc_SPL}(b)) because the grazing flow inhibits the development of acoustic-induced flow during the outflow phase. As SPL increases, the outflow contribution becomes increasingly significant, and at 160\,dB, a two-lobe pattern re-emerges, indicating that the high amplitude pressure fluctuations overcome the blocking effect of the grazing flow.

Finally, the temporal fluctuations of the dissipation rate during the inflow phase in the presence of flow (\fref{fig:visc_SPL} (b)) are higher than in the no-flow case, because of the turbulence entering the orifice. In contrast, the outflow phase shows smoother variations, suggesting that most turbulent structures are dissipated after the inflow, and the flow grazing the walls in the outflow phase contains lower amplitude turbulent fluctuations—similar to the behavior observed in the no-flow case (\fref{fig:visc_SPL} (a)).

\red{
\subsection{Three-dimensional analysis of the dissipation mechanisms at 150 dB}
\label{sec:3D_analysis}

To verify that the two-dimensional parametric analysis provides a meaningful
representation of the dissipation mechanisms, a full three-dimensional
evaluation was performed for one representative configuration, consisting in a volume around the central orifice. The configuration analysed corresponds to SPL = 150 dB at $M=0$ and $M=0.3$.

Figures~\ref{fig:diss_3D_M0} and \ref{fig:diss_3D_M03} summarize the
dissipation by vortex shedding and viscous effects for $M=0$ and $M=0.3$,
respectively. The first four subfigures show iso-surfaces of the normalized
power density $\Pi_g^+$ and the normalized viscous dissipation density
$\Phi^+$ during the inflow and outflow phases. For the vortex-shedding
contribution, both positive and negative iso-surfaces are reported (as
specified in the captions) to emphasize that the acoustic energy conversion
can change sign depending on the phase.


Subfigures (e) and (f) report the phase-dependent acoustic energy conversion rate $\Pi^+$ and viscous dissipation rate $D^+$ for both the 3D case and the corresponding 2D results. The 2D results are scaled to their 3D counterparts performing the integration of the dissipation densities obtained on the 2D slice assuming axial symmetry. This assumption is well satisfied in the absence of grazing flow, where both dissipation contributions are distributed approximately uniformly around the orifice. In the presence of grazing flow, however, the dissipation becomes azimuthally non-uniform, with a dominant contribution localized near the downstream region of the orifice. The scaling therefore represents an averaged reconstruction of the 3D dissipation by considering only the portion of the azimuth where the dissipation is effectively concentrated, which in the present case corresponds to one third of the circumference.

\begin{figure}[htpb]
    \centering
    \includegraphics[width=.85\textwidth]{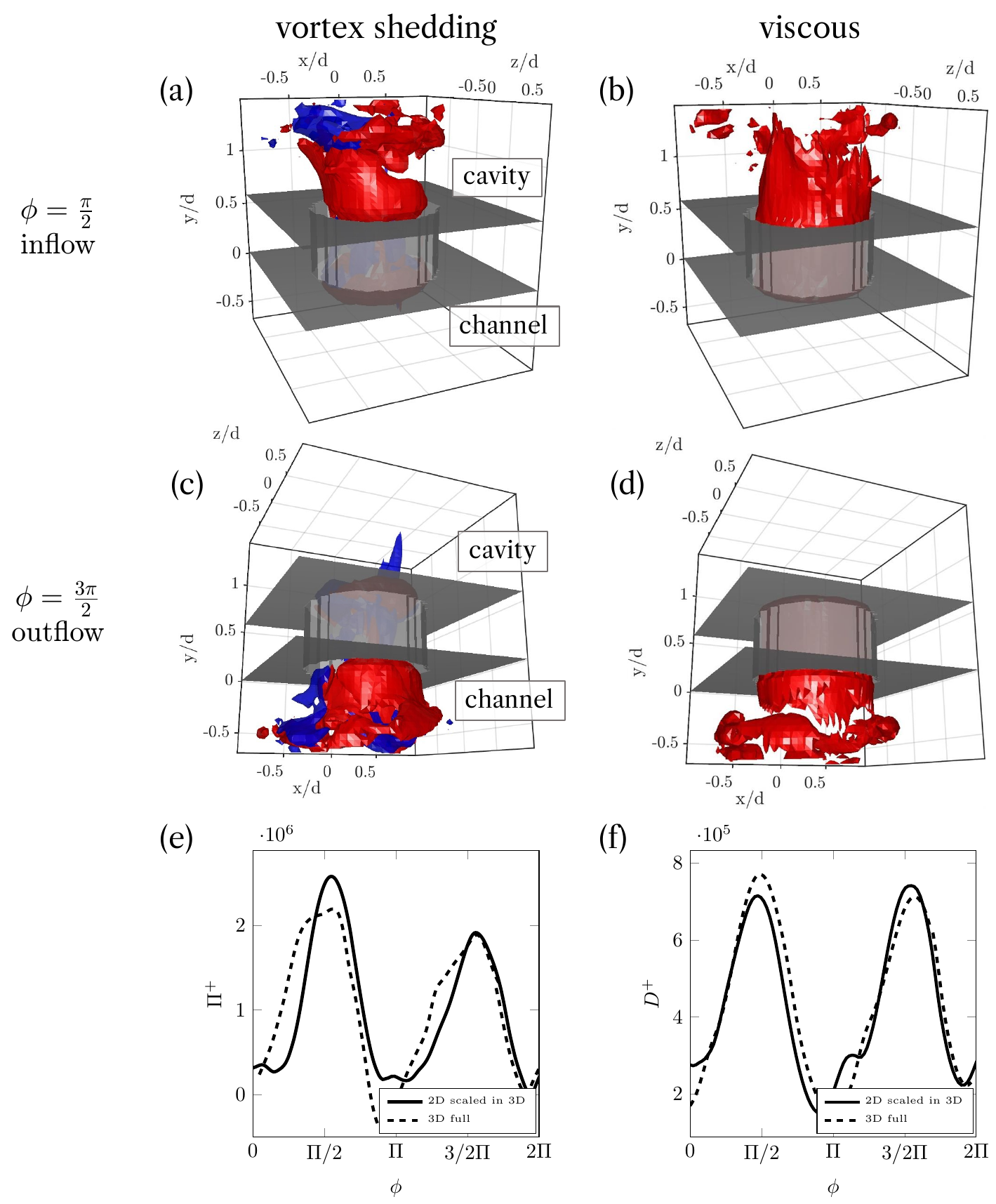}
    \caption{\red{3D dissipation analysis for the case at SPL = 150 dB and M=0; (a) iso-surface of $\Pi_g^+ = \pm 0.1$ in inflow (a) and outflow (c), and iso-surface of $\Phi^+ = 0.06$ in inflow (b) and outflow (d); Dissipation rate as function of the phase and comparison with the scaled 2D results, (e) $\Pi^+$ and (f) $D^+$.} }
    \label{fig:diss_3D_M0}
\end{figure}

\begin{figure}[htpb]
    \centering
    \includegraphics[width=.85\textwidth]{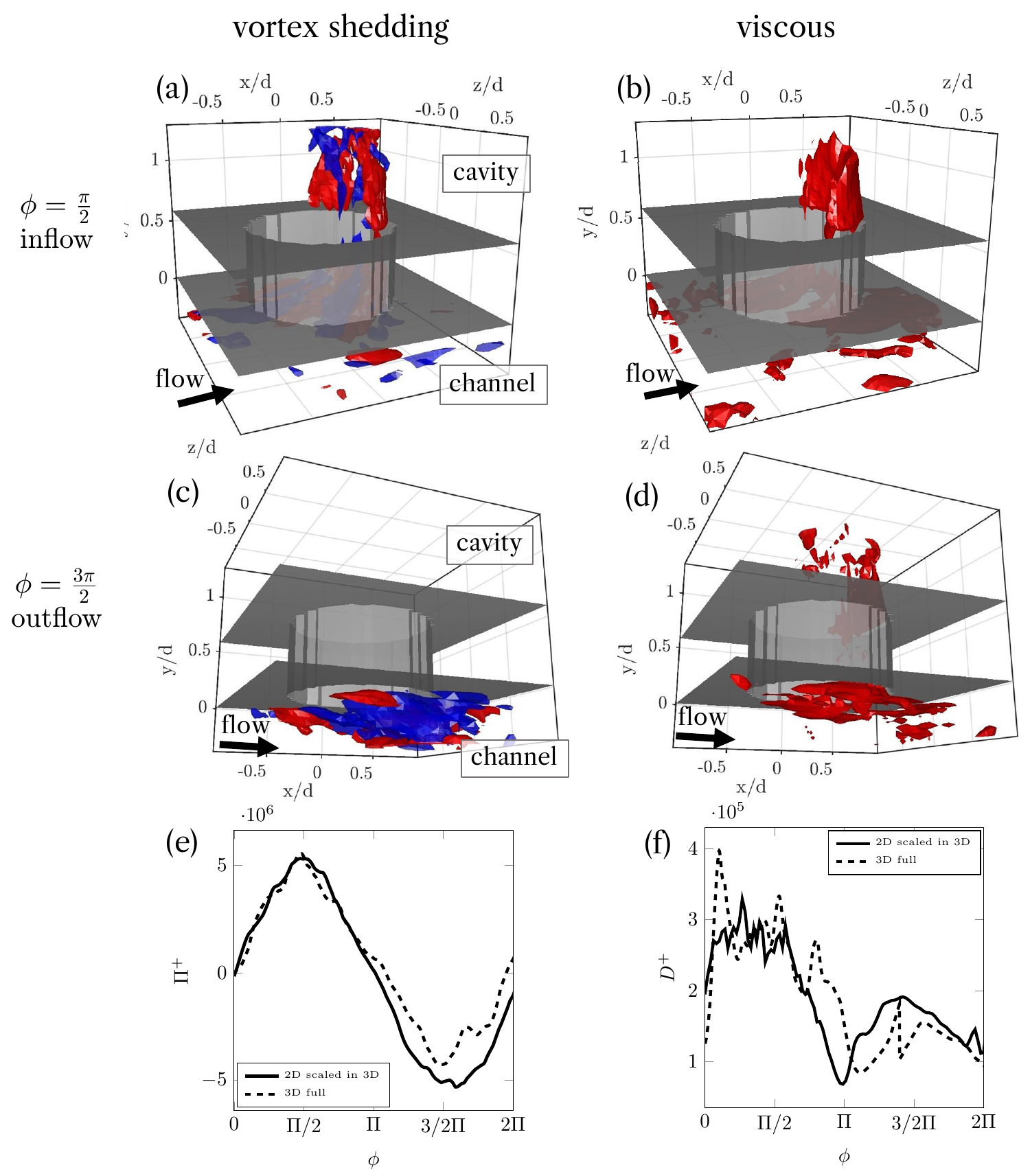}
    \caption{\red{3D dissipation analysis for the case at SPL = 150 dB and M=0.3; (a) iso-surface of $\Pi_g^+ = \pm 5$ in inflow (a) and outflow (c), and iso-surface of $\Phi^+ = 0.1$ in inflow (b) and outflow (d); dissipation rate as function of the phase and comparison with the scaled 2D results, (e) $\Pi^+$ and (f) $D^+$.}}
    \label{fig:diss_3D_M03}
\end{figure}

The comparison shows good agreement between the 2D and 3D results in
terms of trends, phase dependence, and scaled magnitude. In particular,
the characteristic two-lobe behaviour of the vortex-shedding contribution
over the acoustic cycle at $M=0$ is preserved, as well as the phase
dependence observed at $M=0.3$.

Some differences can be observed in the presence of grazing flow. In the
3D results, the negative contribution associated with the outflow phase
is less pronounced than in the 2D case, leading to vortex shedding dissipation contribution that is close to zero and not negative.
In addition, when integrating over the full three-dimensional volume,
the relative contribution of viscous dissipation decreases compared to
the vortex-shedding term. This is due to the fact that viscous dissipation
is confined to a limited near-wall region inside the orifice, whereas the
vortex-shedding contribution extends over a larger portion of the flow
domain. As a result, at 150~dB the 3D evaluation yields a smaller relative
viscous contribution than suggested by the 2D slice analysis.

Overall, the 3D results confirm that the dissipation mechanisms identified
in the 2D parametric study are physically consistent and that the 2D
formulation captures the dominant trends.

}

\red{
\subsection{Energy dissipated per unit time and per unit volume}
\label{sec:budget}

The net acoustic energy per unit time transferred to the vortical field by vortex shedding and the energy dissipated directly by viscous effects, as functions of SPL, are shown in \fref{fig:energy_budget}(a) for both the no-flow and grazing flow cases. All values are normalised by the corresponding impinging acoustic energy upstream of the orifice, $E_i$. The details of the calculation of $E_i$ are reported in the Appendix \ref{appB}.

In the following, in order to use the two dimensional results previously described, the energy quantities are scaled to their
three-dimensional counterparts thus evaluating the net acoustic energy dissipated per unit volume under the hypothesis of axial-symmetry. 
For validation purposes, the full three-dimensional energy evaluation is
reported for the representative configuration described above. The comparison
between the 3D energy values and the scaled 2D estimates shows good
agreement. This confirms that the scaled 2D
formulation captures the dominant trends.

In the absence of grazing flow (red curve, solid squares), the net acoustic energy converted into vorticity increases with SPL: it remains negligible at 130\,dB, becomes slightly larger at 140\,dB, and rises more consistently at 150\,dB before levelling off at 160\,dB. Despite differences in normalisation and the definition of $E_i$, these results are broadly consistent with earlier findings by \citet{tam_microfluid_2000}. Conversely, the viscous dissipation contribution (red curve, empty triangles) dominates at low SPL but its relative importance decreases as the SPL increases, becoming lower than the vortex shedding contribution at 150\,dB.

\begin{figure}[htpb]
    \centering
    \includegraphics[width=.85\textwidth]{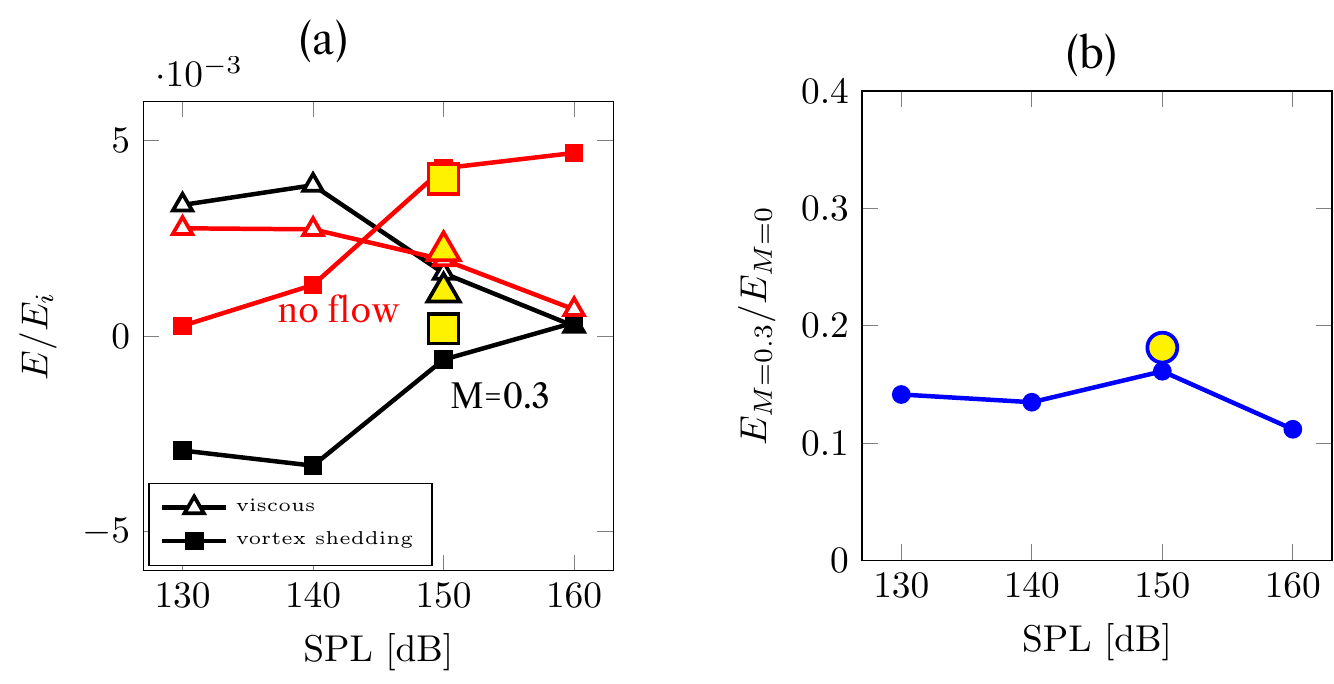}
    \caption{\red{Noise dissipation contributions as function of the SPL, (a) normalized energy per unit time and per unit volume dissipated by viscous and vortex shedding for the no flow case and for the M=0.3, (b) ratio between the total dissipation in the presence and in the absence of grazing flow; the values are obtained for 2D computations and scaled in 3D, yellow markers represent the full 3D computations.}}
    \label{fig:energy_budget}
\end{figure}

When the grazing flow is introduced (black line in \fref{fig:energy_budget}(a)), the relative importance of the dissipation mechanisms change substantially. At low SPL (130 and 140\,dB), the viscous dissipation becomes larger than in the no-flow case. It increases slightly from 130 to 140\,dB, highlighting that, at low to moderately low SPL, grazing flow strongly modifies the dissipation process by increasing the viscous contribution. However, at higher SPL, the viscous dissipation in the grazing flow case approaches the level observed without flow, suggesting that the effect of high amplitude pressure fluctuations becomes predominant over flow-induced effects.

It is worth to note that the net vortex shedding contribution becomes negative up to 150\,dB, indicating that, rather than dissipating acoustic energy, the vortex shedding process contributes to acoustic energy generation. This effect is driven by the outflow phase, where the acoustic-induced velocity exiting the orifice interacts with the grazing flow, producing vorticity that radiates acoustic waves. At 160\,dB, the strong acoustic-induced velocity extends over a larger portion of the orifice diameter, while the quasi-steady vortex is confined to a narrower region. This allows the inflow phase to dominate over the outflow, resulting in a net positive dissipation. However, the magnitude of this positive contribution remains smaller compared to the no-flow case. Overall, these results suggest that, particularly at moderate SPL in the presence of grazing flow, vortex shedding can have a detrimental effect on the liner's noise absorption performance by reducing the net acoustic dissipation.

The ratio of the total dissipated energy (viscous plus vortex shedding)
in the presence of grazing flow relative to the no-flow condition is shown
as a function of SPL in \fref{fig:energy_budget}(b). 
In the presence of grazing flow, the negative contribution associated with
vortex shedding leads to a systematically lower overall dissipation at all
SPL levels compared to the no-flow case, highlighting its detrimental
effect on the liner acoustic performance.
For the present shear-layer conditions, the total energy dissipated by a
single orifice in the presence of grazing flow is approximately 15\% of
the dissipation observed in the no-flow configuration. This ratio remains
nearly constant across the investigated SPL range.

}

\section{Effect of the acoustic source frequency, propagation direction \red{and interference between adjacent orifices}}

\subsection{Effect of the source frequency}

In this section, we examine the influence of the acoustic excitation frequency while keeping the SPL fixed at $150~\mathrm{dB}$. The analysis is carried out both in the absence and in the presence of grazing flow.

\begin{figure}[htpb]
    \centering
    \includegraphics[width=1.\textwidth]{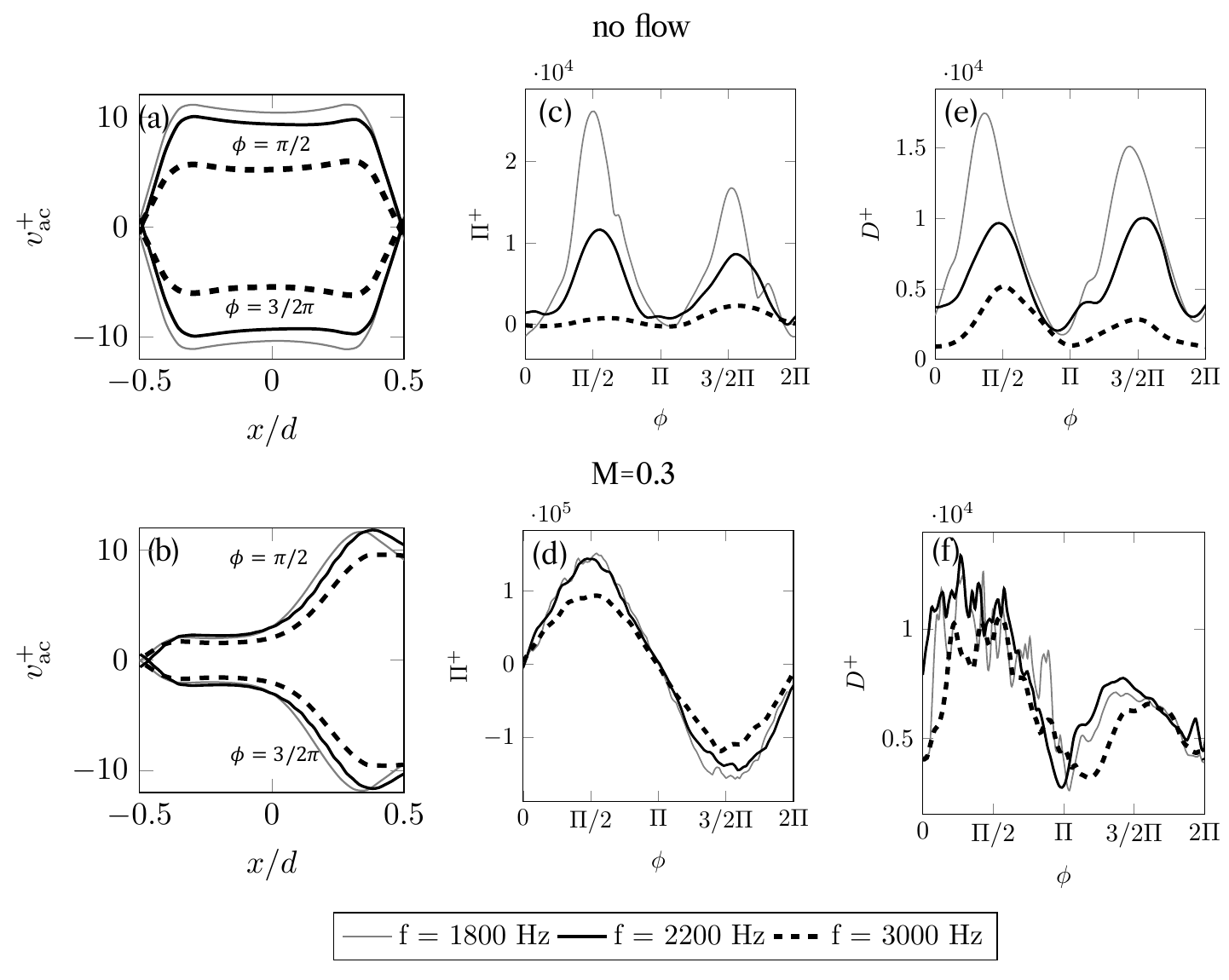}
    \caption{Effect of changing the source frequency on the acoustic-induced velocity profiles (a-b), rate of the acoustic energy dissipation by vortex shedding \red{in viscous units}, \red{$\Pi^+(\phi)$} (c-d), viscous dissipation rate, \red{$D^+(\phi)$} in one cycle considering both orifice edges (e-f). The SPL is fixed at 150 dB, \red{first row} no flow condition, \red{second row} $M=0.3$.}
    \label{fig:effect_f}
\end{figure}

In \fref{fig:effect_f} (a,b), the acoustic-induced velocity profiles \red{in wall units} at the orifice centreline are reported for the inflow and outflow phases \red{for three frequency cases, 1800, 2200 and 3000 Hz}. Without grazing flow, the case at 1800~Hz exhibits the largest acoustic-induced velocity, as this frequency is the closest to the resonant frequency of the liner in quiescent conditions (approximately 1600~Hz). As the frequency increases beyond resonance, the amplitude of the velocity decreases accordingly.

When the grazing flow is introduced, the acoustic-induced velocity profiles at 1800~Hz and 2200~Hz become nearly indistinguishable, with only moderate differences near the downstream edge, which appears slightly more energetic at 2200~Hz. This shift can be attributed to a modification of the liner’s effective resonant frequency caused by the grazing flow. The presence of a quasi-steady recirculating vortex at the upstream corner of the orifice reduces the effective flow area, thereby increasing the resonant frequency.


The rate of dissipation by vortex shedding \red{in wall units}, \red{$\Pi^+$}, in the absence of grazing flow is shown in \fref{fig:effect_f} (c). As discussed in the previous sections, the dissipation remains positive in both inflow and outflow phases, with a slightly larger contribution during the inflow. The latter might be due to the attenuation of acoustic waves as they exit the orifice. The largest dissipation occurs at the lowest frequency (1800~Hz), and it decreases as the forcing frequency moves away from resonance.
When the grazing flow is present (\fref{fig:effect_f} (d)), the dissipation by vortex shedding becomes positive during the inflow and negative during the outflow. The outflow contribution dominates thus resulting in a net negative shedding dissipation. 
The viscous dissipation rate \red{in wall units}, \red{$D^+$}, is reported in \fref{fig:effect_f} (e,f). The frequency dependence is more pronounced without grazing flow (M=0), whereas under grazing flow conditions the flow dynamics dominate, and the influence of the excitation frequency becomes weaker. 

\red{The normalized net acoustic energy dissipated per unit time by viscous effects and vortex shedding as a function of frequency is reported in figure~\ref{fig:EEi_f}(a). Figure~\ref{fig:EEi_f}(b) summarizes the total normalized net acoustic energy per unit time as a function of frequency for both the grazing-flow and no-flow configurations, the results span all the investigated frequencies, from 1200 to 3000~Hz.

\begin{figure}[htpb]
    \centering
    \includegraphics[width=.8\textwidth]{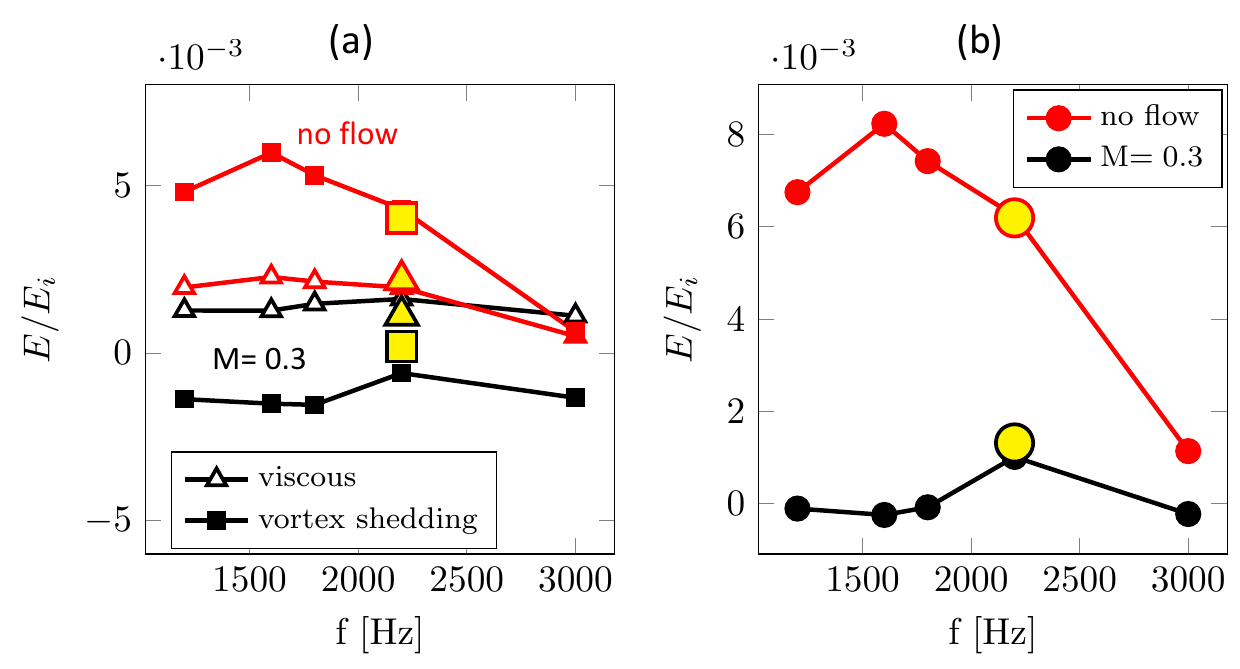}
    \caption{
    \red{Noise dissipation contributions as function of the frequency at SPL=150 dB, (a) normalized energy per unit time and per unit volume scaled in 3D, dissipated by viscous effects and vortex shedding for the no flow case (red) and for the M=0.3 (black), (b) sum of the viscous and shedding contribution; yellow markers represent the full 3D computations.}}
    \label{fig:EEi_f}
\end{figure}

In the absence of grazing flow (figure~\ref{fig:EEi_f}(a), red curves), both dissipation mechanisms remain positive, exhibiting a maximum at 1600~Hz and decreasing at other frequencies. As shown in figure~\ref{fig:EEi_f}(b), the total dissipation peaks at 1600~Hz, which is consistent with the liner’s resonant frequency under quiescent conditions, as predicted by Eq.~\ref{eq:resonance_frequency}.

In the presence of grazing flow, the vortex-shedding contribution (figure~\ref{fig:EEi_f}(a), black squares) remains negative over the entire frequency range, with a local increase in magnitude around 2200~Hz. A moderate increase in viscous dissipation is also observed at 2200~Hz (figure~\ref{fig:EEi_f}(a), black triangles).
At frequencies different than 2200~Hz, the combined effect of acoustic forcing and grazing flow results in a positive viscous contribution that is offset by the negative vortex-shedding term. The magnitude of both contributions decreases as the frequency moves away from resonance due to the reduction of the acoustic-induced velocity. Consequently, the net dissipated energy becomes negligible at frequencies different from 2200~Hz.

In the presence of grazing flow, the frequency at which the maximum total dissipation occurs shifts from 1600~Hz to 2200~Hz (figure~\ref{fig:EEi_f}(b)). This result is consistent with several numerical and experimental studies on acoustic liners \citep{paduano2025impact, Quintino2025, Bonomo2023AProfiles}, which report a shift of the reactance zero crossing toward higher frequencies under grazing flow conditions. This shift is commonly attributed to a modification of the effective porosity and to the interaction between the flow and the acoustic field at the orifice.}

\subsection{Effect of acoustic propagation direction}

\begin{figure}[htpb]
    \centering
    \includegraphics[width=.8\textwidth]{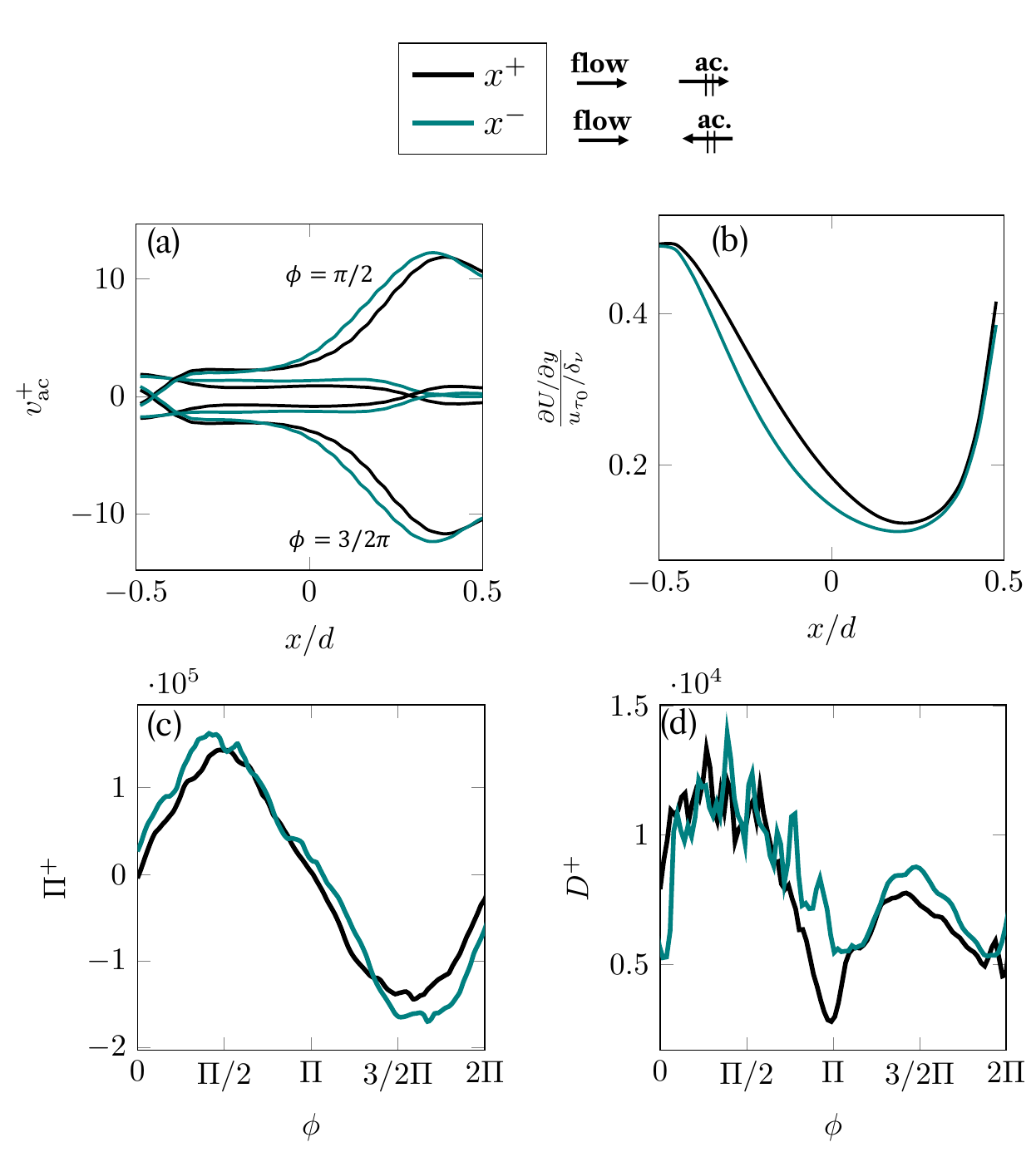}
    \caption{\red{Acoustic wave propagating in the direction opposite to the flow, $M=0.3$ case, forcing frequency equal to \SI{2200}{Hz}, SPL = 150 dB; (a) acoustic-induced velocity profile comparison with the case with the wave propagating in the same direction of the flow, rate of the acoustic energy dissipation by (b) line plots of the
    normalized shear over the aperture of the two orifices at $y/d=0$, (c) vortex shedding, $\Pi^+(\phi)$, and (c) by viscous effects $D^+(\phi)$, comparison with the case with wave propagating in the same flow direction, $x^+$}}
    \label{fig:x+x-}
\end{figure}

In this section, we examine the effect of reversing the acoustic wave
propagation direction while keeping the SPL fixed at
$150~\mathrm{dB}$. Specifically, we consider the case where the acoustic
wave propagates opposite to the grazing flow, denoted as $x^-$, and we compare it with the case where the acoustic wave propagates in the same direction of the flow, $x^+$.

\red{
In \fref{fig:x+x-}(a), the acoustic-induced velocity profile at the
orifice centreline is shown. Compared with the case where the acoustic wave
propagates in the same direction of the grazing flow ($x^+$), only minor differences are
observed. The peak velocity is slightly higher in the $x^-$ case, and shifted towards the center of the orifice. This indicates a modest momentum enhancement through the orifice during both inflow and outflow phases.

In \fref{fig:x+x-}(c–d), the dissipation rates in wall units due to
vortex shedding and viscous effects are reported. For vortex shedding,
both inflow and outflow contributions are marginally larger in the $x^-$
case than in the $x^+$ one. However, the enhanced inflow and
outflow contributions compensate each other over one acoustic cycle, resulting in a negligible net difference.

Viscous dissipation shows a slightly larger contribution in the $x^-$ case,
consistent with the observed thickening of the near-wall region inside the
orifice. The increase is more evident during the inflow phase, where the
dissipation pattern becomes more irregular. During the outflow
phase, the higher exit velocity in the $x^-$ configuration leads to a
modest increase in near-wall viscous dissipation near the downstream lip. This can be seen in \fref{fig:EEi_x+x-_orifices} where a modest enhancement of the viscous dissipation is noticeable with respect to the $x^+$ case.

Overall, reversing the acoustic propagation direction produces only minor modifications of the shear-layer structure and dissipation contributions under otherwise identical boundary-layer and SPL conditions.
This is further confirmed by the total normalized energy dissipated per unit time, reported in \fref{fig:EEi_x+x-_orifices}, which remains nearly
identical for both $x^+$ and $x^-$ configurations.
This supports the view that, for the present configuration, liner
dissipation is primarily governed by the impinging boundary-layer properties and acoustic
amplitude rather than by propagation direction.
Differences in educed impedance when reversing the acoustic propagation direction, reported in some experimental and numerical studies
(e.g.\ \cite{Tam2014, Auregan2004}) may therefore be influenced by modelling assumptions or
streamwise flow-condition variations rather than by a change in flow topology dictated by the acoustic propagation direction \citep{paduano2025impact}.
}

\red{
\subsection{Interference effects between two adjacent orifices}
\label{sec:2orifices}

The interference between two adjacent orifices aligned in the streamwise direction is
investigated in this section. The analysis is conducted on a streamwise plane,
parallel to the plane used for the single-orifice analysis, which intersects two orifices. The objective is to assess how the modification of the
boundary layer induced by the upstream orifice (hereafter referred to as orifice~1)
influences the impinging boundary layer experienced by the downstream orifice (orifice~2),
and how this affects the dissipation mechanisms.
In the present configuration, the variation of the impinging SPL between
the two orifices is modest (less than 1~dB), allowing a comparison of the shear-layer
properties and dissipation mechanisms under nearly iso-SPL conditions.
For the sake of conciseness, the analysis is limited to a single representative
configuration in the presence of grazing flow at $M=0.3$, with an imposed SPL of 150~dB and a forcing frequency of 2200~Hz. 

\begin{figure}[htpb]
    \centering
    \includegraphics[width=1\textwidth]{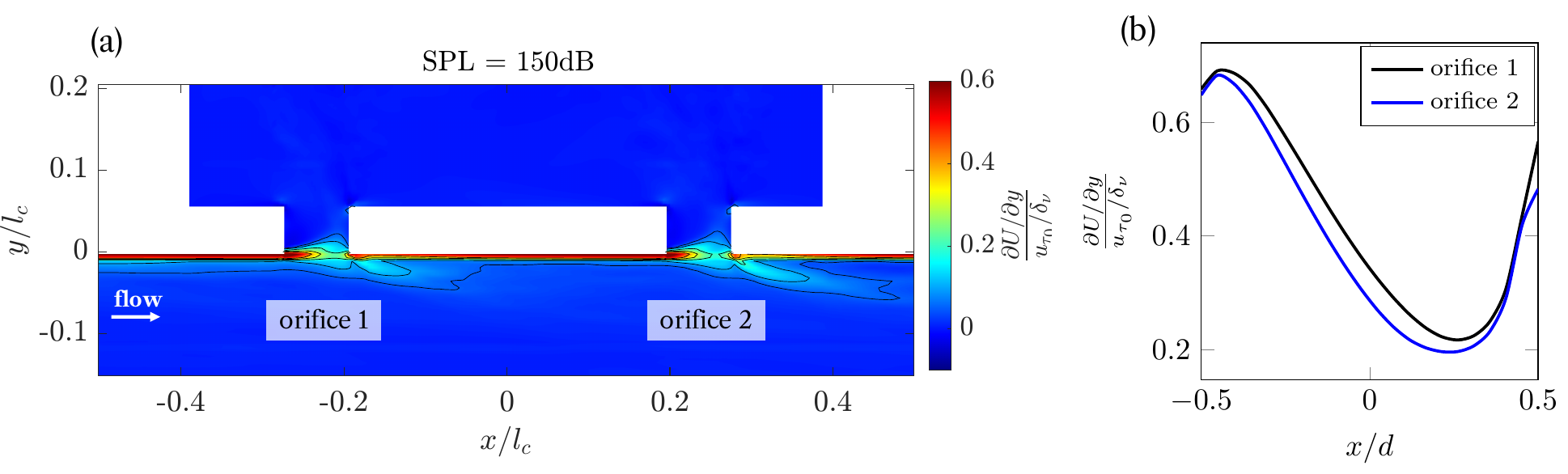}
    \caption{(a) Contours of the normalized shear forming at the mouth of two adjacent
    orifices in the presence of grazing flow at $M=0.3$, SPL = 150~dB and forcing frequency of 2200 Hz. (b) Line plots of the
    normalized shear over the aperture of the two orifices at $y/l_c=0$.}
    \label{fig:shear_2orif}
\end{figure}

As discussed in section~\ref{sec:flowfield}, the acoustic-induced velocity during the
outflow phase generates a jetting-like mechanism at the orifice mouth, which displaces the
near-wall fluid away from the wall and modifies the downstream boundary layer. In the
present configuration, this mechanism leads to an increase of the displacement thickness
$\delta^*$ downstream of orifice~1, of approximately $1\%$.
As a consequence, orifice~2 is exposed to an impinging boundary layer characterized by a
larger displacement thickness and a weaker shear layer. This effect is shown by the
contours of the normalized shear in \fref{fig:shear_2orif}(a) and quantified through the
comparison of the line plots across the apertures of the two orifices in
\fref{fig:shear_2orif}(b).

\begin{figure}[htpb]
    \centering
    \includegraphics[width=1\textwidth]{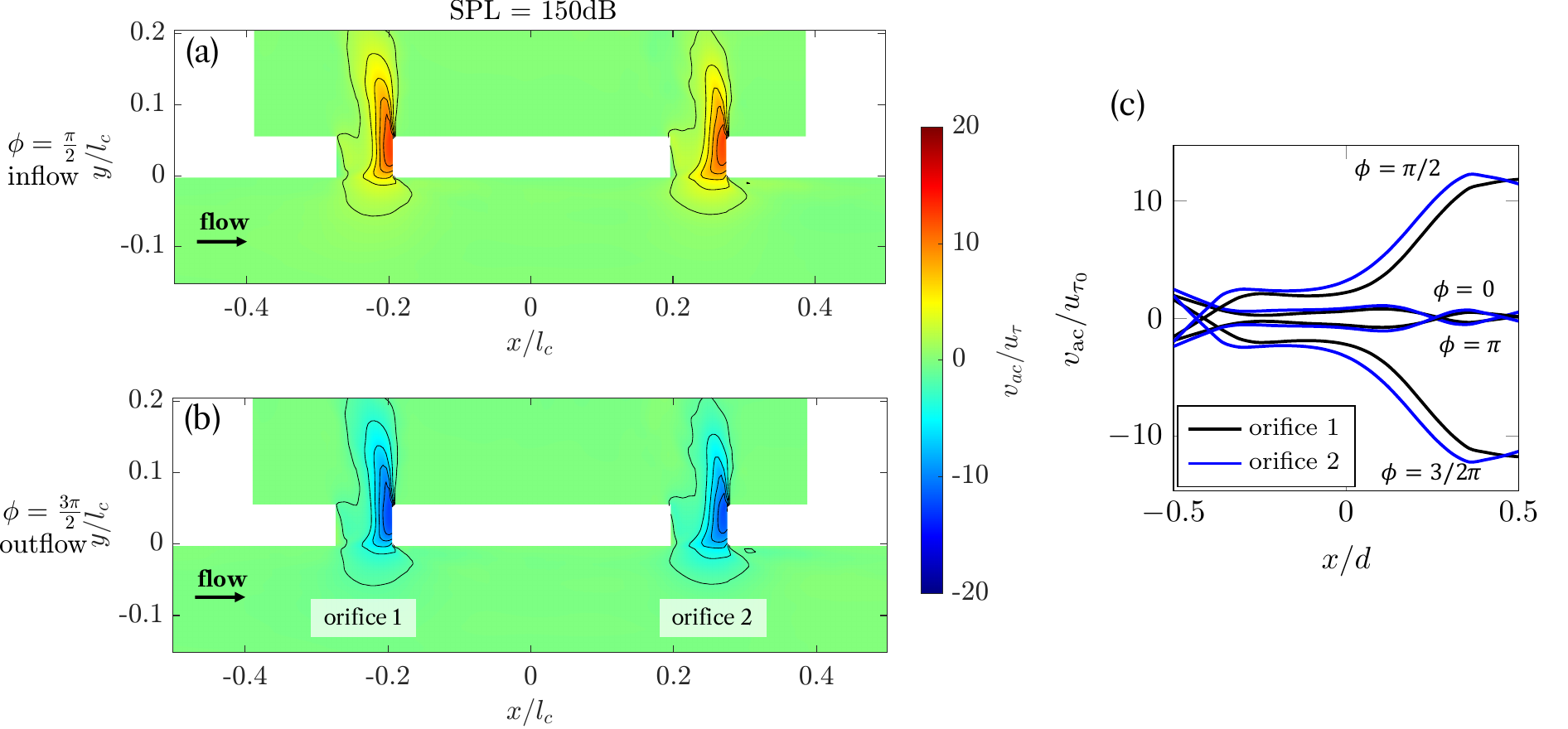}
    \caption{Contours of the wall-normal acoustic-induced velocity at the inflow
    ($\phi=\pi/2$) and outflow ($\phi=3\pi/2$) phases for $M=0.3$ and 150~dB, streamwise plane
    containing two adjacent orifices.}
    \label{fig:Vac_2orif}
\end{figure}

As a consequence, the local porosity of the
orifice 2 increases and the size of the quasi-steady vortex forming within the neck is reduced. This results in a slightly larger acoustic-induced velocity inside orifice~2 and in the
development of a thicker boundary layer along the internal walls of the orifice. These
effects are visible in the contours of the acoustic-induced velocity shown in
\fref{fig:Vac_2orif}(a,b) and in the corresponding line plots extracted at the
centreline of the two orifices at different phases (\fref{fig:Vac_2orif}(c)).
We note that the phase reference used in these plots is defined locally with respect to the passage of the acoustic wave at each orifice. Since the acoustic wave propagates from the upstream to the downstream orifice,
a finite convection time introduces a time delay between the two locations.
As a result, the phase is evaluated based on the local arrival of the acoustic wave at each orifice.

The increase of the acoustic-induced velocity downstream translates into a modification of
the dissipation for the second orifice. In particular, the conversion of acoustic energy into vortical
energy is locally enhanced by the increase in acoustic-induced velocity. This is evidenced by the
contours of the power density transferred from the acoustic field to the vortical field,
$\Pi_g^+$, shown in \fref{fig:shedding_2orif}(a,b), where larger values are observed in
orifice~2 during both the inflow and outflow phases.

\begin{figure}[htpb]
    \centering
    \includegraphics[width=1\textwidth]{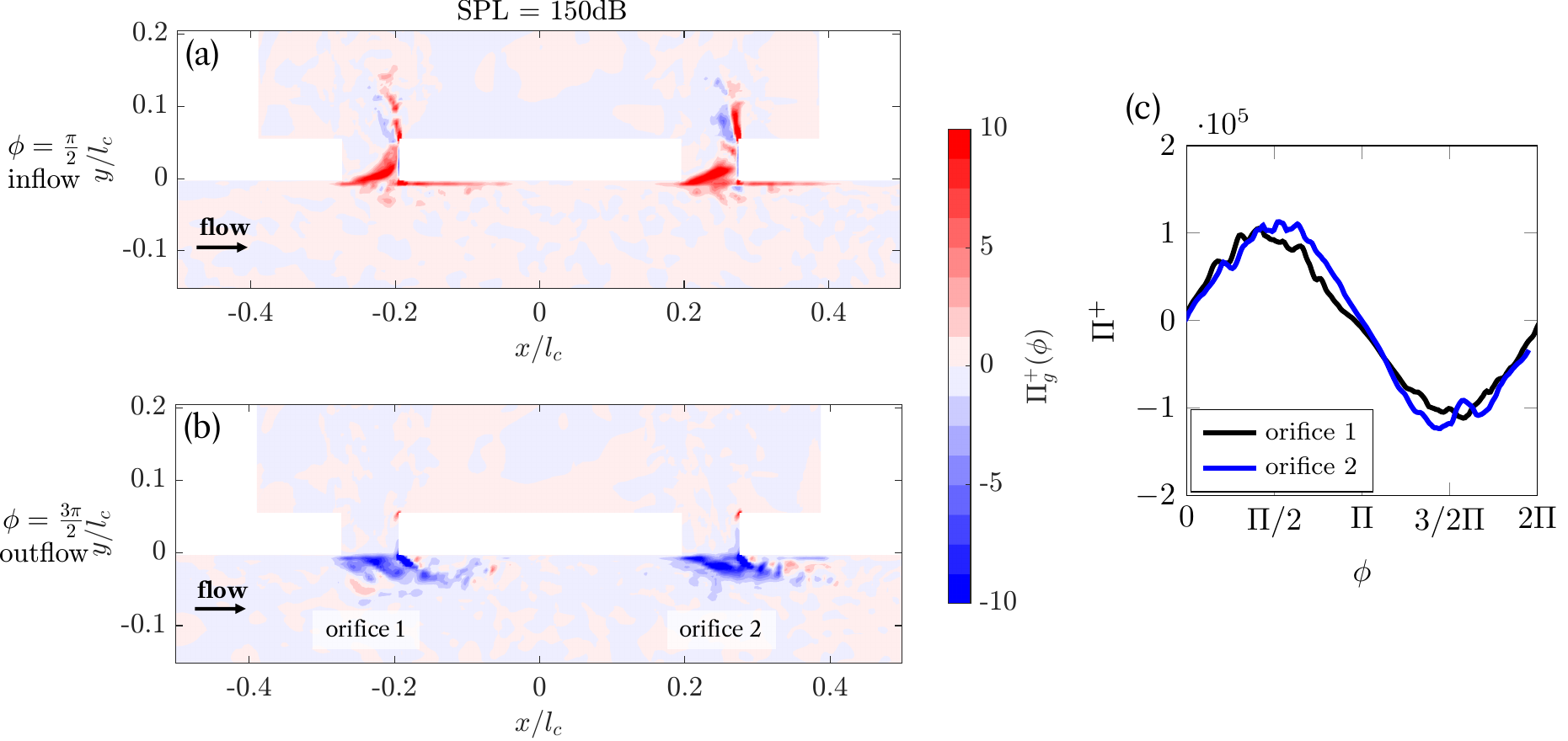}
    \caption{Contours of the normalized power density $\Pi_g^+$ transferred from the
    acoustic field to the vortical field during the inflow (a) and outflow (b) phases for
    $M=0.3$, 150~dB, and a forcing frequency of 2200~Hz, streamwise plane containing two
    orifices. (c) Phase-averaged acoustic energy transfer rate by vortex shedding,
    $\Pi^+(\phi)$, over one acoustic cycle, comparison between orifice~1 and orifice~2.}
    \label{fig:shedding_2orif}
\end{figure}

The phase-averaged dissipation rates integrated over the regions surrounding the two
orifices are reported in \fref{fig:shedding_2orif}(c). While a modest increase in the
instantaneous values during both inflow and outflow is observed for orifice~2, the net
cycle-averaged contribution remains nearly identical to that of orifice~1 due to the
phase-dependent sign changes of $\Pi^+(\phi)$. This is evident looking at \fref{fig:EEi_x+x-_orifices} where the dissipations contributions of the two orifices are compared.

A more pronounced difference is observed when considering viscous dissipation at the
orifice walls. The increase in acoustic-induced velocity in orifice~2 leads to an
enhancement of the viscous dissipation rate $D^+$ during the inflow phase, while negligible
changes are observed during the outflow phase. As a result, the cycle-averaged viscous
dissipation in orifice~2 is approximately $5\%$ higher than in orifice~1 (\fref{fig:EEi_x+x-_orifices}).

\begin{figure}[htpb]
    \centering
    \includegraphics[width=1\textwidth]{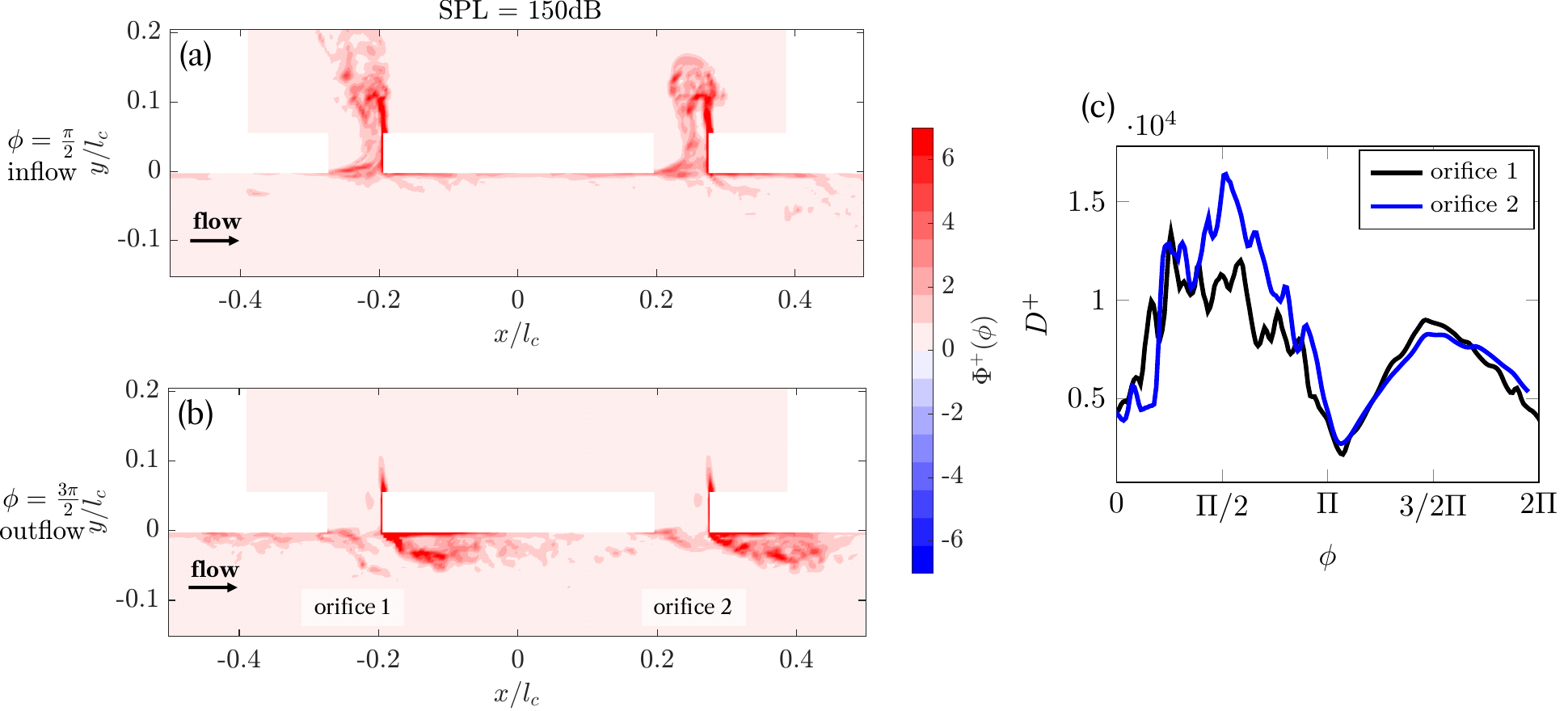}
    \caption{Contours of the viscous dissipation density during the inflow (a) and outflow
    (b) phases for $M=0.3$, 150~dB, and a forcing frequency of 2200~Hz, streamwise plane
    containing two orifices. (c) Phase-averaged viscous dissipation rate, $D^+(\phi)$, over
    one acoustic cycle, comparison between orifice~1 and orifice~2.}
    \label{fig:viscous_2orif}
\end{figure}

These results indicate that a modification of the impinging boundary layer,
such as that induced by the presence of an upstream orifice affects the shear
layer at the orifice mouth, the in-orifice flow topology, and the magnitude of the two
dissipation mechanisms. In the present configurations, these modifications
remain modest, as the variation in displacement thickness between the two
orifices is limited. Nevertheless, the analysis highlights the sensitivity
of the dissipation process to the properties of the impinging boundary layer.

\begin{figure}[htpb]
    \centering
    \includegraphics[width=.6\textwidth]{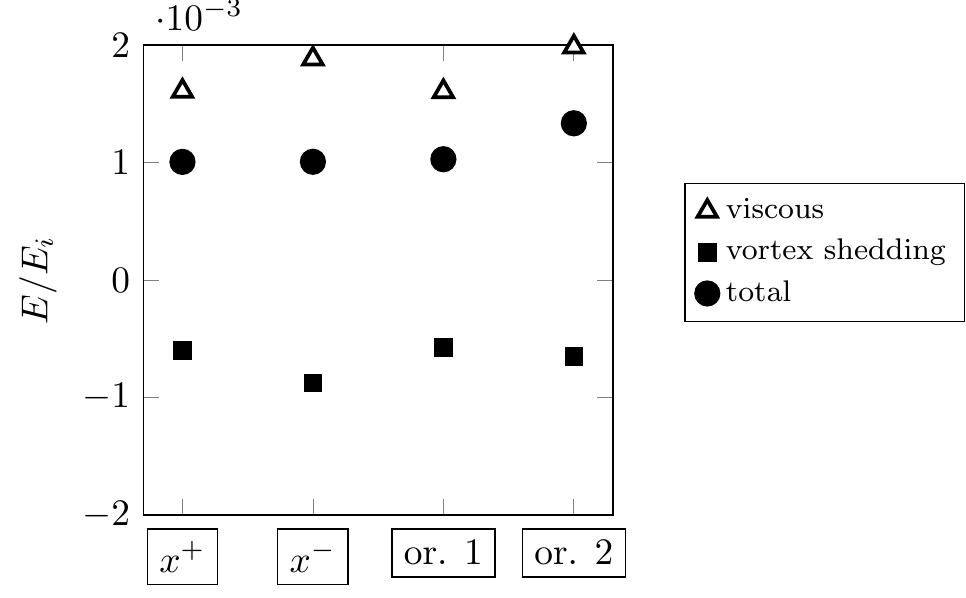}
    \caption{Noise dissipation contributions, normalized energy per unit time and per unit volume scaled in 3D dissipated by viscous and vortex shedding at SPL=150 dB, $M=0.3$, comparison between $x^+$, $x^-$ (central orifice), and the two orifice dissipation contributions from the streamwise plane analysis that intersects two adjacent orifices.}
    \label{fig:EEi_x+x-_orifices}
\end{figure}

This observation is particularly relevant for finite multi-cavity liners,
where the boundary layer evolves
progressively along the streamwise direction. In such configurations,
cumulative modifications of the impinging boundary layer may lead to
more pronounced variations in local dissipation characteristics. For a finite acoustic liner composed of multiple cavities, previous studies (e.g.\
\cite{paduano2025impact}) have shown a pronounced streamwise decrease in SPL (up to
10~dB) accompanied by a significant increase in the displacement thickness (up to 14$\%$ from the first to the eleventh cavity of the liner). In such configurations, the local dissipation is expected to depend on
the combined effects of the local SPL and the local boundary-layer properties. 
The present analysis could therefore be performed on finite liners composed of multiple cavities to fully characterize the streamwise distribution of the dissipation.

}

\section{Concluding remarks}
\noindent

The present study investigates the dissipation mechanisms of an acoustic liner subjected to grazing turbulent flow through high-fidelity lattice-Boltzmann very-large-eddy simulations.  
A two-dimensional domain centred on a single orifice of a single cavity liner was analysed to isolate the physical mechanisms responsible for acoustic energy dissipation, independently from flow development effects.  
The analysis covered a range of SPLs from 130 to 160~dB and frequencies between \red{1200} and 3000~Hz. \red{To validate the two-dimensional parametric study, for one representative condition (SPL = 150 dB, frequency 2200 Hz, with and without flow), the analysis are conducted on a three-dimensional domain encompassing one orifice.}
The main findings are summarised below, addressing the key research questions posed in the introduction.

The presence of a grazing turbulent flow alters the flow topology inside the orifice and, through that modification, the mechanisms of acoustic energy dissipation. In the no-flow configuration, the acoustically induced velocity field is almost symmetric across the orifice and both inflow and outflow phases contribute positively to the conversion of acoustic energy into vortical motion and to viscous losses. By contrast, grazing flow generates a near-wall shear layer above the orifice and produces a quasi-steady recirculating vortex occupying the upstream half of the orifice. This vortex reduces the effective open area of the orifice and confines the acoustic-induced motion to the downstream half. The altered topology concentrates the regions of strong vorticity production and wall shear at the downstream half of the orifice and modifies the spatial distribution of the dissipation rate.

In the absence of grazing flow, both the inflow and outflow phases contribute positively to the dissipation of acoustic energy, with vortex shedding dominating at high SPL and viscous effects prevailing in the linear regime. The topological change introduced by the grazing flow has two direct and measurable consequences for the dissipation budget. First, the contribution of vortex shedding becomes strongly phase dependent: the inflow phase shows enhanced conversion of acoustic energy into vorticity because turbulent structures from the grazing flow are entrained into the orifice, whereas the outflow phase acts as an acoustic source. During the outflow, the acoustic-induced velocity interacts with the grazing flow and produces vortical structures (akin to those produced by a jet in cross-flow) that radiate acoustic energy; consequently the net shedding contribution (integrated over a full cycle) reduces and may become negative at low–moderate SPL. Second, viscous dissipation at the orifice walls increases where the grazing flow pushes fluid toward the downstream lip: at low SPL this wall contribution is larger than in the no-flow case but limited to the inflow phase. At sufficiently high SPL the viscous term approaches values comparable with the quiescent configuration. At the highest SPLs considered, in fact, inflow and outflow contributions to the viscous dissipation become more symmetric and the overall asymmetries between upstream and downstream orifice lip and inflow and outflow phases are attenuated, approaching the condition obtained in the absence of flow.
However, the reduced net contribution from shedding in the presence of grazing flow remains a principal factor lowering the liner’s net energy absorption compared with the no-flow case.

In sum, the grazing flow modifies where and in which phase energy is exchanged between acoustics and fluid motion. At low SPL, viscous dissipation dominates during the inflow phase and it is concentrated near the downstream lip of the orifice, whereas the outflow phase provides a negligible contribution to the viscous dissipation. The dominant mechanism by which the grazing flow reduces the liner’s net absorption is the transformation of the outflow phase from a dissipative to a generative mechanism of acoustic energy via vortex shedding, only partially compensated by the increased viscous losses at the downstream wall. 

Additional analyses demonstrated that the excitation frequency and the direction of acoustic wave propagation also influence the liner response.  
At constant SPL, the dissipation reaches its maximum near the resonant frequency of the system.  
In the absence of flow, this occurs around 1800~Hz, while in the presence of grazing flow the resonant frequency shifts to approximately 2200~Hz owing to a reduced effective porosity induced by the quasi-steady vortex. 

\red{The effect of the streamwise development of the flow was analysed for one
representative case by extracting a two-dimensional plane containing two
consecutive orifices. The downstream orifice experiences the influence of
the upstream orifice, which leads to a slightly weaker shear layer at the orifice mouth. This modification results in a modest increase in the acoustic dissipation.}
Similarly, when the acoustic wave propagates opposite to the grazing flow, the shear layer above the orifice is weakened, promoting deeper acoustic penetration into the cavity and slightly enhancing overall dissipation.

Future work shall focus on the streamwise evolution of the dissipation along multi-orifice liners, where the grazing flow development and the mutual interaction between cavities are expected to further influence the local acoustic dissipation. Ongoing work is dedicated to assess, from a dissipation standpoint, how variations in orifice geometry, such as the edge shape, modify the viscous and vortex-shedding mechanisms.
Such insights may guide the design of advanced liners, including meta-structured orifices or flow-control strategies that mitigate outflow-induced acoustic generation.

\section*{Declaration of Interests}

The authors report no conflict of interest.

\section*{Funding Sources}

The work of F. Scarano, A. Paduano and F. Avallone is co-funded by the European Union (ERC, LINING, 101075903). Views and opinions expressed are however those of the author(s) only and do not necessarily reflect those of the European Union or the European Research Council. Neither the European Union nor the granting authority can be held responsible for them.

\begin{appen}

\section{Grid convergence study, validation of the baseline cases with reference experiments and impedance results in the presence and in the absence of grazing flow}\label{appA}

\subsection{Flow validation}

\red{

The incoming turbulent boundary layer for the case with grazing flow is presented and the convergence of the mesh is reported. Three different mesh refinements are reported: 20, 40 and 80 voxels/mm.

\begin{figure}[htpb]
    \centering
    \includegraphics[width=1.\textwidth]{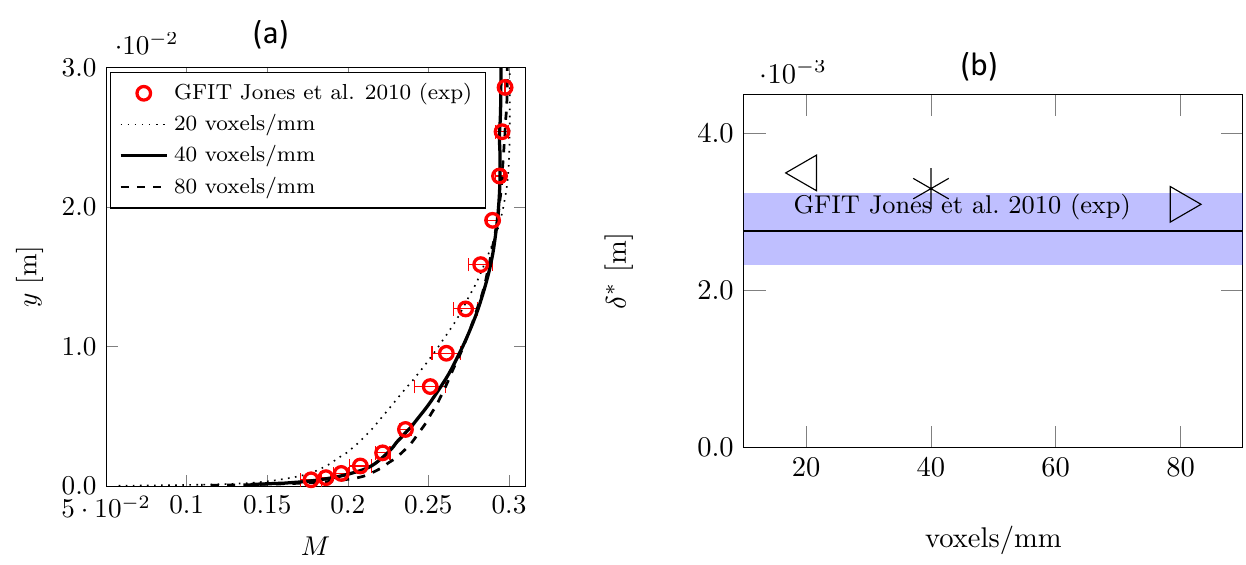}
    \caption{(a) Streamwise Mach number profile of the incoming turbulent grazing flow for three grid resolutions compared with the data by \cite{Jones2010}; (b) displacement thickness as function of the mesh and comparison with the results by \cite{Jones2010}.}
    \label{fig:validation_Mprofile_deltastar}
\end{figure}

The Mach number profile for the three grid resolutions is plotted in figure \ref{fig:validation_Mprofile_deltastar}(a) and compared against the experimental data of \cite{Jones2010}. The results show that only minor variations are present between the two finest grids. 
The convergence of the mesh is also reported in terms of boundary layer displacement thickness, $\delta^*$. The coarser grid shows a higher value of $\delta^*$, which approaches the value of the reference experiment of \cite{Jones2010} for the two finest grids.

The grid convergence study is reported for the streamwise velocity mean and variance $<u^{2}>$, the profiles are presented in wall-units in \fref{fig:validation_uvarplus}.
The value of the friction velocity, $u_\tau$, was computed from the wall shear stress calculated using the  wall model described in Section \ref{Sec:setup}. 
The mean velocity profile is compared against the law of the wall using constants $\kappa=0.4$ and $B=5.0$ and the experimental results by \cite{DeGraaff2000} at similar Reynolds number.
Figure \ref{fig:validation_uvarplus} (b) presents the streamwise Reynolds stress for different mesh sizes, compared with experimental data reported by \citet{Vallikivi2015} and \cite{DeGraaff2000} at similar Reynolds numbers. 
Overall a good agreement between experiments and simulations is found for the two finest grid resolutions.


\begin{figure}[htpb]
    \centering
    \includegraphics[width=1.\textwidth]{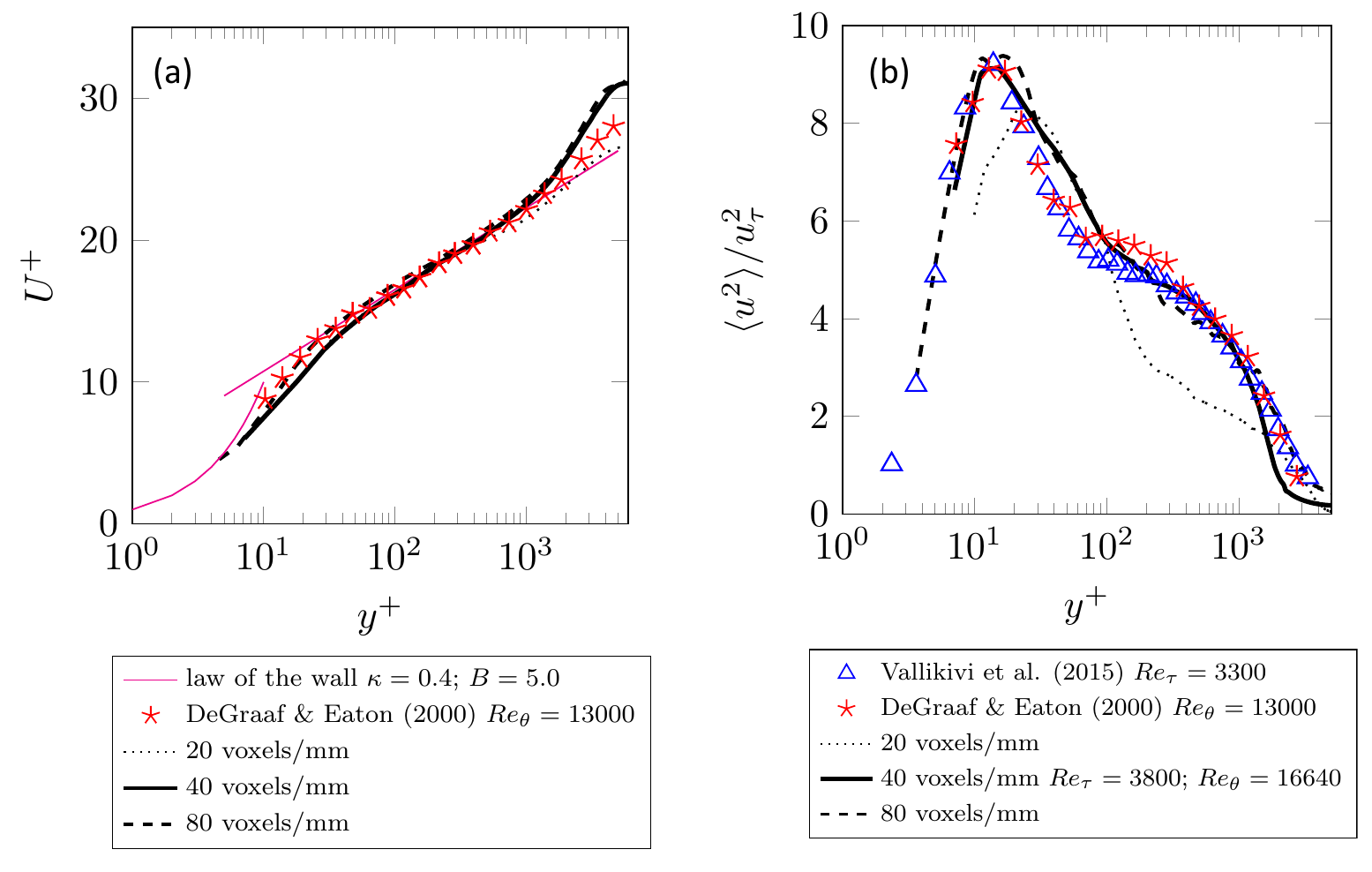}
    \caption{Streamwise mean velocity (a) and variance (b)  profiles in wall units and comparison with log-law and experimental data of \cite{DeGraaff2000} and \cite{Vallikivi2015} for similar Reynolds numbers.}
    \label{fig:validation_uvarplus}
\end{figure}

The friction coefficient as function of $Re_{\theta}$ is reported for the simulation case at 40 voxels/mm in \fref{fig:validation_Cf_retheta}.
The results show a good match with the the semi-empirical relation of Coles-Fernholtz \citep{Nagib, FERNHOLZ1996245} valid for smooth walls and with the oil film interferometry measurements (OFI) by \cite{Osterlund} obtained for similar values of Reynolds numbers. 

\begin{figure}[htpb]
    \centering
    \includegraphics[width=.5\textwidth]{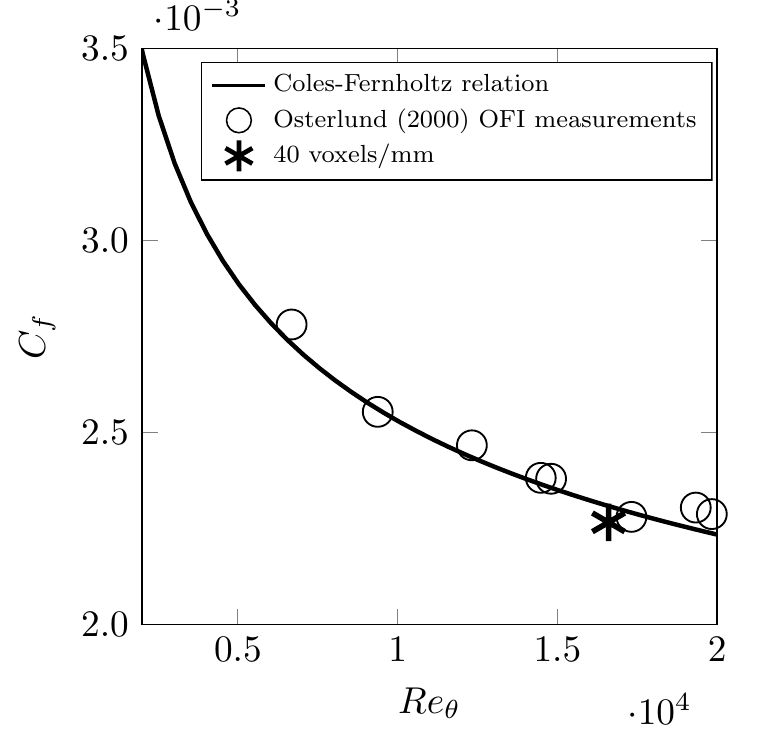}
    \caption{Friction coefficient as function of the Reynolds number based on the momentum thickness, $Re_{\theta}$, comparison with Coles-Fernholtz \cite{Nagib, FERNHOLZ1996245} and the experimental results by \cite{Osterlund}.}
    \label{fig:validation_Cf_retheta}
\end{figure}

\subsection{Statistical convergence}

Figure~\ref{fig:conv_stat} shows the statistical convergence analysis performed in the
single-orifice domain. The convergence is evaluated at a point located within the boundary
layer in proximity of the orifice, $y/d \cong -0.9$, where the interaction between the acoustic forcing and
the grazing flow is expected to be the strongest. The vertical velocity component is considered
without spatial averaging in order to assess local statistical convergence.
The convergence behaviour is shown for two representative SPLs,
130~dB and 160~dB, at 2200 Hz. As expected, convergence is slightly improved at lower SPL, where the
acoustic forcing induces weaker flow perturbations. Nevertheless, the convergence remains
acceptable for both cases, considering the time-resolved nature of the simulations.
Increasing the number of simulated acoustic cycles would further improve statistical
convergence; however, this would require a longer computational domain and significantly
higher computational cost, especially given the number of configurations analysed in the
present study. 

\begin{figure}[ht]
    \centering
    \includegraphics[width=\textwidth]{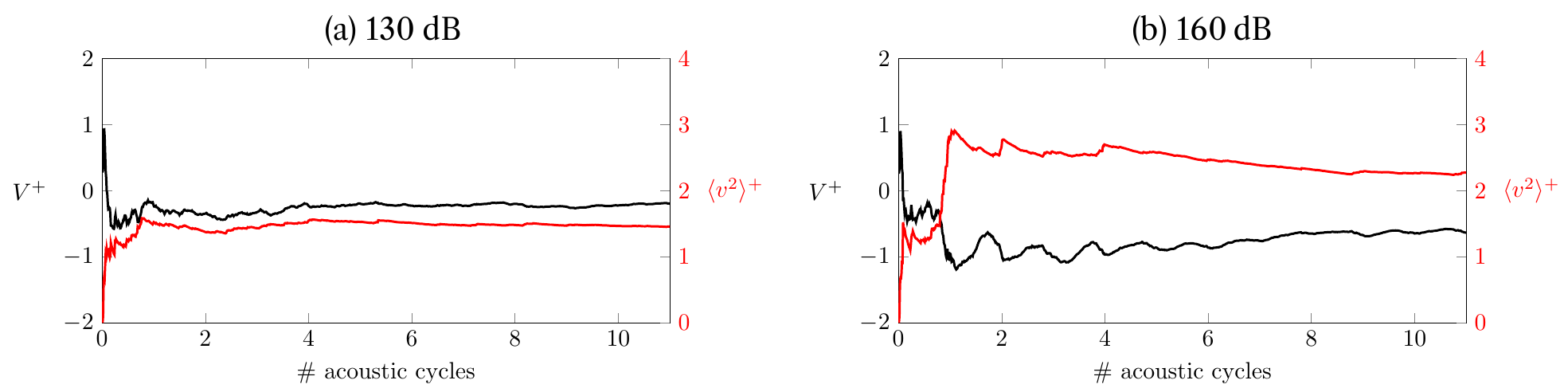}
    \caption{Convergence of the mean and standard deviation of the vertical velocity, $y^+ \cong 250$ $(y/d \cong -0.9)$, taken upstream of the orifices (a) 130 dB and (b) 160 dB both at $M=0.3$ and 2200 Hz.}
    \label{fig:conv_stat}
\end{figure}

}
\red{
\subsection{Computational cost}
To provide the reader with an indication of the computational effort required to build the present numerical database, an overview of the associated cost is given below in table \ref{tab:mesh_dof}. To optimise computational resources, coarser simulations were first carried out to develop the flow field within the channel; their solutions were then used to initialise the fine-grid acoustic simulations. Simulations were performed on 10 compute nodes (480 solver processes in total) connected via InfiniBand, using single precision and AVX2 vectorisation. 
\begin{table}the
\centering
\caption{Number of voxels and degrees of freedom (DOF) for the different mesh resolutions.}
\label{tab:mesh_dof}
\begin{tabular}{lccc}
\hline
Mesh resolution & Number of voxels & DOF, $19N_\mathrm{voxels}$ & CPU time [h]\\
\hline
Medium & $75{,}368{,}079$ & $1{,}431{,}993{,}501$ & 17,224\\
Fine & $475{,}359{,}916$ & $9{,}031{,}838{,}404$ &49,337\\
VeryFine & $2{,}998{,}179{,}779$ & $56{,}965{,}415{,}801$ &68,016\\
\hline
\end{tabular}
\end{table}
}
\subsection{Impedance}

The grid convergence is verified also for the impedance for the simulations without flow at the three frequencies at a fixed SPL = 130 dB. The computed impedance is compared with experimental data obtained at the GFIT facility (GIT-95M test section) \citep{Jones2004}, where impedance was extracted using a two-dimensional finite element method (2D-FEM).
In the present numerical framework, the acoustic impedance is evaluated using the in-situ technique, originally introduced by \citet{Dean1974AnDucts}, which relies on pressure measurements acquired at two distinct locations within the liner: at the face sheet and at the backplate of the cavity. This method, often referred to as the two-microphone approach, provides a local estimation of impedance and is applicable under the assumptions that: (i) the acoustic wavelength is significantly larger than the cavity width; (ii) the cavity behaves as a locally reactive element due to sufficiently thick side walls; and (iii) acoustic waves entering the cavity are fully reflected at the backplate. This method is particularly suited to the current study, as only a single cavity is simulated. 

Time-resolved pressure signals are sampled at discrete locations to reconstruct the transfer function between face sheet and backplate. 
Specifically, pressure at the face sheet is recorded every $30^\circ$ along a circular contour of radius 1$d$ centred on each orifice, while the backplate pressure is sampled at the orifice centres. The resulting transfer functions are then averaged across all spatial locations to improve robustness.
The impedance values finally averaged out of the $7$ orifices.

Following the formulation in \citet{Dean1974}, \citet{Manjunath2018} and \citet{Avallone2019}, the normalized acoustic impedance $Z_f$ is obtained via:
\begin{equation}
    Z_{f} = \frac{Z}{Z_0} = -i \tilde{H}_{fb} \frac{1}{\sin(kd_c)},
    \label{eq:insitu-equation}
\end{equation}
where $Z_0$ is the characteristic impedance of air, $\tilde{H}_{fb} = \tilde{p}_{f} / \tilde{p}_{b}$ is the complex pressure transfer function between face sheet and backplate, $d_c$ is the cavity depth, and $k = \omega / c_0$ is the acoustic wavenumber, with $\omega$ the angular frequency and $c_0$ the speed of sound.

\begin{figure}[htpb]
    \centering
    \includegraphics[width=1.\textwidth]{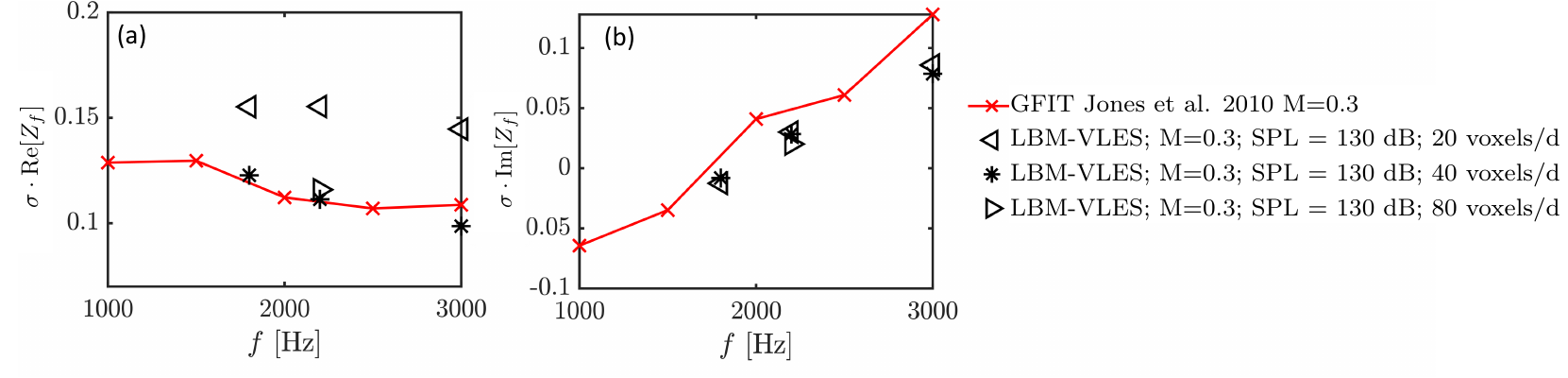}
    \caption{Grid convergence study for the impedance as function of the source frequency (a) acoustic resistance and (b) reactance for the three grid resolutions compared with experimental results by \cite{Jones2004} at $M=0$. The results are scaled with porosity for comparison.}
    \label{fig:validation_impedance}
\end{figure}

\red{Surface data for the estimation of the impedance are sampled at a frequency of $20$~kHz. The flow field is, on the other hand, sampled such to have $720$ point per wavelength. For each configurations, acoustic waves with $10$ acoustic periods are considered based on the findings from previous studies from the \citet{Manjunath2018,Avallone2019} and DNS results by \citet{Zhang2016}. Data are sampled after convergence of the unsteady field, which is in general obtained after no more than $2$ acoustic periods. }

The two components of the impedance, the resistance, $\text{Re[$Z_f$]}$, and the reactance, $\text{Im[$Z_f$]}$, are shown in figure \ref{fig:validation_impedance}(a-b); they are both scaled with the porosity, $\sigma$, to allow comparison with the reference experiments in the absence of grazing flow.  
As for the turbulent boundary layer, the case with resolution equal to $40$~voxels/$d$ shows converging results and good agreement with the experimental results. 
\red{The stronger dependence of the acoustic resistance on grid resolution is primarily due to the fact that the grazing flow modifies mainly the resistive part of the impedance, whereas the reactive part is largely controlled by the resonator geometry i.e. height. In particular, refining the mesh near the wall improves the resolution within the boundary layer and the near-orifice flow field, which directly affects the resistance. 
This behaviour can be interpreted in two complementary ways. First, following the theoretical derivation by  \cite{PantonResonantresonators1975}, a Helmholtz resonator can be modelled as a mass–spring–damper system. In that framework, the liner reactance is dominated by the term $\cot(k d_c)$ (with $k$ the acoustic wavenumber and $d_c$ the cavity depth), while grazing-flow effects enter mainly through secondary terms (e.g., via orifice end corrections). Second, consistent with the semi-empirical model of \citet{Yu2008ValidationData}, the reactance depends predominantly on SPL and geometry, whereas grazing flow primarily influences the end-correction and resistive terms, hence affecting the resistance more strongly than the reactance.}

\subsection{Impedance as function of SPL}

In figure \fref{fig:Acoustic_impedance} the impedance for the cases with ($\textrm{M}\!=\!0.3$) and without ($\textrm{M}\!=\!0$) grazing turbulent flow for varying SPL of the grazing acoustic wave at a frequency of $2200$~Hz are compared.
The two components of the acoustic impedance, resistance and reactance, averaged out of the $7$ orifices and scaled by the liner porosity $\sigma$, are plotted.

\begin{figure}[htpb]
    \centering
    \includegraphics[width=0.75\textwidth]{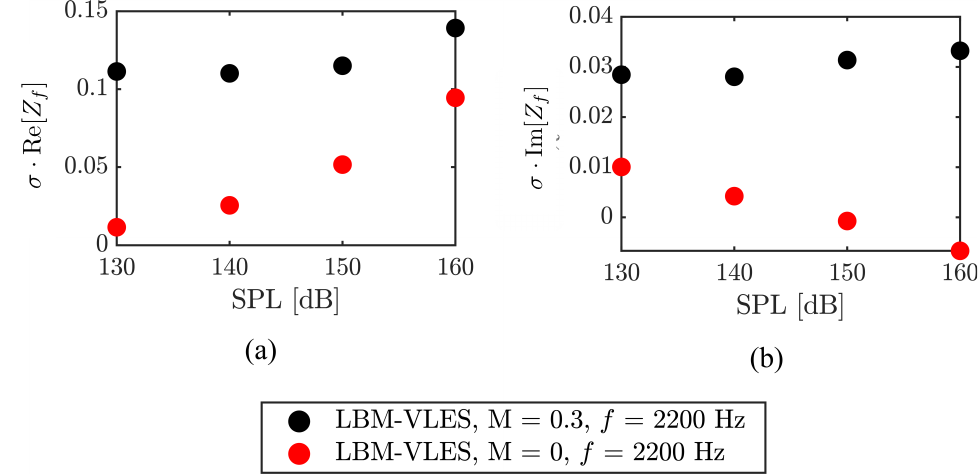}
    \caption{(left) Resistance and (right) reactance components of impedance scaled with the liner porosity $\sigma$. The frequency of the acoustic wave is 2200~Hz, while the amplitude varies between 130 and 160~dB.}
    \label{fig:Acoustic_impedance}
\end{figure}

The results show that, with increasing SPL, the resistance grows more significantly in the absence of flow compared to the case with grazing turbulent flow. Specifically, while the resistance exhibits an exponential increase without flow, it remains nearly constant up to $150$~dB in the presence of flow, only starting to increase at higher SPL. Additionally, it is observed that, in the presence of grazing flow, the amplitude of the resistance is comparable to the maximum value attained in the no-flow case. The grazing flow also affects the reactance: for $\textrm{M}\!=\!0$, it decreases with increasing SPL, whereas for $\textrm{M}\!=\!0.3$, it slightly increases with SPL.
These results clearly demonstrate that the acoustic impedance is significantly altered by the presence of a grazing turbulent flow. This behavior may be related to modifications in the acoustic-induced flow topology and the associated dissipation mechanisms.

\section{Evaluation of the impinging acoustic energy}\label{appB}
The acoustic power used to normalize the dissipation terms in \fref{fig:energy_budget} and \ref{fig:EEi_f} is computed by integrating the fluctuating pressure field along the entire domain height, as illustrated in \fref{fig:Ei_scheme}. This differs from the definition adopted by \citet{tam_microfluid_2000}, where the orifice surface was used as the reference dimension.


\red{

\begin{figure}[ht]
\centering
\includegraphics[width=.8\textwidth]{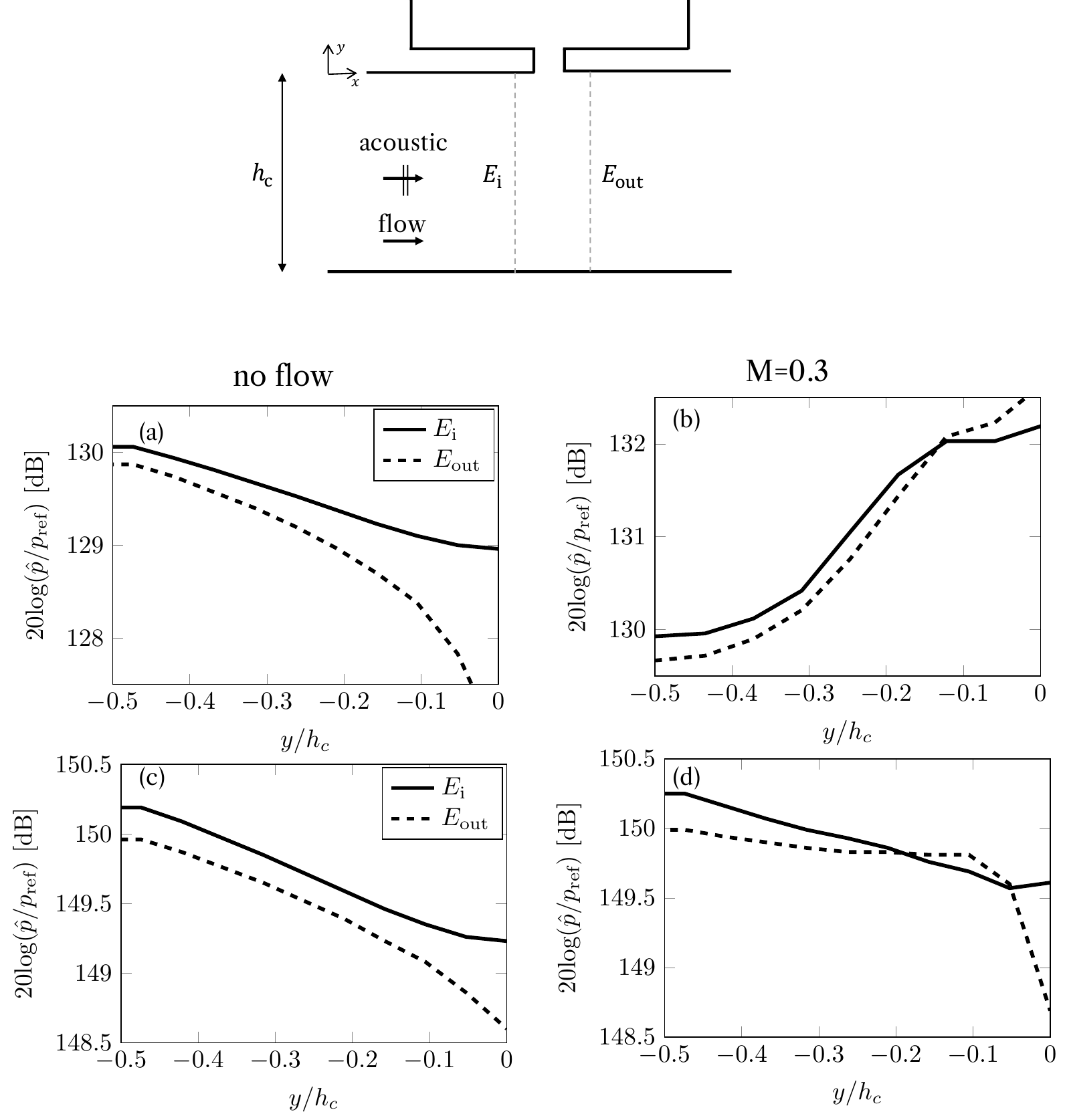}
\caption{Methodology for evaluating the impinging acoustic energy $E_\text{i}$ [W/m] and downstream energy $E_\text{out}$ along the domain height. Energy distributions in [dB] are reported for (a,b) SPL = 130 dB without and with grazing flow, and (c,d) SPL = 150 dB.}
\label{fig:Ei_scheme}
\end{figure}

In the two-dimensional configuration, the acoustic pressure associated with the
incident wave varies along the wall-normal direction. Denoting by $\hat{p}_i(y)$ the incident pressure amplitude at a given
wall-normal location $y$, the local time-averaged acoustic intensity of a harmonic plane
wave is evaluated as
\begin{equation}
\overline{I}_i(y) = \frac{\hat{p}_i^{\,2}(y)}{2\rho c},
\end{equation}
where $\rho$ is the fluid density and $c$ is the speed of sound. The impinging acoustic
quantity denoted $E_{\text{i}}$ is then obtained by
integrating the local acoustic intensity along the wall-normal direction on an upstream
control plane normal to the streamwise direction,
\begin{equation}
E_{\text{i}}
= \int_{0}^{h_c} \overline{I}_i(y)\,\mathrm{d}y
= \int_{0}^{h_c} \frac{\hat{p}_i^{\,2}(y)}{2\rho c}\,\mathrm{d}y.
\end{equation}
This quantity has units of $\mathrm{W/m}$ and represents the
time-averaged acoustic power per unit span, i.e.\ an acoustic energy flux per unit time and per unit span. For the three-dimensional evaluation, the corresponding impinging acoustic power is obtained by multiplying
this value by the spanwise extent of the 3D computational domain, yielding the total impinging acoustic power in $\mathrm{W}$.
This three-dimensional value is used to normalize the dissipation terms
reported in figures~\ref{fig:energy_budget} and~\ref{fig:EEi_f}, as well as
in the full 3D analysis.

The same formulation can be applied on a downstream control plane to evaluate the transmitted acoustic power $E_{\text{out}}$. 
}
With this definition, the impinging power $E_\text{i}$ upstream of the orifice and the outgoing $E_\text{out}$ downstream of the orifice can be consistently compared \red{to highlight the modification of the pressure field due to the presence of the orifice}. Shifting the evaluation planes within two orifice diameters upstream or downstream does not affect the results, confirming the robustness of the method.


Representative wall-normal profiles of the \red{pressure fluctuation amplitude} (normalized with the reference pressure and reported in dB) are shown in \fref{fig:Ei_scheme} for SPL $=130$ and $150$ dB for the sake of conciseness. In the absence of grazing flow (subfigures (a,c)), the distribution resembles an acoustic boundary layer: energy peaks at the domain centreline and decreases towards the wall. 
Downstream of the orifice, the profiles exhibit the expected attenuation due to dissipation.

When a grazing flow is introduced at 130 dB (subfigure (b)), the profiles are markedly altered. The acoustic energy increases towards the wall, reflecting the contribution of turbulence-induced pressure fluctuations in the boundary layer. Downstream of the orifice, the energy decreases in at the centre of the domain but slightly increases near the wall due to the interaction of the shear layer with the orifice jet.

At 150 dB with grazing flow (subfigure (d)), the profiles more closely resemble the no-flow case: the acoustic source dominates over the background turbulence, yielding a higher acoustic-to-aerodynamic fluctuation ratio, in agreement with \citet{Scarano_decompositions}. Interestingly, downstream of the orifice the energy distribution exhibits a local maximum not at the wall but around $y/h_c \approx -0.1$, suggesting that the stronger jetting effect at this SPL penetrates deeper into the grazing flow and locally amplifies the pressure fluctuations.

\section{Acoustic-induced velocity normalized with respect to the lumped element model of the Helmholtz resonator}\label{appC}

\begin{figure}[ht]
\centering
\includegraphics[width=.55\textwidth]{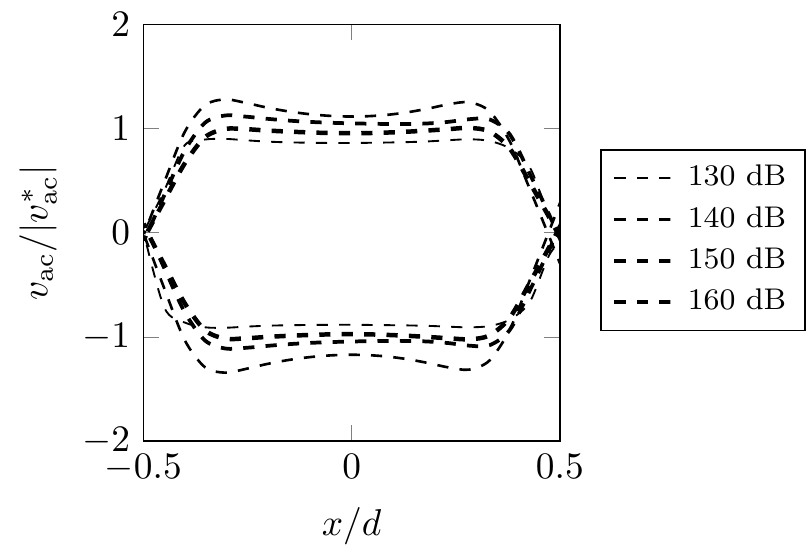}
\caption{\red{Spatial distribution along the diameter at half face-sheet thickness of the non-dimensional acoustic-induced vertical velocity normalized with the lumped element velocity. Thicker lines represent larger SPL values. Inflow and outflow phases are reported.}}
\label{fig:vac_vstar}
\end{figure}

\red{To give additional confidence that the wall-modelled LB–VLES framework correctly
reproduces both the order of magnitude and the peak value of the
acoustic-induced velocity, the profiles at the centre of the orifice for the no-flow case are normalized with respect to the lumped element model of the Helmholtz resonator \citep{Morse1968}:
\begin{equation}
v^*_{ac} = \frac{\hat{p}}{\rho \omega (\tau+0.8d)}
\frac{1}{\sqrt{ \left [(\omega_H/\omega)^2 - 1 \right ]^2 + (\omega_H/\omega Q)^2 }},
\end{equation}
where $\hat{p}$ is the pressure fluctuation amplitude, $\omega$ and $\omega_H$ are the forcing and resonant frequencies, $Q$ is the quality factor (taken as $10$), $\rho$ is the density, and $d$ and $\tau$ are the diameter and thickness of the orifice, respectively. This expression gives theoretical peak velocities for Helmholtz resonator in the absence of grazing flow. The values obtained are $3.6$, $9.5$, $30.2$, and $67.6\,\mathrm{m/s}$ for the four tested SPLs.

The normalized acoustic velocity, reported in \fref{fig:vac_vstar} is observed to approach unity across the
investigated SPL range, indicating that the numerical framework accurately
captures the expected oscillatory velocity amplitude inside the orifice.
This agreement provides additional confidence that the wall model correctly
predicts the near-wall gradients associated with the acoustic motion within the orifice.}

\red{
\section{Integration region for viscous dissipation}\label{appD}

The integration region used for the viscous dissipation in 3D is illustrated in \fref{fig:integration_viscous} (a). In the three-dimensional formulation,
this region forms a toroidal volume surrounding the internal wall of the
orifice, corresponding to a near-wall band where viscous stresses are
dominant (\fref{fig:integration_viscous} (b)).

\begin{figure}[htpb]
    \centering
    \includegraphics[width=1.\textwidth]{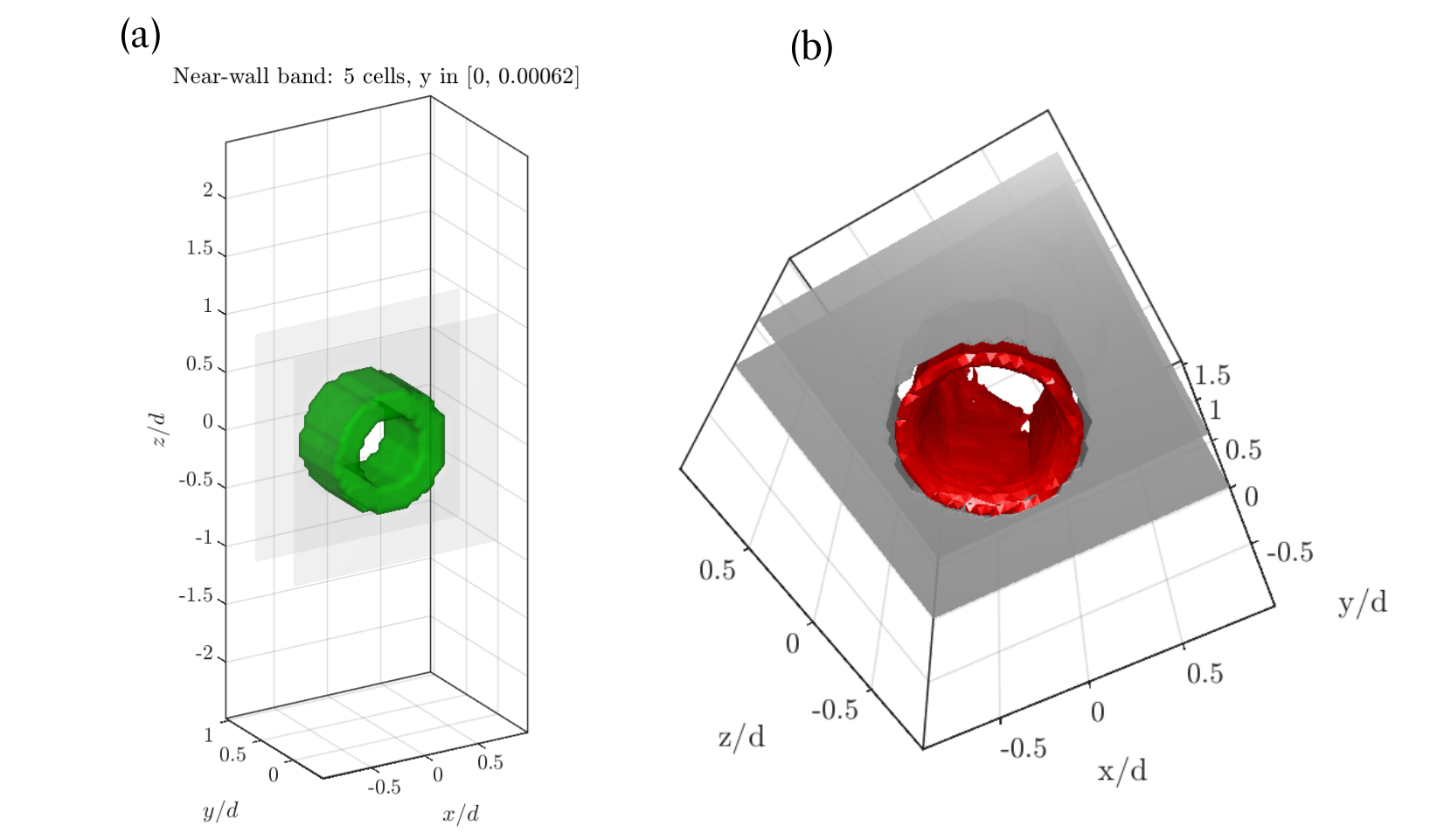}
    \caption{\red{(a) 3D region where the shear stress is integrated close to the orifice internal wall, (b) iso-surface $\Phi^+=0.1$ of the stress tensor at M=0 from another perspective.}}
    \label{fig:integration_viscous}
\end{figure}

}

\end{appen}\clearpage

\newpage
\bibliographystyle{jfm}
\bibliography{jfm}

@inproceedings{Avallone2021Acoustic-inducedLayer,
    title = {{Acoustic-induced velocity in a multi-orifice acoustic liner grazed by a turbulent boundary layer}},
    year = {2021},
    booktitle = {AIAA Aviation and Aeronautics Forum and Exposition, AIAA AVIATION Forum 2021},
    author = {Avallone, F. and Casalino, D.},
    publisher = {American Institute of Aeronautics and Astronautics Inc, AIAA},
    isbn = {9781624106101},
    doi = {10.2514/6.2021-2169}
}

@techreport{Dean1974AnDucts,
    title = {{An in situ method of wall acoustic impedance measurement in flow ducts}},
    year = {1974},
    booktitle = {Journal of Sound and Vibration},
    author = {Dean, P D},
    number = {1},
    pages = {97--130},
    volume = {34}
}

@article{Quintino2025,
    title = {{Comparison of Impedance Eduction Test Rigs with Different Boundary-Layer Profiles}},
    year = {2025},
    journal = {AIAA Journal},
    author = {Quintino, N.T. and Bonomo, L.A. and Cordioli, J.A. and Jones, M.G. and Howerton, B.M. and Nark, D.M and Avallone, F.},
    doi = {10.2514/1.J065173},
}

@article{Tam2010,
    title = {{A computational and experimental study of resonators in three dimensions}},
    year = {2009},
    journal = {Journal of Sound and Vibration},
    author = {Tam, C.K.W. and Ju, H. and Jones, M.G. and Watson, W.R. and Parrott, T.L.},
    pages = {5164--5193},
    volume = {329},
    url = {www.elsevier.com/locate/jsvi https://arc-aiaa-org.tudelft.idm.oclc.org/doi/abs/10.2514/6.2009-3171},
    doi = {10.2514/6.2009-3171}
}

@article{RONCEN2025119058,
title = {Revisiting nonlinear impedance in acoustic liners},
journal = {Journal of Sound and Vibration},
volume = {608},
pages = {119058},
year = {2025},
issn = {0022-460X},
doi = {https://doi.org/10.1016/j.jsv.2025.119058},
url = {https://www.sciencedirect.com/science/article/pii/S0022460X25001324},
author = {Rémi Roncen}
}

@article{Dean1974,
    title = {{An in situ method of wall acoustic impedance measurement in flow ducts}},
    year = {1974},
    journal = {Journal of Sound and Vibration},
    author = {Dean, P.D.},
    number = {1},
    month = {5},
    pages = {97-IN6},
    volume = {34},
    publisher = {Academic Press},
    doi = {10.1016/S0022-460X(74)80357-3}
}

@inproceedings{Mann2013,
    title = {{Characterization of Acoustic Liners Absorption using a Lattice-Boltzmann Method}},
    year = {2013},
    booktitle = {19th AIAA/CEAS Aeroacoustics Conference},
    author = {Mann, A. and Perot, F. and Kim, M.-S. and Casalino, D.},
    pages = {2013--2271},
    url = {http://arc.aiaa.org/doi/10.2514/6.2013-2271},
    isbn = {978-1-62410-213-4},
    doi = {10.2514/6.2013-2271}
}

@inproceedings{Manjunath2018,
    title = {{Characterization of Liners using a Lattice-Boltzmann Solver}},
    year = {2018},
    booktitle = {2018 AIAA/CEAS Aeroacoustics Conference},
    author = {Manjunath, P. and Avallone, F. and Casalino, D. and Ragni, D. and Snellen, M.},
    month = {6},
    pages = {2018--4192},
    url = {https://arc.aiaa.org/doi/10.2514/6.2018-4192},
    isbn = {978-1-62410-560-9},
    doi = {10.2514/6.2018-4192}
}

@inproceedings{Jones2004,
    title = {{Design and Evaluation of Modifications to the NASA Langley Flow Impedance Tube}},
    year = {2004},
    booktitle = {10th AIAA/CEAS Aeroacoustics Conference},
    author = {Jones, M. and Watson, W. and Parrott, T.L. and Smith, C.},
    month = {5},
    pages = {2004--2837},
    publisher = {American Institute of Aeronautics and Astronautics},
    url = {http://arc.aiaa.org/doi/10.2514/6.2004-2837},
    address = {Reston, Virigina},
    isbn = {978-1-62410-071-0},
    doi = {10.2514/6.2004-2837}
}

@incollection{Motsinger1991,
    title = {{Design and performance of duct acoustic treatment}},
    year = {1991},
    booktitle = {Aeroacoustics of Flight Vehicles: Theory and Practice. Volume 2: Noise Control},
    author = {Motsinger, R.E. and Kraft, R.E.},
    editor = {Hubbard, H.H.},
    chapter = {14},
    month = {8},
    pages = {165--206},
    publisher = {NASA},
    url = {https://ntrs.nasa.gov/search.jsp?R=19920005565}
}

@article{Yakhot1992,
    title = {{Development of turbulence models for shear flows by a double expansion technique}},
    year = {1992},
    journal = {Physics of Fluids A},
    author = {Yakhot, V. and Orszag, S.A. and Thangam, S. and Gatski, T.B. and Speziale, C.G.},
    number = {7},
    month = {7},
    pages = {1510--1520},
    volume = {4},
    publisher = {American Institute of PhysicsAIP},
    url = {http://aip.scitation.org/doi/10.1063/1.858424},
    doi = {10.1063/1.858424},
    issn = {08998213}
}

@inproceedings{Zhang2016a,
    title = {{Direct numerical investigation of acoustic liners with single and multiple orifices grazed by a Mach 0.5 boundary layer}},
    year = {2016},
    booktitle = {46th AIAA Fluid Dynamics Conference},
    author = {Zhang, Q. and Bodony, D.J.},
    month = {6},
    publisher = {American Institute of Aeronautics and Astronautics},
    url = {http://arc.aiaa.org/doi/10.2514/6.2016-3626},
    address = {Reston, Virginia},
    isbn = {978-1-62410-436-7},
    doi = {10.2514/6.2016-3626}
}

@inproceedings{Hazir2017,
    title = {{Effect of Temperature Variations on the Acoustic Properties of Engine Liners}},
    year = {2017},
    booktitle = {23rd AIAA/CEAS Aeroacoustics Conference},
    author = {Hazir, A.G. and Casalino, D.},
    pages = {2017--3874},
    url = {https://arc.aiaa.org/doi/10.2514/6.2017-3874},
    isbn = {978-1-62410-504-3},
    doi = {10.2514/6.2017-3874}
}

@inproceedings{Jones2010,
    title = {{Effects of Flow Profile on Educed Acoustic Liner Impedance}},
    year = {2010},
    booktitle = {16th AIAA/CEAS Aeroacoustics Conference},
    author = {Jones, Michael and Watson, Willie and Nark, Douglas},
    month = {6},
    pages = {2010--3763},
    publisher = {American Institute of Aeronautics and Astronautics},
    url = {https://arc.aiaa.org/doi/10.2514/6.2010-3763},
    address = {Reston, Virigina},
    isbn = {978-1-60086-955-6},
    doi = {10.2514/6.2010-3763}
}

@article{Zhang2006,
    title = {{Efficient kinetic method for fluid simulation beyond the Navier-Stokes equation}},
    year = {2006},
    journal = {Physical Review E - Statistical, Nonlinear, and Soft Matter Physics},
    author = {Zhang, R. and Shan, X. and Chen, H.},
    number = {4},
    month = {10},
    pages = {046703},
    volume = {74},
    publisher = {American Physical Society},
    url = {https://journals.aps.org/pre/abstract/10.1103/PhysRevE.74.046703},
    doi = {10.1103/PhysRevE.74.046703},
    issn = {15393755},
    arxivId = {physics/0604191}
}

@article{Chen2004,
    title = {{Expanded analogy between Boltzmann kinetic theory of fluids and turbulence}},
    year = {2004},
    journal = {Journal of Fluid Mechanics},
    author = {Chen, H and Orszag, S.A. and Staroselsky, I. and Succi, S.},
    month = {11},
    pages = {301--314},
    volume = {519},
    publisher = {Cambridge University Press},
    url = {https://doi.org/10.1017/S0022112004001211},
    doi = {10.1017/S0022112004001211},
    issn = {00221120}
}

@article{Tam2014,
    title = {{Experimental validation of numerical simulations for an acoustic liner in grazing flow: Self-noise and added drag}},
    year = {2014},
    journal = {Journal of Sound and Vibration},
    author = {Tam, C.K.W. and Pastouchenko, N.N. and Jones, M.G. and Watson, W.R.},
    number = {13},
    month = {6},
    pages = {2831--2854},
    volume = {333},
    publisher = {Academic Press},
    url = {https://www.sciencedirect.com/science/article/pii/S0022460X1400131X},
    doi = {10.1016/J.JSV.2014.02.019},
    issn = {0022-460X}
}

@article{Spillere2020,
    title = {{Experimentally testing impedance boundary conditions for acoustic liners with flow: Beyond upstream and downstream}},
    year = {2020},
    journal = {Journal of Sound and Vibration},
    author = {Spillere, A.M.N. and Bonomo, L.A. and Cordioli, J.A. and Brambley, E..},
    month = {12},
    pages = {115676},
    volume = {489},
    publisher = {Academic Press},
    url = {https://linkinghub.elsevier.com/retrieve/pii/S0022460X2030506X},
    doi = {10.1016/j.jsv.2020.115676},
    issn = {0022460X}
}

@article{Winkler2021,
    title = {{High fidelity modeling tools for engine liner design and screening of advanced concepts:}},
    year = {2021},
    journal = {International Journal of Aeroacoustics},
    author = {Winkler, J. and Mendoza, J.M. and Reimann, C.A. and Homma, K. and Alonso, J.S.},
    number = {0},
    month = {8},
    pages = {1--31},
    volume = {0},
    publisher = {SAGE PublicationsSage UK: London, England},
    url = {https://journals.sagepub.com/doi/full/10.1177/1475472X211023884},
    doi = {10.1177/1475472X211023884}
}

@article{Teixeira1998,
    title = {{Incorporating Turbulence Models into the Lattice-Boltzmann Method}},
    year = {1998},
    journal = {International Journal of Modern Physics C},
    author = {Teixeira, C.M.},
    number = {08},
    month = {12},
    pages = {1159--1175},
    volume = {09},
    publisher = {World Scientific Publishing Company},
    doi = {10.1142/S0129183198001060}
}

@article{GABARD201630,
title = {Boundary layer effects on liners for aircraft engines},
journal = {Journal of Sound and Vibration},
volume = {381},
pages = {30-47},
year = {2016},
issn = {0022-460X},
doi = {https://doi.org/10.1016/j.jsv.2016.06.032},
author = {Gwénaël Gabard}
}

@article{Qian1992LatticeEquation,
    title = {{Lattice bgk models for navier-stokes equation}},
    year = {1992},
    journal = {EPL},
    author = {Qian, Y. H. and D’Humi{\`{e}}res, D. and Lallemand, P.},
    number = {6},
    volume = {17},
    doi = {10.1209/0295-5075/17/6/001},
    issn = {12864854}
}

@inproceedings{Yu2008ValidationData,
    title = {{Validation of Goodrich perforate liner impedance model using NASA langley test data}},
    year = {2008},
    booktitle = {14th AIAA/CEAS Aeroacoustics Conference (29th AIAA Aeroacoustics Conference)},
    author = {Yu, J. and Ruiz, M. and Kwan, H. W.},
    isbn = {9781563479397},
    doi = {10.2514/6.2008-2930}
}

@article{Shan2006a,
    title = {{Kinetic theory representation of hydrodynamics: a way beyond the Navier–Stokes equation}},
    year = {2006},
    journal = {Journal of Fluid Mechanics},
    author = {Shan, X. and Yuan, X.-F. and Chen, H.},
    pages = {413--441},
    volume = {550},
    publisher = {Cambridge University Press},
    doi = {10.1017/S0022112005008153}
}

@article{Chen1998a,
    title = {{Lattice Boltzmann method for fluid flows}},
    year = {1998},
    journal = {Annual Review of Fluid Mechanics},
    author = {Chen, S. and Doolen, G.D.},
    number = {1},
    month = {1},
    pages = {329--364},
    volume = {30},
    publisher = {Annual Reviews 4139 El Camino Way, P.O. Box 10139, Palo Alto, CA 94303-0139, USA},
    doi = {10.1146/annurev.fluid.30.1.329},
    issn = {0066-4189}
}

@inproceedings{Avallone2019,
    title = {{Lattice-Boltzmann Very Large Eddy Simulation of a Multi-Orifice Acoustic Liner with Turbulent Grazing Flow}},
    year = {2019},
    booktitle = {25th AIAA/CEAS Aeroacoustics Conference},
    author = {Avallone, F. and Manjunath, P. and Ragni, D. and Casalino, D.},
    month = {5},
    pages = {2019--2542},
    publisher = {American Institute of Aeronautics and Astronautics},
    url = {https://arc.aiaa.org/doi/10.2514/6.2019-2542},
    address = {Reston, Virginia},
    isbn = {978-1-62410-588-3},
    doi = {10.2514/6.2019-2542}
}

@inproceedings{Auregan2004,
    title = {{Measurement of liner impedance with flow by an inverse method}},
    year = {2004},
    booktitle = {10th AIAA/CEAS Aeroacoustics Conference},
    author = {Aur{\'{e}}gan, Y. and Leroux, M. and Pagneux, V.},
    pages = {2004--2838},
    volume = {1},
    publisher = {American Institute of Aeronautics and Astronautics Inc.},
    url = {https://arc-aiaa-org.tudelft.idm.oclc.org/doi/abs/10.2514/6.2004-2838},
    address = {Manchester, Great Britain},
    isbn = {1563477130},
    doi = {10.2514/6.2004-2838}
}

@article{Leon2019,
    title = {{Near-wall aerodynamic response of an acoustic liner to harmonic excitation with grazing flow}},
    year = {2019},
    journal = {Experiments in Fluids},
    author = {L{\'{e}}on, Olivier and M{\'{e}}ry, Fabien and Piot, Estelle and Conte, Claudia},
    number = {9},
    month = {9},
    pages = {144},
    volume = {60},
    publisher = {Springer Verlag},
    url = {http://link.springer.com/10.1007/s00348-019-2791-5},
    doi = {10.1007/s00348-019-2791-5},
    issn = {0723-4864}
}

@article{Zhang2012,
    title = {{Numerical investigation and modelling of acoustically excited flow through a circular orifice backed by a hexagonal cavity}},
    year = {2012},
    journal = {Journal of Fluid Mechanics},
    author = {Zhang, Q. and Bodony, D.J.},
    month = {2},
    pages = {367--401},
    volume = {693},
    publisher = {Cambridge University Press},
    url = {http://www.journals.cambridge.org/abstract_S0022112011005374},
    doi = {10.1017/jfm.2011.537}
}

@article{Zhang2016,
    title = {{Numerical investigation of a honeycomb liner grazed by laminar and turbulent boundary layers}},
    year = {2016},
    journal = {Journal of Fluid Mechanics},
    author = {Zhang, Qi and Bodony, Daniel J.},
    month = {4},
    pages = {936--980},
    volume = {792},
    publisher = {Cambridge University Press},
    url = {https://www.cambridge.org/core/product/identifier/S0022112016000793/type/journal_article},
    doi = {10.1017/jfm.2016.79},
    issn = {0022-1120}
}

@article{Myers1980,
    title = {{On the acoustic boundary condition in the presence of flow}},
    year = {1980},
    journal = {Journal of Sound and Vibration},
    author = {Myers, M.K. K},
    number = {3},
    month = {8},
    pages = {429--434},
    volume = {71},
    publisher = {Academic Press},
    doi = {10.1016/0022-460X(80)90424-1},
    issn = {0022460x}
}

@article{Chen2014,
    title = {{Recovery of Galilean invariance in thermal lattice Boltzmann models for arbitrary Prandtl number}},
    year = {2014},
    journal = {International Journal of Modern Physics C},
    author = {Chen, H. and Gopalakrishnan, P. and Zhang, R.},
    number = {10},
    month = {10},
    pages = {1450046},
    volume = {25},
    publisher = {World Scientific Publishing Co. Pte Ltd},
    doi = {10.1142/S0129183114500466},
    issn = {01291831},
    arxivId = {1403.2357}
}

@article{DeGraaff2000,
    title = {{Reynolds-number scaling of the flat-plate turbulent boundary layer}},
    year = {2000},
    journal = {Journal of Fluid Mechanics},
    author = {De Graaff, D.B. and Eaton, J.K.},
    month = {11},
    pages = {319--346},
    volume = {422},
    publisher = {Cambridge University Press},
    doi = {10.1017/S0022112000001713},
    issn = {00221120}
}

@article{FERNHOLZ1996245,
title = {The incompressible zero-pressure-gradient turbulent boundary layer: An assessment of the data},
journal = {Progress in Aerospace Sciences},
volume = {32},
number = {4},
pages = {245-311},
year = {1996},
issn = {0376-0421},
doi = {https://doi.org/10.1016/0376-0421(95)00007-0},
url = {https://www.sciencedirect.com/science/article/pii/0376042195000070},
author = {H.H. Fernholz and P.J. Finleyt}
}

@article{Nagib,
author = {Nagib, Hassan and Chauhan, Kapil and Monkewitz, Peter},
year = {2007},
month = {01},
pages = {755-70},
title = {Approach to an asymptotic state for zero pressure gradient turbulent boundary layers},
volume = {365},
journal = {Philosophical transactions. Series A, Mathematical, physical, and engineering sciences},
doi = {10.1098/rsta.2006.1948}
}

@phdthesis{Osterlund,
author = {Osterlund, Jens},
year = {2000},
month = {07},
school ={KTH},
pages = {},
title = {Experimental Studies of Zero Pressure-Gradient Turbulent Boundary Layer Flow}
}

@book{Morse1968,
    title = {{Theoretical Acoustics}},
    year = {1968},
    author = {Morse, P.M. and Ingard, K.U.},
    publisher = {Princeton University Press},
    isbn = {0691024014}
}

@article{Casalino2017,
    title = {{Turbofan Broadband Noise Prediction Using the Lattice Boltzmann Method}},
    year = {2017},
    journal = {AIAA Journal},
    author = {Casalino, D. and Hazir, A.G. and Mann, A.},
    number = {2},
    month = {9},
    pages = {1--20},
    volume = {56},
    url = {https://arc.aiaa.org/doi/10.2514/1.J055674},
    doi = {10.2514/1.J055674}
}

@article{Vallikivi2015,
    title = {{Turbulent boundary layer statistics at very high Reynolds number}},
    year = {2015},
    journal = {Journal of Fluid Mechanics},
    author = {Vallikivi, M. and Hultmark, M. and Smits, A.J.},
    pages = {371--389},
    volume = {779},
    publisher = {Cambridge University Press},
    doi = {10.1017/jfm.2015.273}
}

@article{Shur2021,
    title = {{Unsteady Simulations of Sound Propagation in Turbulent Flow Inside a Lined Duct}},
    year = {2021},
    journal = {AIAA Journal},
    author = {Shur, M. and Strelets, M. and Travin, A. and Suzuki, T. and Spalart, P.},
    month = {5},
    pages = {1--17},
    publisher = {American Institute of Aeronautics and Astronautics},
    url = {https://arc.aiaa.org/doi/10.2514/1.J060181},
    doi = {10.2514/1.J060181},
    issn = {0001-1452}
}

@techreport{Baumeister1975,
    title = {{Visual study of the effect of grazing flow on the oscillatory flow in a resonator orifice}},
    year = {1975},
    author = {Baumeister, K.J. and Rice, E.J.},
    month = {11},
    institution = {NASA Lewis Research Center },
    address = {Cleveland, OH, United States}
}

@article{yoshikawa_experimental_2012,
	title = {Experimental examination of vortex-sound generation in an organ pipe: {A} proposal of jet vortex-layer formation model},
	volume = {331},
	copyright = {https://www.elsevier.com/tdm/userlicense/1.0/},
	issn = {0022460X},
	shorttitle = {Experimental examination of vortex-sound generation in an organ pipe},
	url = {https://linkinghub.elsevier.com/retrieve/pii/S0022460X12000739},
	doi = {10.1016/j.jsv.2012.01.026},
	language = {en},
	number = {11},
	urldate = {2024-06-14},
	journal = {Journal of Sound and Vibration},
	author = {Yoshikawa, Shigeru and Tashiro, Hiromi and Sakamoto, Yumiko},
	month = may,
	year = {2012},
	pages = {2558--2577},
}

@article{tabata_three-dimensional_2021,
	title = {Three-dimensional numerical analysis of acoustic energy absorption and generation in an air-jet instrument based on {Howe}'s energy corollary},
	volume = {149},
	issn = {0001-4966, 1520-8524},
	url = {https://pubs.aip.org/jasa/article/149/6/4000/1059316/Three-dimensional-numerical-analysis-of-acoustic},
	doi = {10.1121/10.0005133},
	language = {en},
	number = {6},
	urldate = {2024-05-15},
	journal = {The Journal of the Acoustical Society of America},
	author = {Tabata, Ryoya and Matsuda, Rei and Koiwaya, Toshiaki and Iwagami, Sho and Midorikawa, Hiroko and Kobayashi, Taizo and Takahashi, Ki'nya},
	month = jun,
	year = {2021},
	pages = {4000--4012},
}

@article{howe_dissipation_1980,
	title = {The dissipation of sound at an edge},
	volume = {70},
	copyright = {https://www.elsevier.com/tdm/userlicense/1.0/},
	issn = {0022460X},
	url = {https://linkinghub.elsevier.com/retrieve/pii/0022460X80903089},
	doi = {10.1016/0022-460X(80)90308-9},
	language = {en},
	number = {3},
	urldate = {2024-05-15},
	journal = {Journal of Sound and Vibration},
	author = {Howe, M.S.},
	month = jun,
	year = {1980},
	pages = {407--411},
}

@article{howe_absorption_1984,
	title = {On the {Absorption} of {Sound} by {Turbulence} and {Other} {Hydrodynamic} {Flows}},
	volume = {32},
	issn = {0272-4960, 1464-3634},
	url = {https://academic.oup.com/imamat/article-lookup/doi/10.1093/imamat/32.1-3.187},
	doi = {10.1093/imamat/32.1-3.187},
	language = {en},
	number = {1-3},
	urldate = {2024-05-23},
	journal = {IMA Journal of Applied Mathematics},
	author = {Howe, M. S.},
	year = {1984},
	pages = {187--209},}

@article{howe_contributions_1975,
	title = {Contributions to the theory of aerodynamic sound, with application to excess jet noise and the theory of the flute},
	volume = {71},
	copyright = {https://www.cambridge.org/core/terms},
	issn = {0022-1120, 1469-7645},
	url = {https://www.cambridge.org/core/product/identifier/S0022112075002777/type/journal_article},
	doi = {10.1017/S0022112075002777},
	language = {en},
	number = {4},
	urldate = {2024-07-05},
	journal = {Journal of Fluid Mechanics},
	author = {Howe, M. S.},
	month = oct,
	year = {1975},
	pages = {625--673},
}

@article{tam_microfluid_2000,
	title = {Microfluid {Dynamics} and {Acoustics} of {Resonant} {Liners}},
	volume = {38},
	issn = {0001-1452, 1533-385X},
	url = {https://arc.aiaa.org/doi/10.2514/2.1132},
	doi = {10.2514/2.1132},
	language = {en},
	number = {8},
	urldate = {2024-05-20},
	journal = {AIAA Journal},
	author = {Tam, Christopher K. W. and Kurbatskii, Konstantin A.},
	month = aug,
	year = {2000},
	pages = {1331--1339},
}

@article{Zhang_Bodony_2012, title={Numerical investigation and modelling of acoustically excited flow through a circular orifice backed by a hexagonal cavity}, volume={693}, DOI={10.1017/jfm.2011.537}, journal={Journal of Fluid Mechanics}, author={Zhang, Qi and Bodony, Daniel J.}, year={2012}, pages={367–401}}

@article{tang_piv_2024,
	title = {{PIV} measurements of coherent vortices and turbulence production inside acoustic liner cavity with offset slit},
	volume = {154},
	issn = {08941777},
	url = {https://linkinghub.elsevier.com/retrieve/pii/S0894177724000268},
	doi = {10.1016/j.expthermflusci.2024.111157},
	language = {en},
	urldate = {2024-04-10},
	journal = {Experimental Thermal and Fluid Science},
	author = {Tang, Yuchao and Wang, Peng and Liu, Yingzheng},
	month = may,
	year = {2024},
	pages = {111157},
}

@inproceedings{Lyu2024,
title={Canonical correlation decomposition of numerical and experimental data for observable diagnosis},
author = {B. Lyu},
booktitle={30th AIAA/CEAS Aeroacoustics Conference},
pages={AIAA 2024-3206},
year={2024}
}

@article{Scarano_decompositions,
title = {Filtering acoustic from hydrodynamic velocity using modal decomposition methods on an acoustic liner under grazing turbulent flow},
journal = {Journal of Sound and Vibration},
volume = {625},
pages = {119568},
year = {2026},
issn = {0022-460X},
doi = {https://doi.org/10.1016/j.jsv.2025.119568},
url = {https://www.sciencedirect.com/science/article/pii/S0022460X25006418},
author = {Francesco Scarano and Benshuai Lyu and Angelo Paduano and Francesco Avallone}
}

@techreport{Mallat1989ARepresentation,
    title = {{A Theory for Multiresolution Signal Decomposition: The Wavelet Representation}},
    year = {1989},
    booktitle = {IEEE TRANSACTIONS ON PATTERN ANALYSIS AND MACHINE INTELLIGENCE},
    author = {Mallat, Stephane G},
    number = {7},
    volume = {I},
    doi = {http://repository.upenn.edu/cis reports/668}
}

@article{Hughes2011TheIntegration,
    title = {{The Promise and Challenges of Ultra High Bypass Ratio Engine Technology and Integration}},
    year = {2011},
    journal = {AIAA Aero Sciences Meeting},
    author = {Hughes, Chris},
    pages = {0--11},
    url = {www.nasa.gov http://hdl.handle.net/2060/20110011737}
}

@article{Casalino2018TurbofanMethod,
    title = {{Turbofan broadband noise prediction using the lattice boltzmann method}},
    year = {2018},
    journal = {AIAA Journal},
    author = {Casalino, D. and Hazir, A. and Mann, A.},
    number = {2},
    volume = {56},
    doi = {10.2514/1.J055674},
    issn = {00011452}
}

@techreport{Motsinger19914Treatment,
    title = {{4 Design and Performance of Duct Acoustic Treatment}},
    year = {1991},
    author = {Motsinger, R E and Kraft, R E}
}

@article{PantonResonantresonators1975,
    author = {Panton, Ronald L. and Miller, John M.},
    title = {Resonant frequencies of cylindrical Helmholtz resonators},
    journal = {The Journal of the Acoustical Society of America},
    volume = {57},
    number = {6},
    pages = {1533-1535},
    year = {1975},
    month = {06},
    issn = {0001-4966},
    doi = {10.1121/1.380596},
    url = {https://doi.org/10.1121/1.380596},
    eprint = {https://pubs.aip.org/asa/jasa/article-pdf/57/6/1533/11595089/1533\_1\_online.pdf},
}

@article{paduano2025impact,
  title={On the impact of the turbulent grazing flow development on the acoustic response of an acoustic liner},
  author={Paduano, Angelo and Scarano, Francesco and Cordioli, Julio and Casalino, Damiano and Avallone, Francesco},
  journal={arXiv preprint arXiv:2507.22714},
  year={2025}
}

@techreport{Bonomo2023AProfiles,
    title = {{A Comparison of Impedance Eduction Test Rigs with Different Flow Profiles}},
    year = {2023},
    author = {Bonomo, Lucas A and Quintino, Nicolas T and Cordioli, Julio A and Avallone, Francesco and Jones, Michael G and Howerton, Brian M and Nark, Douglas M},
    doi = {https://doi.org/10.2514/6.2023-3346}
}

@techreport{Melling1973TheLevels,
    title = {{The acoustic impedance of perforates at medium and high sound pressure levels}},
    year = {1973},
    booktitle = {Journal of Sound attd Vibration (I 973)},
    author = {Melling, T H},
    number = {1},
    pages = {1--65},
    volume = {29}
}

@inproceedings{Shahzad2023DirectLiners,
    title = {{Direct Numerical Simulation of a Turbulent Boundary Layer over Acoustic Liners}},
    year = {2023},
    author = {Shahzad, Haris and Hickel, Stefan and Modesti, Davide},
    month = {6},
    publisher = {American Institute of Aeronautics and Astronautics (AIAA)},
    doi = {10.2514/6.2023-3887}
}

@article{KennethBaumeister1975NASAOrifice,
    title = {{NASA TM X-3288 2. Government Accession No. 4. Title and Subtitle visual study of the effect of grazing flow on the oscillatory flow in a resonator orifice}},
    year = {1975},
    author = {Baumeister, Kenneth J. and Rice, Edward J}
}

@article{Leon2019Near-wallFlow,
    title = {{Near-wall aerodynamic response of an acoustic liner to harmonic excitation with grazing flow}},
    year = {2019},
    journal = {Experiments in Fluids},
    author = {L{\'{e}}on, Olivier and M{\'{e}}ry, Fabien and Piot, Estelle and Conte, Claudia},
    number = {9},
    month = {9},
    volume = {60},
    publisher = {Springer Verlag},
    doi = {10.1007/s00348-019-2791-5},
    issn = {14321114}
}

@article{schmidt_guide_2020,
	title = {Guide to {Spectral} {Proper} {Orthogonal} {Decomposition}},
	volume = {58},
	issn = {0001-1452, 1533-385X},
	url = {https://arc.aiaa.org/doi/10.2514/1.J058809},
	doi = {10.2514/1.J058809},
	language = {en},
	number = {3},
	urldate = {2023-04-12},
	journal = {AIAA Journal},
	author = {Schmidt, Oliver T. and Colonius, Tim},
	month = mar,
	year = {2020},
	pages = {1023--1033},
}

@article{schoder_postprocessing_2020,
	title = {Postprocessing of {Direct} {Aeroacoustic} {Simulations} {Using} {Helmholtz} {Decomposition}},
	volume = {58},
	issn = {0001-1452, 1533-385X},
	url = {https://arc.aiaa.org/doi/10.2514/1.J058836},
	doi = {10.2514/1.J058836},
	language = {en},
	number = {7},
	urldate = {2024-01-09},
	journal = {AIAA Journal},
	author = {Schoder, Stefan and Roppert, Klaus and Kaltenbacher, Manfred},
	month = jul,
	year = {2020},
	pages = {3019--3027},
}

@article{unnikrishnan_pressure_2020,
	title = {A pressure decomposition framework for aeroacoustic analysis of turbulent jets},
	volume = {81},
	issn = {09977546},
	url = {https://linkinghub.elsevier.com/retrieve/pii/S0997754619303553},
	doi = {10.1016/j.euromechflu.2020.01.006},
	language = {en},
	urldate = {2023-07-19},
	journal = {European Journal of Mechanics - B/Fluids},
	author = {Unnikrishnan, S. and Gaitonde, Datta V.},
	month = may,
	year = {2020},
	pages = {41--61},
}

@techreport{Camelier1976JetCrossflow,
  author       = {Camelier, Jean and Karamcheti, Krishnamurty},
  title        = {An experimental study of the structure and acoustic field of a jet in a cross stream},
  institution  = {NASA},
  number       = {NASA CR-162464},
  year         = {1976},
  address      = {Washington, D.C.},
  url          = {https://ntrs.nasa.gov/citations/19800007611},
  note         = {NASA Contractor Report 162464}
}

@techreport{Stimpert1973CrossflowNoise,
  author       = {Stimpert, D. L. and Fogg, R. G.},
  title        = {Effect of crossflow velocity on the generation of lift fan jet noise in VTOL aircraft},
  institution  = {NASA},
  number       = {NASA CR-114571},
  year         = {1973},
  address      = {Washington, D.C.},
  url          = {https://ntrs.nasa.gov/citations/19730013202},
  note         = {NASA Contractor Report 114571}
}

@article{Shahzad_2025AIAA,
author = {Shahzad, Haris and Hickel, Stefan and Modesti, Davide},
title = {Direct Numerical Simulation of a Turbulent Boundary Layer over Acoustic Liners},
journal = {AIAA Journal},
volume = {63},
number = {11},
pages = {4650-4661},
year = {2025},
doi = {10.2514/1.J065204},

URL = { 
    
        https://doi.org/10.2514/1.J065204
    
    

},
eprint = { 
    
        https://doi.org/10.2514/1.J065204
    
    

}

}


\end{document}